\documentclass[a4paper,11pt]{article}
\usepackage{jheppub}

\usepackage[most]{tcolorbox}

\usepackage{amsmath}

\usepackage{esint}
\usepackage{babel}
\usepackage{color}
\usepackage{enumitem}

\usepackage{subfigure}

\usepackage{float}

\usepackage{amsfonts}
\usepackage{amssymb}
\usepackage{amsthm}
\usepackage{enumitem}
\usepackage{shuffle}
\usepackage{comment}
\usepackage{cancel}
\usepackage{blkarray}
\usepackage{multirow}
\usepackage{subfigure}
\usepackage{longtable}
\usepackage{savesym}
\savesymbol{div}
\usepackage{physics}
\restoresymbol{TXF}{div}
\usepackage{tikz}
\usepackage{tikz-3dplot}
\usepackage{xcolor}
\usetikzlibrary{patterns, decorations.markings, arrows}
\usetikzlibrary{arrows.meta}
\usepackage{mathabx}
\usepackage{mathtools}
\usepackage[export]{adjustbox}

\theoremstyle{definition}
\newtheorem{proposal}{Proposal}

\usepackage[numbers,sort&compress]{natbib}
\usepackage{hyperref}
\allowdisplaybreaks
\hypersetup{
colorlinks=true,
linkcolor=blue,
linktocpage=true,
citecolor=violet
}

\usepackage{xargs}
\usepackage[colorinlistoftodos,prependcaption,textsize=tiny]{todonotes}

\newcommand{\q}[1]{``#1''}

\newcommand{\la}{\langle}
\newcommand{\ra}{\rangle}

\definecolor{mybrown}{RGB}{171,6,44}
\definecolor{mygreen}{RGB}{0,128,0}
\definecolor{pinegreen}{RGB}{62,134,103}

\def\laddonepic{{\begin{tikzpicture}[baseline=(current bounding box.south)]

    \draw[mybrown, thick] (0,0) -- (0.3,0);

    \draw[fill=white] (0,0) circle (2pt);
    \draw (0,0) -- (45:2pt);
    \draw (0,0) -- (135:2pt);
    \draw (0,0) -- (-45:2pt);
    \draw (0,0) -- (-135:2pt);

   \filldraw (0.3,0) circle (2pt);
\end{tikzpicture}}}

\def\laddtwopic{{\begin{tikzpicture}[baseline=(current bounding box.south)]

    \draw[mybrown, thick] (0,0) -- (0.3,0) -- (0.6,0);

    \draw[fill=white] (0,0) circle (2pt);
    \draw (0,0) -- (45:2pt);
    \draw (0,0) -- (135:2pt);
    \draw (0,0) -- (-45:2pt);
    \draw (0,0) -- (-135:2pt);

   \filldraw (0.3,0) circle (2pt);
    \filldraw (0.6,0) circle (2pt);
\end{tikzpicture}}}

\def\laddthreepic{{\begin{tikzpicture}[baseline=(current bounding box.south)]

    \draw[mybrown, thick] (0,0) -- (0.3,0) -- (0.6,0)--(0.9,0);

    \draw[fill=white] (0,0) circle (2pt);
    \draw (0,0) -- (45:2pt);
    \draw (0,0) -- (135:2pt);
    \draw (0,0) -- (-45:2pt);
    \draw (0,0) -- (-135:2pt);

   \filldraw (0.3,0) circle (2pt);
    \filldraw (0.6,0) circle (2pt);
     \filldraw (0.9,0) circle (2pt);
\end{tikzpicture}}}

\def\laddladdpic{{\begin{tikzpicture}[baseline=(current bounding box.south)]
    \draw[mybrown, thick] (0,0) -- (0.3,0) -- (0.6,0);
    \filldraw (0,0) circle (2pt);
    \draw[fill=white] (0.3,0) circle (2pt);
    \draw (0.3,0) -- ++(45:2pt);
    \draw (0.3,0) -- ++(135:2pt);
    \draw (0.3,0) -- ++(-45:2pt);
    \draw (0.3,0) -- ++(-135:2pt);
    \filldraw (0.6,0) circle (2pt);
\end{tikzpicture}}}

\def\looppic{{\begin{tikzpicture}[baseline=(current bounding box.center)]
    \coordinate (A) at (0,0);
    \coordinate (B) at (0,0.3);
    \coordinate (C) at (-0.2,0.15); 
    
    \draw[mybrown, thick] (A) -- (B);
    \draw[mybrown, thick] (B) -- (C);
    \draw[mybrown, thick] (C) -- (A);

    \draw[fill=white] (C) circle (2pt);
    \draw (C) -- ++(45:2pt);
    \draw (C) -- ++(135:2pt);
    \draw (C) -- ++(-45:2pt);
    \draw (C) -- ++(-135:2pt);

    \filldraw (B) circle (2pt);

    \filldraw (A) circle (2pt);
\end{tikzpicture}}}

\begin{document}

\preprint{MPP-2025-144, LAPTH-023/25}

\title{Geometric Landau Analysis and Symbol Bootstrap}

\author[a]{Dmitry Chicherin,}
\author[b]{Johannes Henn,}
\author[b]{Elia Mazzucchelli,}
\author[c]{Jaroslav Trnka,}
\author[b]{Qinglin Yang,}
\author[b]{Shun-Qing Zhang}

\affiliation[a]{LAPTh, Université Savoie Mont Blanc, CNRS, B.P. 110, F-74941 Annecy-le-Vieux, France}

\affiliation[b]{Max-Planck-Institut für Physik, Werner-Heisenberg-Institut, Boltzmannstr. 8, 85748 Garching, Germany}

\affiliation[c]{Center for Quantum Mathematics and Physics (QMAP), University of California, Davis, 95616 CA, USA}

\emailAdd{chicherin@lapth.cnrs.fr}
\emailAdd{henn@mpp.mpg.de}
\emailAdd{eliam@mpp.mpg.de}
\emailAdd{trnka@ucdavis.edu}
\emailAdd{qlyang@mpp.mpg.de}
\emailAdd{sqzhang@mpp.mpg.de}

\abstract{
We investigate how the positive geometry framework for loop integrands in $\mathcal{N}{=}4$ super Yang-Mills theory constrains the structure of the integrated answers. 
This is done in the context of a geometric expansion of Wilson loops with a Lagrangian insertion, called {negative geometries}, extending ideas previously used for scattering amplitudes related to the Amplituhedron.
The procedure we adopt combines the knowledge of all maximal codimension boundaries of the geometry, which characterize all possible leading singularities of the integral, with a geometrically informed Landau analysis.
The interplay between geometry and Landau analysis arises from associating Landau diagrams to geometric boundaries. The boundary structure of the geometry then determines which solutions to the Landau equations are spurious and which ones are physical, that is, which singularities are actually present in the integral. 
This method allows us to efficiently determine the symbol alphabet of the associated integral, and serves as a starting point for the symbol bootstrap. We successfully implement this procedure and compute the six-point two-loop and five-point three-loop ladder negative geometries at the symbol level. We also present the conjectural alphabet for ladder negative geometries at two loops for all multiplicities. These are finite integrals that serve as building blocks for the Wilson loop with Lagrangian insertion, and therefore provide insights into the function space of the latter.}

\maketitle
\graphicspath{{figs/}}

\newpage

\section{Introduction}
\label{sec:intro}

Scattering amplitudes hold a significant position in todays high-energy physics research, presenting themselves as a vital bridge between the theoretical framework of quantum field theory (QFT) and experimental results from collider experiments. In recent years, there have been remarkable advances in the study of mathematical structures in amplitudes, especially in planar {maximally supersymmetric Yang-Mills theory}, also known as $\mathcal{N}=4$ sYM theory~\cite{Arkani-Hamed:2008owk,Arkani-Hamed:2022rwr,Travaglini:2022uwo}. Amplitudes in this theory are dual to correlators of null-polygonal Wilson loops~\cite{Alday:2007hr} in the same theory, and they therefore present additional symmetries as {dual conformal invariance}~\cite{Drummond:2008vq,Drummond:2009fd}. Also, scattering amplitudes in $\mathcal{N}=4$ sYM exhibit rich combinatorial and geometric structures beyond the traditional Feynman integral expansion derived from the Lagrangian description. For instance, the integrand of the amplitude for any number of particles and loops can in principle be calculated via on-shell recursion relations~\cite{Arkani-Hamed:2010zjl}, or as the canonical form of a positive geometry called the \textit{Amplituhedron} \cite{Arkani-Hamed:2013jha}. Moreover, their leading singularities have a combinatorial description arising from contour integral representation on the Grassmannian~\cite{Arkani-Hamed:2009ljj,Arkani-Hamed:2012zlh}, and the singularity structure of the scattering amplitudes themselves, that is after integration, can be described in terms of cluster algebras~\cite{Goncharov:2010jf,Golden:2013xva,Golden:2014xqa}.

In all of this fascinating progress, the concept of the Amplituhedron has significantly deepened our understanding of amplitudes in planar $\mathcal{N}=4$ sYM~\cite{Arkani-Hamed:2013jha,Arkani-Hamed:2013kca,Arkani-Hamed:2017vfh,Damgaard:2019ztj,Ferro:2022abq}, see also \cite{Bern:2015ple,Ferro:2018vpf,Arkani-Hamed:2018rsk,Lukowski:2020dpn,Herrmann:2020qlt,Dian:2021idl,Arkani-Hamed:2021iya,Dian:2022tpf,Paranjape:2022ymg,Brown:2023mqi,Arkani-Hamed:2023epq,Even-Zohar:2024nvw,Ferro:2024vwn,Flieger:2025ekn,Even-Zohar:2025ydi,Koefler:2025zvy} for recent work. The Amplituhedron gives a purely geometric description of the amplitude's integrand in the perturbative regime, directly in four spacetime dimensions. The geometric dictionary posits that the integrand is the \textit{canonical  form} of a positive geometry, namely the Amplituhedron. The canonical form is the unique logarithmic form with prescribed poles on the geometry's boundaries. 
Analogous geometric objects have been discovered in different quantum field theories, such as the Associahedron in bi-adjoint $\phi^3$ theory~\cite{Arkani-Hamed:2017mur,Arkani-Hamed:2023mvg,Arkani-Hamed:2023jry}, the Amplituhedron in ABJM theory~\cite{He:2022cup,He:2023rou}, the Correlahedron~\cite{Eden:2017fow,He:2024xed,He:2025rza}, and in cosmology~\cite{Arkani-Hamed:2017fdk,Arkani-Hamed:2024jbp}. The rich combinatorics, and the intersection of complex, real and tropical geometry unraveled by these geometric objects gave rise to a new research area in mathematics, called \textit{Positive Geometry} \cite{Arkani-Hamed:2017tmz,Herrmann:2022nkh,Ranestad:2025qay,Fevola:2025yzx}. They turn out to be very useful in computing sYM amplitudes at integrand level, which are rational functions in the external and loop kinematics.

Furthermore, in order to obtain the physical quantities of interest, such as scattering amplitudes, one has to integrate these rational functions over the loop momenta on the whole Minkowski space. The integration step generally presents various difficulties, from the more technical issue on how to actually perform the loop momentum integration, to more conceptual one of how to handle divergences and to express and simplify integrated results by suitable transcendental functions. Many advanced tools for Feynman integrals and transcendental functions, such as canonical differential equation method~\cite{Henn:2013pwa,Henn:2014qga}, symbol alphabets~\cite{Goncharov:2010jf,Duhr:2011zq}, and the bootstrap program~\cite{Dixon:2011pw,Dixon:2014xca,Dixon:2014iba,Drummond:2014ffa,Dixon:2015iva,Caron-Huot:2016owq,Dixon:2016nkn,Drummond:2018caf,Caron-Huot:2019vjl,Caron-Huot:2019bsq,Caron-Huot:2020bkp} have been widely exploited recently to overcome these difficulties. Any progress in loop momentum integration will significantly advance our understanding and study of experimentally observable theories, e.g.,  scattering amplitudes in Quantum Chromodynamics (QCD).

As previously noted, compared to QCD, 
$\mathcal{N}{=}4$ sYM possesses enhanced symmetries and additional geometric features. 
Consequently, in all computed scattering amplitude data, its results consistently exhibit simpler and more elegant structures. $\mathcal{N}{=}4$ sYM scattering amplitudes conjecturally always yield uniform-weight transcendental functions. At equivalent multiplicities, they depend on fewer kinematical variables and singularity loci. However, this theory still provides crucial insights for QCD amplitude computations and research \cite{Henn:2020omi}. 
Through computations within the sYM framework, including specific cases to be analyzed in this work, we can develop necessary techniques and tools for future studies of QCD scattering amplitudes with analogous complexity. Moreover, in many special instances, our results will directly constitute components of the physical quantities under investigation.

Therefore, in this work we will mainly focus on  integration step under the $\mathcal{N}{=}4$ sYM background. Especially, we will explore how the positive geometry description can help with loop  integration, whose full power remains widely unexplored. First steps in this direction were taken in  ref. \cite{Dennen:2015bet,Dennen:2016mdk,Prlina:2017azl,Prlina:2017tvx}, where the authors used the geometry of the Amplituhedron to refine the Landau analysis used to determine the branch cut structure of amplitudes in planar $\mathcal{N}=4$ sYM. The main goal of the present work is to further explore these ideas for a different class of physical quantities that have close analogies to QCD amplitudes. More precisely, we apply the ideas presented in the references above to quantities related to the correlator of an $n$-sided null-polygonal Wilson loop with single Lagrangian insertion in planar $\mathcal{N}{=}4$ sYM and its novel geometric building blocks, called \textit{negative geometries}~\cite{Arkani-Hamed:2021iya}, which are potentially useful beyond $\mathcal{N}{=}4$ sYM.

Let us briefly review the Wilson-loop's structure and its decomposition into negative geometries. We denote by $W_n(x_1,\cdots,x_n)$ the $n$-sided light-like polygonal Wilson loop, with cusps at the locations $x_i$, which are points in dual momentum space. We define 
\begin{equation}\label{F_n}
F_n(x_1,\ldots,x_n;x_0) := \pi^2\frac{\langle{W_n(x_1,\cdots,x_n) {\cal L}(x_0)\rangle}}{\langle W_n(x_1,\ldots,x_n)\rangle} \,,
\end{equation}
where $\mathcal{L}(x_0)$ denotes the Lagrangian inserted at a point $x_0$.
This observable has been considered in a series of previous works \cite{Alday:2011ga,Alday:2012hy,Alday:2013ip,Engelund:2011fg,Engelund:2012re,Henn:2019swt,Chicherin:2022bov,Chicherin:2022zxo,Chicherin:2024hes,Carrolo:2025pue,Abreu:2024yit} and enjoys a number of ideal properties. 
Firstly, it is finite at each order in perturbation theory. Secondly, as a consequence of the duality between Wilson loops and scattering amplitudes \cite{Alday:2007hr,Drummond:2007aua,Brandhuber:2007yx}, the integrand of $F_n$ at $L$ loops is equal to the integrand at $(L{+}1)$ loops of the maximally-helicity-violating (MHV) amplitude's logarithm, after treating $x_0$ as an integration variable. 
After carrying out the loop integrations, the loop correction to \eqref{F_n} has a structure 
reminiscent of that of non-MHV scattering amplitudes, such as remainder and ratio functions \cite{Drummond:2008vq}. That is, at each perturbative order $L$ and at any multiplicity $n$, it is a linear combination of uniform weight transcendental functions with certain rational prefactors~\cite{Brown:2025plq}. Furthermore, \eqref{F_n} seems to have a surprising interplay with the all-plus $n$-point scattering amplitude in YM theory, without dual conformal invariance. Therefore, $F_n$ serves as an ideal playground for important methods and results discovered in the method of canonical differential equations, such as pentagon/hexagon functions~\cite{Chicherin:2017dob,Gehrmann:2018yef,Henn:2021cyv,Henn:2024ngj,Liu:2024ont,Abreu:2024fei,Henn:2025xrc}, which is the state-of-art in perturbative calculations in general gauge theories, i.e. those without dual conformal symmetry.

Moreover, the positive geometry dictionary provided by the Amplituhedron, relevant for scattering amplitudes, also applies to integrand of $F_n$.
However, compared to the amplitudes setting, the integrand of $F_n$, at a fixed loop-order, is not directly related to the canonical form of an individual positive geometry, but it is rather a linear combination of such. In fact, in \cite{Arkani-Hamed:2021iya} the authors decomposed the $L$-loop integrand of the amplitude in terms of basic building blocks called negative geometries, each of which has a geometric description very similar to that of MHV loop Amplituhedra. Each $L$-loop negative geometry is represented by a connected graph with $L$ vertices. The amplitude is obtained by summing over all such graphs.
This construction extends to the logarithm of the amplitude, and hence to the integrand of $F_n$, where the sum now reduces to only connected graphs. It is important to note that the decomposition into negative geometries is unrelated to the standard sum over Feynman diagrams: each negative geometry yields an integrand which is non-planar in dual space $x_i$. Moreover, each individual term in such an expansion, i.e. each individual negative geometry, is free of (infrared) divergences after integration.
Therefore, to each connected graph with $L+1$ vertices we can associate a transcendental function obtained by integrating $L$ loop variables from the canonical form of the associated negative geometry. We call such functions \textit{integrated negative geometries}.

On the one hand, evaluating integrated negative geometries provides perturbative data for the full observable $F_n$. On the other hand, because the integrals are finite, they provide a playground to test the relationship between positive geometries and integrals arising from integrating their canonical forms. Integrated negative geometries usually contain {\it leading singularities} (LS) as prefactors, which are rational functions in the external data $x_i$, and transcendental functions accompanying each LS. Recently, \cite{Brown:2025plq} provided a full classification of LS of $F_n$ at all loops and all points. The LS are maximal codimension residues of the integrand, and can therefore be defined also for individual negative geometries. The complete classification of LS of a negative geometry is a complicated task in general, but in this work we provide some results for specific graph topologies. In particular, we provide a full classification for negative geometries corresponding to ladder-type graphs.

After having determined the leading singularities, the next step is to compute the transcendental functions that accompany each of them. To achieve this, we adopt the {\it symbol bootstrap} strategy to deal with the integration step. A symbol bootstrap relies on knowledge of both the kinematic prefactors and the singular loci of the integral to be computed. The symbol alphabet of the integral can then be constructed from its singular loci, based on an ansatz for the functions space that we expect the integral to belong to. Understanding the singular loci, as well as the study of the symbols of transcendental functions are necessary steps for this calculation. These topics have attracted significant attention in recent years \cite{Arkani-Hamed:2019rds,Drummond:2019qjk,Henke:2019hve,Drummond:2019cxm,Caron-Huot:2020bkp,Chicherin:2020umh,He:2021esx,He:2021non,He:2021eec,Yang:2022gko,Dlapa:2023cvx,He:2023umf,Jiang:2024eaj,Aliaj:2024zgp,Pokraka:2025ali,Bossinger:2025rhf}.
As an example, the symbol bootstrap procedure showed overwhelming power in calculating scattering amplitudes at six and seven points, and also form factors at three points, determining results for these cases at astonishingly high loop orders \cite{Dixon:2011pw,Brandhuber:2012vm,Dixon:2014xca,Dixon:2014iba,Drummond:2014ffa,Dixon:2015iva,Caron-Huot:2016owq,Dixon:2016nkn,Drummond:2018caf,Caron-Huot:2019vjl,Caron-Huot:2019bsq,Caron-Huot:2020bkp,Dixon:2020bbt,Dixon:2022rse, Dixon:2022xqh,Dixon:2023kop,Basso:2024hlx}.

The understanding of singular loci of integrals of our interest comes from the {\it geometric Landau analysis}, which combines the original Landau analysis \cite{Landau:1959fi} with the geometric information provided by the Amplituhedron, or rather by the negative geometries. In the context of amplitudes, the connection between the Amplituhedron and the Landau analysis has been explored in~\cite{Dennen:2015bet,Dennen:2016mdk,Prlina:2017azl,Prlina:2017tvx}. The idea from these works is the following. A full list of amplitude's singularities is recovered by the Landau analysis on their {\it Landau diagrams}. The latter correspond geometrically to various codimension boundaries of the Amplituhedron. Furthermore, due to the chiral property of N$^k$MHV amplitudes, all Landau loci from this process will be refined by checking the vanishing of the integrand's numerator, and finally be partitioned into \textit{physical} and \textit{spurious} ones. Here spurious means that the singularity is absent in the integrated quantity. Geometrically speaking, this information is encoded by checking the real dimension of the Landau equation's vanishing locus intersected with the Amplituhedron. Then, every physical singular locus yields a symbol letter. Therefore, this procedure enables us to construct an ansatz for the bootstrap procedure.

In this paper we apply the geometric Landau analysis procedure for bootstrapping integrated negative geometries in $\mathcal{N}=4$ sYM. There are two main features that should be distinguished between our setting and that of amplitudes. Firstly, the kinematic dependence of negative geometries includes $n$ external momenta, as well as a distinguished unintegrated loop variable. Therefore, even if the underlying negative geometries are defined in terms of the MHV Amplituhedron, many of its properties, such as the maximal codimension boundaries, rather resemble that of other helicity sectors of the Amplituhedron. 
Secondly, the negativity condition of the multiloop propagators now allows for non-planar configurations in the Landau diagrams. This morally explains why individual geometries can be expected to be non-planar objects in some sense, and this is reflected by their singularities, as we will see for the ladder at five points and three loops. We therefore refine the geometric selection rule via on-shell diagrams in the context of negative geometries.
This is a modified version of the proposal made in~\cite{Prlina:2017azl}. The modification takes into account that negative geometries can have maximal codimension boundaries with helicity weight higher than MHV. We use and test this conjectural procedure against our computed examples, which we now explain.

After the geometric Landau analysis described previously yields an ansatz for the symbol alphabet of the loop integral, we can then bootstrap the symbol of the negative geometries. Known results for integrated negative geometries include all negative graphs at four points and two loops \cite{Arkani-Hamed:2021iya} and the ladder and triangle at five points and two loops \cite{Chicherin:2024hes}. In this work, we determine the integrated negative geometries in $\mathcal{N}=4$ sYM at the symbol level. In particular we compute the ladder at six points and two loops, and at five points and three loops. 
Since ladder geometries serve as building blocks in the negative geometry expansion of the Wilson loop with Lagrangian insertion, they also provide insights into the relevant function space for that observable.
More generally, our study provides insights into the interplay between the positive geometry framework and integrated results.
We also extect that it will facilitate future calculations of higher-loop Feynman integrals and the understanding of their singularity structures.

The bootstrap method for integrated ladders integrated negative geometries boils down to constructing the function space from the symbol alphabet, imposing the absence of unphysical singularities and consistency with lower-point perturbative data, and finally imposing a differential equation associated with the d’Alembertian operator. For the six-point two-loop ladder, we find that its alphabet is a subset of the alphabet of the full two-loop observable \cite{Carrolo:2025pue}. Moreover, this ladder localizes in the two-loop planar hexagon function space \cite{Henn:2025xrc}. For the five-point three-loop ladder, as compared to the two-loop approximation \cite{Chicherin:2024hes}, we observe that an extension of the two-loop planar pentagon alphabet \cite{Gehrmann:2015bfy} is required, reflecting the non-planar nature of individual negative geometries.

The outline of this paper is as follows. Section~\ref{sec:prel} is dedicated to a review of notions necessary in this work. In Section~\ref{sec:lead sing}, we pass to the study of leading singularities, and in particular we classify all leading singularity configurations for negative geometries with an underlying topology of a ladder. In Section~\ref{sec:landa and schubert}, we perform the Landau analysis to determine the symbol alphabet for the integrals associated to ladders. We take advantage of the underlying geometric description to obtain a selection rule for physical singularities. Section~\ref{sec: bootstrapping integrated negative geometries} contains the explicit computations of some integrated negative geometries, in particular, the six-point, two-loop and five-point, three-loop ladders. The main body ends with a discussion and outlook, in Section~\ref{sec:conclusions}. The appendices contain ancillary discussions: Appendix~\ref{app:ladder proof} contains the proof of the classification result for leading singularities, Appendix~\ref{app: All leading singularity values at two loops} contains the leading singularity values of ladders with marked points in the interior, Appendix \ref{sec: sel rules and bipartite} is a detailed discussion for identifying physical singularities from geometric Landau analysis, Appendix~\ref{app:integrand} is about the integrand of six-point ladders, Appendix~\ref{app:da} reviews the action of the d’Alembertian operator of ladders and Appendix~\ref{app:super} contains some details about a special class of Landau singularities of ladders. 

\section{Preliminaries}
\label{sec:prel}

We begin by reviewing basic concepts and notation that will be used throughout this work.

\subsection{Momentum twistors} 
For describing scattering amplitudes in planar $\mathcal{N}=4$ sYM theory and the underlying Amplituhedron geometry \cite{Arkani-Hamed:2013jha} we will use momentum twistor variables~\cite{Hodges:2009hk}. Dual momentum variables are related to cyclically ordered external momenta $p_i \in \mathbb{R}^{3,1}$ via $p_i = x_i - x_{i-1}$. Momentum conservation $\sum_{i=1}^n p_i =0$ translates to dual points forming a closed $n$-gon, i.e. $x_{n+1} = x_1$. The (complexified) kinematic data for $n$-point massless planar scattering amplitudes is encoded in $n$ complex vectors $Z_i\in \mathbb{C}^4$, called \textit{momentum twistors}, where $i = 1, \dots,n$ labels the external particles. These transform in the fundamental representation of ${\rm SL}(4)$ corresponding to action of the dual conformal symmetry.
The (up to scalar unique) ${\rm SL}(4)$-invariant multilinear form is given by the twistor bracket
\begin{equation}\label{four_bracket}
    \langle ijkl \rangle := \det(Z_i Z_j Z_k Z_l ) = \epsilon_{IJKL} Z^{I}_i Z^{J}_j Z^{K}_k Z^{L}_l \ ,
\end{equation}
where $\epsilon_{IJKL}$ is the four-dimensional Levi-Civita tensor and in the determinant's argument we stack together the vectors $Z_i$'s to form a $4 \times 4$ matrix. Furthermore, due to the dual conformal invariance (DCI) ~\cite{Drummond:2008vq,Drummond:2009fd}, we can regard the $Z_i$'s as points in the projective space $\mathbb{P}^{3}$. In order to describe integrands for amplitude's $L$-loop corrections, we need to introduce (dual) loop momenta $y_\ell \in \mathbb{R}^{3,1}$, $\ell=1,\cdots,L$. This is achieved by associating a line $AB_\ell$ in $\mathbb{P}^3$ to $y_\ell$, where the notation means that $A_\ell, B_\ell \in \mathbb{P}^3$ and $AB_\ell$ is the unique line passing through $A_\ell$ and $B_\ell$. This formulation naturally yields the \textit{Grassmannian} ${\rm Gr}(k,n)$, the space of $k$-dimensional subspaces of $\mathbb{C}^{n}$~\cite{Arkani-Hamed:2012zlh}.

It is useful for our purposes to introduce the \textit{infinity bitwistor}
\begin{equation}
    I_{\infty}:=\left(\begin{matrix}
        0&0 & 0 & 1\\0&0& 1 & 0\\
    \end{matrix}\right) \,,
\end{equation}
which is dual to the point at infinity in dual momentum space. Then, the bracket $\langle ab \, I_\infty\rangle$ is equal to the usual bracket between the spinor helicity variables $\lambda_a^\alpha$ and $\lambda_b^\beta$.

We will also frequently use the shorthand notation
\begin{equation}
\begin{aligned}
&(ab)\cap(cde):=Z_a\langle bcde\rangle-Z_b\langle acde\rangle \, , \\
&(abc)\cap(def):=(ab)\langle cdef\rangle-(ac)\langle bdef\rangle+(bc)\langle adef\rangle \, , \\
&    \langle a(bc)(de)(fg)\rangle:=\langle abde\rangle\langle acfg\rangle-\langle acde\rangle\langle abfg\rangle \, ,
\end{aligned}
\end{equation}
where these formulas reflect intersections in $\mathbb{P}^3 $, with $(ab)$ representing the line through $Z_a$ and $Z_b$, and $(abc)$ representing the plane containing $Z_a$, $Z_b$ and $Z_c$.

\subsection{The MHV loop Amplituhedron} In this work we use the topological definition of the Amplituhedron via sign-flips that was introduced in \cite{Arkani-Hamed:2017vfh}. We parameterize the external kinematics of $n \geq 4$ massless particles by a real $4 \times n$ matrix $Z$, representing an element in the configuration space ${\rm Gr}(4,n)/(\mathbb{C}^{*})^{n}$.
Moreover, there is an additional symmetry of the Amplituhedron called \textit{twisted cyclic symmetry}, such that the ${\rm  MHV}$ Amplituhedron is invariant under the transformation
\begin{equation}\label{twist_symm}
    Z_1 \mapsto Z_2, \ Z_2 \mapsto Z_3,  \ \dots \ , \ Z_n \mapsto -Z_1 \ .
\end{equation}
The first defining condition of the geometry is that of \textit{external positivity}, which requires $Z$ to be an element in the positive Grassmannian, i.e.
\begin{equation}\label{external_positivity}
    \langle ijkl    \rangle > 0 \ , \qquad \forall \, i<j<k<l \ . 
\end{equation}
Then, the \textit{$n$-point $L$-loop MHV Amplituhedron} $\mathcal{A}_{n}^{(L)}$, where we omit the other parameters,
which we fix to $m=4$ and $k=0$ in this paper, is a semialgebraic set in ${\rm Gr}_{\mathbb{R}}(2,4)^{L}$, parameterized by \textit{loop variables} $AB_\ell \in {\rm Gr}_{\mathbb{R}}(2,4)$ for $\ell =1, \dots, L$, and carved out by two type of inequalities. First, the loop positivity conditions for every $1 \leq \ell \leq L$:
\begin{equation}\label{loop_positivty}
\begin{aligned}
    &\langle AB_\ell \, 12 \rangle > 0 \ , \dots, \langle AB_\ell \, n-1n \rangle > 0 \ , \ \langle AB_\ell \, 1 n \rangle > 0 \ \text{and} \\
    & \text{the sequence } (\langle AB_\ell \, 12 \rangle , \dots, \langle AB_\ell \, 1n \rangle) \text{ has exactly two sign flips, ignoring zeros.}
\end{aligned}
\end{equation}
The sign-flip condition can be equivalently imposed fixing any other index instead of $1$, but one has to take into account the twisted cyclic symmetry \cite{Arkani-Hamed:2017vfh}.
Second, the \textit{mutual positivity} conditions between loops:
\begin{equation}\label{mutual_positivity}
    \langle AB_\ell \, AB_{\ell'} \rangle > 0 \ , \qquad \forall \, \ell \neq \ell' \ .
\end{equation}
The Amplituhedron $\mathcal{A}^{(L)}_n$ is the Euclidean closure in ${\rm Gr}_{\mathbb{R}}(2,4)^{L}$ of the space defined above, and it is conjectured to be a positive geometry \cite{Arkani-Hamed:2017tmz} with canonical form
\begin{equation}
     \prod_{\ell=1}^{L} \langle AB_\ell \, d^{2}A_\ell \rangle \, \langle AB_\ell \, d^{2}B_\ell \rangle \, \, \Omega_n^{(L)} \ ,
\end{equation}
where $\Omega_n^{(L)}$ is a rational function in the twistor coordinates of the loop variables $AB_\ell$ and the $Z_i$, called the \textit{canonical function}. The rational function $\Omega_n^{(L)}$ is is equal to the $n$-point $L$-loop MHV integrand in $\mathcal{N}=4$ sYM. In fact, the BCFW recursion relation for the amplitude is believed to yield tilings of the Amplituhedron~\cite{Arkani-Hamed:2012zlh}. The latter statement has been recently proven at tree level in~\cite{Even-Zohar:2021sec,Even-Zohar:2023del,Galashin:2024ttp}. On the other hand, the fact that the Amplituhedron is a positive geometry has been proven only for the one-loop MHV Amplituhedron~\cite{Ranestad:2024svp}, which coincides with the tree Amplituhedron for the parameters $k=m=2$. It was shown that at $L=2$ the Amplituhedron is a \emph{weighted positive geometry} \cite{Dian:2022tpf} to accommodate the geometry of internal boundaries. The precise geometric description of the Amplituhedron for higher $L$ is an open question.

\subsection{Negative geometries} 

The authors of ref. \cite{Arkani-Hamed:2021iya} found that the integrand $\widetilde{\Omega}^{(L)}_n$ for the $L$-loop correction to the logarithm of the $n$-point amplitude in $\mathcal{N}=4$ sYM, which is related to the integrand of the Wilson loop with Lagrangian insertion by Wilson-loop/amplitude duality \cite{Alday:2007hr,Drummond:2007aua,Brandhuber:2007yx}, can be constructed from geometric objects very similar to the Amplituhedron. These are called \textit{negative geometries}, since they are essentially MHV loop Amplituhedra, but with all \textit{mutual positivity} conditions replaced by a certain number of \textit{mutual negativity} conditions $\langle AB_\ell \, AB_{\ell'} \rangle < 0$ between loop variables. Moreover, each of these geometries has a canonical form, which has the nice property that if one integrates all but one loop variables, the resulting quantity is infrared (IR) finite.

Let us briefly introduce this construction, following the graphical notation of \cite{Arkani-Hamed:2021iya}. Firstly, each loop line is represented by a black node, so there are $L \geq 1$ in total. Then, the mutual positivity condition between a pair of loops is represented by a dashed blue edge joining the corresponding nodes. Hence, the $L$-loop Amplituhedron geometry corresponds to a complete graph with $L$ nodes (note that in this notation there is no reference to the number of points $n$). 
The mutual negativity conditions $\la AB_\ell \, AB_{\ell'}\ra < 0$ are represented by a red edge. By additivity of the canonical forms with respect to triangulations, see \cite{Arkani-Hamed:2017tmz}, the following equation holds true: 
\begin{equation}
\label{eq:linkrel}
	\begin{tabular}{cc}
	 \includegraphics[scale=.86]{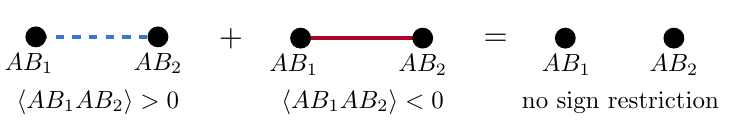}
	\end{tabular}
\end{equation}
where the geometry on the right-hand side is the Cartesian product of two one-loop Amplituhedra. This relation can be used iteratively on the edges of the complete positive graph to obtain an expansion only in \textit{negative geometries}, i.e. graphs with only negative edges and no positive edge. The decomposition translates directly at the level of canonical forms, and gives a decomposition of the amplitude's integrand as a sum over forms of negative geometries:
\begin{equation}\label{omega_in_neg_geom}
    \Omega^{(L)}_n = \sum_{\Gamma} (-1)^{E(\Gamma)} \, \Omega_{n,\Gamma}^{(L)} \ ,
\end{equation}
where the summation runs over all graphs $\Gamma$ with $L$ nodes and any number of negative edges, and $E(\Gamma)$ denotes the number of edges of $\Gamma$. The rational function $\Omega_{n,\Gamma}^{(L)}$ is then the canonical function of a \textit{negative geometry} defined as the $L$-fold product of one-loop Amplituhedra  $\mathcal{A}^{(1)}_n$ with the extra conditions $\langle AB_\ell \, AB_{\ell'} \rangle < 0 $, whenever $\ell - \ell'$ is an edge of $\Gamma$. In order to obtain the integrand for the $L$-loop correction to the logarithm of the $n$-point amplitude $\widetilde{\Omega}_n^{(L)}$, we take the logarithm of \eqref{omega_in_neg_geom}. By the exponential formula for generating functions, the sum in \eqref{omega_in_neg_geom} is reduced to the sum over \textit{connected graphs}:
\begin{equation}\label{omega_tilde_in_neg_geom}
    \widetilde{\Omega}^{(L)}_n = \sum_{\Gamma \, \text{connected}} (-1)^{E(\Gamma)} \, \Omega_{n,\Gamma}^{(L)} \ .
\end{equation}
Because we want to integrate all loop variables except one, namely $AB$, we usually mark a node in each graph $\Gamma$ that appears in \eqref{omega_tilde_in_neg_geom}. The sum can then be taken over connected graphs with a marked node, but each graph must be weighted by a symmetry factor associated with its automorphism group \cite{Arkani-Hamed:2021iya}. We emphasize that the expansion in eq.~\eqref{omega_tilde_in_neg_geom} has a priori nothing to do with the expansion into Feynman diagrams. This can be seen, for example, by the fact that the integrands $\Omega_{n,\Gamma}^{(L)}$ involve non-planar propagator's configurations.

One of the key properties of the expansion in negative geometries is that the mutual negativity condition between a pair of loop variables prevents their localization to collinear configurations. This in turn allows for the definition of an IR finite quantity for \textit{every} individual connected negative geometry. This is done by performing all but one integrations over the loop momenta. More specifically, to a given connected graph $\Gamma$ with $L+1$ vertices labeled by $ AB_0, AB_1, \dots , AB_L$, we associate the integral.
That is, for a given connected graph $\Gamma$ with ${L+1}$ nodes one of which is marked, corresponding to the loop variable $AB$, we associate the integral
\begin{equation}\label{F_integral}
    F_{n,\Gamma}^{(L)} := -\int_{AB_1}  \dots \int_{AB_{L}} \widetilde{\Omega}_{n,\Gamma}^{(L+1)} \ ,
\end{equation}
where $\widetilde{\Omega}_{n,\Gamma}^{(L+1)}$ is the canonical form of the $n$-point negative geometry associated to $\Gamma$. As we mentioned, this integral in eq.~\eqref{F_integral} is perfectly finite, and it is a function of the unintegrated loop $AB$ and the external momentum twistors $Z$.

\subsection{Leading singularities}

In the following, we will be interested in the leading singularities (LS) of the integral (\ref{F_integral}). We define the {\it value of leading singularity}  of \eqref{F_integral} as the maximal residue obtained by integrating \eqref{F_integral} along any contour homotopic to a topological torus enclosing $4L$ poles in the loop variables $AB_1$, \dots, $AB_{L}$.
From a geometric point of view, an LS is the residue of the canonical form $\widetilde{\Omega}_{n,\Gamma}^{(L+1)}$ on any four-dimensional boundary obtained by localizing all variables $AB_1, \dots, AB_{L}$. The following remark is very important for what follows. Some (homologically inequivalent) contours can yield the same LS value and, correspondingly, many different ordered residues of $\widetilde{\Omega}_{n,\Gamma}^{(L+1)}$ can yield the same result. We therefore distinguish between the value of LS, i.e. the rational function obtained from \eqref{F_integral} by contour integration or a sequence of residues, and the LS \textit{configuration}, which refers to the solution for the loop lines $AB_1$, \dots, $AB_{L}$ obtained by solving the contour integration or the ordered residues. Then, according to our terminology, different LS configurations can yield the same value of LS.

In general, integrated negative geometries \eqref{F_integral} admit a decomposition 
\begin{equation}\label{F_integral_in_LS}
    F^{(L)}_{n,\Gamma} = \sum_{s} \Omega_{n,s,\Gamma} \, f_{n,s,\Gamma}^{(L)} \ ,
\end{equation}
for a negative geometry with topology $\Gamma$. Generally, the coefficients $\Omega_{n,s,\Gamma}$ are linearly independent combinations of the leading singularities of $F^{(L)}_{n,\Gamma}$, and are rational functions of the twistor brackets in the $Z_i$'s and $(AB)_0$. In particular, they are dual conformal invariants when accompanied with the standard measure $\langle d^2 A \, AB \rangle \langle d^2 B \, AB \rangle$ on ${\rm Gr}(2,4)$. Studying $\Omega_{n,s,\Gamma}$ is then the first step towards understanding the structure of \eqref{F_integral_in_LS}, and it is crucial for computational purposes. In practice, one wishes to find a linear basis for the space of LS of \eqref{F_integral_in_LS}. This knowledge is then used in the bootstrap procedure. 

The enumeration and evaluation of LS for individual negative geometries can be a very complicated task, at increasing number of loops $L$. However, a full classification for the $n$-point $L$-loop contribution to the Wilson loop with Lagrangian insertion has been carried out in~\cite{Brown:2025plq}. In particular, the linear space spanned by the LS values of the Wilson loop at $L \geq 2$ is generated by all Kermit forms \eqref{four_invariant} and \eqref{six_invariant}, which at $n \geq 4$ points has dimension $(n-1)(n-2)^{2}(n-3)/12$, see \cite{Brown:2025plq}.
As we will show, a complete classification is also possible for certain simple graph topologies, such as ladders. As an illustration, in Section~\ref{sec:lead sing} we present the classification of all LS configurations of \eqref{F_integral_in_LS} with $\Gamma$ being a \textit{ladder} graph with marked node $(AB)_0$ at one end as in Fig.~\ref{fig:laddersgeneral}.

It is worth noting that, in different contexts, the term ``leading singularity" can have different meanings, such as the leading solution branch in the Landau analysis, as will be reviewed in the following. In this paper, for simplicity, we specifically refer to the prefactors in the decomposition \eqref{F_integral_in_LS}, which correspond to linear combinations of maximal (keeping the dependence in $AB$) residues of the negative geometry's integrand.

\subsection{Landau analysis}Physical singularities in kinematic space of certain Feynman integrals can be obtained systematically by Landau analysis, or more precisely as solutions to the Landau equations (\cite{Landau:1959fi,Eden:1966dnq}, for most recent studies, see~\cite{Mizera:2021icv,Hannesdottir:2021kpd,Fevola:2023fzn,Fevola:2023kaw,Helmer:2024wax,Hannesdottir:2024hke,Caron-Huot:2024brh,He:2024fij,Correia:2025yao}). Suppose we are dealing with a Feynman integral in Feynman 
parametrization 
\begin{equation}\label{Landau_int}
    \int\prod_{i=1}^L{\rm d}^Dk_i\int_{\mathcal{C}}{{\rm d}^\nu\alpha} \, \frac{\mathcal{N}}{\mathcal{D}^\nu} \ ,
\end{equation}
where the denominator reads $\mathcal{D}=\sum_{\ell=1}^\nu\alpha_\ell \, q_\ell^2$ and $q_\ell$ are the loop momenta running through the propagators (for simplicity, we consider massless propagators here) . In general, the numerator $\mathcal{N}$ is a function of $k_i$'s and $q_\ell$'s. The integration is performed along the contour $\mathcal{C}=\{\sum_\ell \alpha_\ell=1,\ \alpha_\ell \geq 0\}$. The singularities of this integral can be deduced from the Landau equations \cite{Landau:1959fi}:
\begin{equation}\label{Land_eq}
\begin{aligned}
    \alpha_\ell \, q_\ell^2 & =0 \ , \ \ \ \ \ \ \ \text{(cut condition)} \ ,\\
    \sum_{\ell \, \in \, \text{each loop}}\alpha_\ell \, q_\ell^\mu & =0 \ , \ \ \ \ \text{(pinch condition)}\ .
\end{aligned}
\end{equation}
In general kinematics, these equations are too constraining to have any non-trivial solutions other than $\alpha_\ell=0$. Physical singularities are then loci of these equations involving the external kinematics $p_\ell$, such that non-trivial solutions of $\alpha_\ell$'s from \eqref{Land_eq} are allowed. A solution branch is called {\it leading Landau singularity}, if it comes from a branch in which all $\alpha_\ell \neq0$. More generally, a sub$^m$-leading Landau singularity is a leading Landau singularity of a sub-topology of the original integral, obtained by shrinking $m$ of its propagators, i.e. $m$ out of all $\alpha_\ell $'s only satisfy $\alpha_\ell =0$. When performing the Landau analysis for an individual integral, one should go through all its sub$^m$-topologies for $m\geq0$, and gather all sub$^m$-leading Landau singularities for a complete prediction of physical singularities. We will present examples of how to perform the Landau analysis in Section \ref{sec:landa and schubert}.

It is worth mentioning that this procedure may omit the so-called \textit{second-type Landau singularities} \cite{FLNP:1962,Eden:1966dnq,Fevola:2023kaw}, which are related to most general scaling of $\alpha_i$ in the Landau equation \eqref{Land_eq}.
However, in $\mathcal{N}=4$ sYM theory, this type of singularity is empirically observed to be absent \cite{Dennen:2015bet,Dennen:2016mdk,Prlina:2017azl,Prlina:2017tvx,Prlina:2018ukf,Lippstreu:2022bib,Lippstreu:2023oio,He:2024fij}. It is also worth mentioning that the Landau analysis described above is blind to the presence of the numerator $\mathcal{N}$ in \eqref{Landau_int}. This is another downside of the Landau analysis, as $\mathcal{N}$ may prevent the appearance of some singularities in the integrated result. In general, one may expect that the vanishing of $\mathcal{N}$ on a Landau singularity prevents the latter from contributing to the integral. However, this is not always the case \cite{Prlina:2017tvx}, and it is a crucial point to be discussed in this paper.

\subsection{Symbols} 
In this paper, we focus on Chen iterated integrals with dlog kernels \cite{Chen:1977oja}. Recall that the total differential of a weight-$w$ iterated integral $\mathcal{F}^{(w)}$ has a general form as
\begin{equation}
    {\rm d}\mathcal{F}^{(w)}=\sum_i\mathcal{F}_i^{(w{-}1)}{\rm d}\log x_i \ , 
\end{equation}
whose {\it symbol} \cite{Goncharov:2010jf, Duhr:2011zq} can be iteratively defined as 
\begin{equation}
    \mathcal{S}\left(\mathcal{F}^{(w)}\right)=\sum_i\mathcal{S}\left(\mathcal{F}_i^{(w{-}1)}\right)\otimes x_i \ . 
\end{equation}
In general, we can write the symbol of $\mathcal{F}^{(w)}$ as a length-$w$ tensor product,
\begin{equation}
    \mathcal{S}\left(\mathcal{F}^{(w)}\right)=\sum_I c_I\, a_1^{I}\otimes\cdots\otimes a_w^{I} \ .
\end{equation}
The entries $a_j^{I}$ are called {\it symbol letters}, which are functions of the external kinematics. The set of symbol letters for an integral is called the {\it symbol alphabet}. In general, the coefficients $c_I$ are also functions of the external kinematics. In the particular case in which the $c_I$'s are rational numbers, $\mathcal{F}^{(w)}$ is called a {\it pure function}. The symbol letters in our case will be functions of dual conformal invariants formed by cross-ratios of $\{x_i;x_0\}_{i=1,\cdots,n}$. Note that due to the insertion point $x_0$, there exists a non-trivial cross ratio even for $n=4$. This is to be compared to scattering amplitudes in $\mathcal{N}=4$ sYM theory, whose remainder/ratio functions become non-trivial from $n=6$ only.

Since symbol bootstrap showed overwhelming power in calculating scattering amplitudes and Feynman integrals, various symbology methods have been developed to detect symbol letters before loop integration. Symbol letters are closely related to the physical singularities of $\mathcal{F}^{(w)}$. Actually, a (Landau) singularity is also called a rational letter, if it appears non-trivially as one symbol letter in the alphabet of the integral we are interested in. However, singularities are not enough to recover all symbol letters. 
Generally, given a set of physical singularities $W_r$ and their multiplicative combinations $\prod_r W_r$, there are many ways to write a factorization
\begin{equation}\label{fact_sr}
    \frac{a{+}\sqrt{\Delta}}{a{-}\sqrt{\Delta}} \ ,\quad \text {with}\quad (a{+}\sqrt{\Delta})(a{-}\sqrt{\Delta}) \ \propto \ \prod_r W_r \ ,
\end{equation}
where $\sqrt{\Delta}$ is a certain square root, usually generated by sub$^m$-leading Landau singularities for some $m$,  
and $a$ is a rational function of kinematic variables to be determined \cite{Heller:2019gkq,Zoia:2021zmb,Henn:2024ngj}. 
Different factorizations in \eqref{fact_sr} result in different {\it algebraic letters}. 
In our computations, we rely on the package \texttt{Effortless.m} \cite{repoEffortless} to construct symbol letters from Landau singularities.

\section{Classification of leading singularities for negative geometries}
\label{sec:lead sing}

\begin{figure}[t]
    \centering
    \begin{tikzpicture}[scale=0.8]

    \draw[mybrown, line width=2pt] (0,0) -- (3,0); 
    \draw[mybrown, line width=2pt] (6,0)--(9,0);
    \draw[mybrown, line width=2pt, dashed] (6,0)--(3,0);

    \draw[fill=white] (0,0) circle (4pt);
    \draw (0,0) -- (45:4pt);
    \draw (0,0) -- (135:4pt);
    \draw (0,0) -- (-45:4pt);
    \draw (0,0) -- (-135:4pt);

   \filldraw (3,0) circle (4pt);
    \filldraw (6,0) circle (4pt);
     \filldraw (9,0) circle (4pt);
    \node[anchor=north] at (0,-0.2) {\small{$(AB)_0$}};
     \node[anchor=north] at (3,-0.2) {\small{$(AB)_1$}};
    \node[anchor=north] at (6,-0.2) {\small{$(AB)_{L{-}1}$}};
        \node[anchor=north] at (9,-0.2) {\small{$(AB)_{L}$}};
\end{tikzpicture}
    \caption{Graph associated to the ladder negative geometry at $L$ loops.}
    \label{fig:laddersgeneral}
\end{figure}
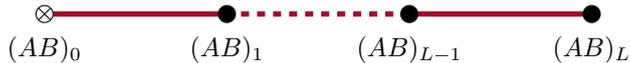

In this section, following ref. \cite{Brown:2025plq}, we discuss the coefficients in the expansion \eqref{F_integral_in_LS}, when $\Gamma$ is of a ladder topology as in Fig.~\ref{fig:laddersgeneral}. In subsection~\ref{subsec:ladderleadingsing}, we present a classification of all LS configurations for ladder negative geometries, which works for ladders at all points $n \geq 4$ and all loops $L \geq 1$. They are line configurations that show maximal cut solutions of loop momenta. 
Maximal residues of the ladder integrand from these solutions yield non-trivial LS values, which we will discuss in subsection~\ref{section_LS_L1}.

\subsection{All-loop leading singularity configurations of ladders}
\label{subsec:ladderleadingsing}
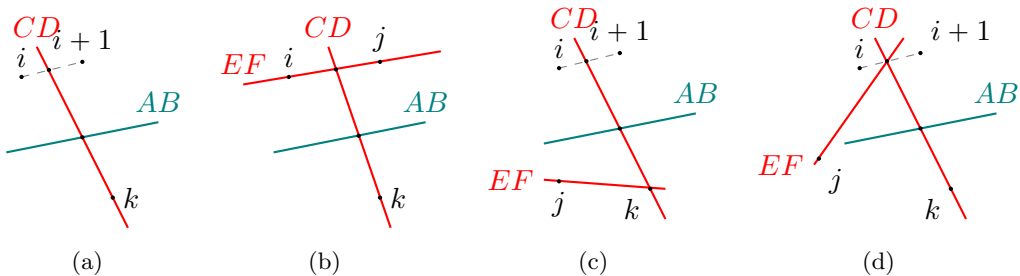
\begin{figure}[t]
\centering
\subfigure[]{
\begin{tikzpicture}[scale=0.4]
    \coordinate (i) at (1,-2);
    \coordinate (i_plus) at (0,0);
    \coordinate (j) at (-2,2);
    \coordinate (j_plus) at (0,2.5);
    \coordinate (C) at (-1.5,3);
    \coordinate (D) at (1.5,-3);
    \coordinate (A) at (-2.5,-0.5);
    \coordinate (B) at (2.5,0.5);
    \coordinate (R) at (-1.1,2.23);


    \draw[dashed, gray] (j) -- (j_plus);

    \draw[red, thick] (C) -- (D);
    \node[above, red] at (C) {$CD$};

    \draw[teal, thick] (A) -- (B);
    \node[above, teal] at (B) {$AB$};

    \fill (i) circle (2pt);
    \fill (i_plus) circle (2pt);
    \fill (j) circle (2pt);
    \fill (j_plus) circle (2pt);
    \fill (R) circle (2pt);

    \node[right] at (i) {$k$};
    \node[above] at (j) {$i$};
    \node[above] at (j_plus) {$i+1$};
    \end{tikzpicture}}
\subfigure[]
{  
     \begin{tikzpicture}[scale=0.4]
    
    \coordinate (i) at (1,-2);
    \coordinate (P) at (0.3,0.05);
    \coordinate (Q) at (-0.45,2.25);
    \coordinate (j) at (-2,2);
    \coordinate (k) at (1,2.5);
    \coordinate (C) at (-0.7,3);
    \coordinate (D) at (1.35,-3);
    \coordinate (E) at (-3.5,1.75);
    \coordinate (F) at (3,2.85);
    \coordinate (A) at (-2.5,-0.5);
    \coordinate (B) at (2.5,0.5);



    \draw[red, thick] (C) -- (D);
    \node[above, red] at (C) {$CD$};

    \draw[red, thick] (E) -- (F);
    \node[above, red] at (E) {$EF$};

    \draw[teal, thick] (A) -- (B);
    \node[above, teal] at (B) {$AB$};

    \fill (i) circle (2pt);
    \fill (j) circle (2pt);
    \fill (k) circle (2pt);
    \fill (P) circle (2pt);
    \fill (Q) circle (2pt);

    \node[right] at (i) {$k$};
    \node[above] at (j) {$i$};
    \node[above] at (k) {$j$};

     \end{tikzpicture}}
\subfigure[]{
    \begin{tikzpicture}[scale=0.4]

    \coordinate (i) at (1,-2);
    \coordinate (i_plus) at (0,0);
    \coordinate (j) at (-2,2);
    \coordinate (j_plus) at (0,2.5);
    \coordinate (l) at (-2,-1.75);
    \coordinate (C) at (-1.5,3);
    \coordinate (D) at (1.5,-3);
    \coordinate (E) at (-2.5,-1.7);
    \coordinate (F) at (1.5,-2);
    \coordinate (A) at (-2.5,-0.5);
    \coordinate (B) at (2.5,0.5);
    \coordinate (R) at (-1.1,2.23);


    \draw[dashed, gray] (j) -- (j_plus);

    \draw[red, thick] (C) -- (D);
    \node[above, red] at (C) {$CD$};

    \draw[red, thick] (E) -- (F);
    \node[left, red] at (E) {$EF$};

    \draw[teal, thick] (A) -- (B);
    \node[above, teal] at (B) {$AB$};

    \fill (i) circle (2pt);
    \fill (i_plus) circle (2pt);
    \fill (j) circle (2pt);
    \fill (j_plus) circle (2pt);
    \fill (l) circle (2pt);
    \fill (R) circle (2pt);

    \node[below left] at (i) {$k$};
    \node[above] at (j) {$i$};
    \node[above] at (j_plus) {$i+1$};
    \node[below] at (l) {$j$};

    \end{tikzpicture}}  
\subfigure[]{
    \begin{tikzpicture}[scale=0.4]

    \coordinate (i) at (1,-2);
    \coordinate (i_plus) at (0,0);
    \coordinate (j) at (-2,2);
    \coordinate (j_plus) at (0,2.5);
    \coordinate (k) at (-3.34,-1);
    \coordinate (C) at (-1.5,3);
    \coordinate (D) at (1.5,-3);
    \coordinate (E) at (-3.5,-1.2);
    \coordinate (F) at (-0.55,3);
    \coordinate (A) at (-2.5,-0.5);
    \coordinate (B) at (2.5,0.5);
    \coordinate (R) at (-1.1,2.23);


    \draw[dashed, gray] (j) -- (j_plus);

    \draw[red, thick] (C) -- (D);
    \node[above, red] at (C) {$CD$};

    \draw[red, thick] (E) -- (F);
    \node[left, red] at (E) {$EF$};

    \draw[teal, thick] (A) -- (B);
    \node[above, teal] at (B) {$AB$};

    \fill (i) circle (2pt);
    \fill (i_plus) circle (2pt);
    \fill (j) circle (2pt);
    \fill (j_plus) circle (2pt);
    \fill (k) circle (2pt);
    \fill (R) circle (2pt);

    \node[below left] at (i) {$k$};
    \node[above] at (j) {$i$};
    \node[above right] at (j_plus) {$i+1$};
    \node[below right] at (k) {$j$};

\end{tikzpicture}}
\caption{LS configurations for ladders at one loop and at two loops.
}
\label{LS_L1}
\end{figure}

From now on, we denote by $AB=(AB)_0$, $(AB)_1=CD$, $(AB)_2=EF$, $(AB)_3=GH$, \dots the $L+1$ loop lines, ordered as on the ladder graph. Moreover, we focus on the line configurations that involve all multiloop cuts
\begin{equation}\label{ladder_cuts}
    \langle AB CD \rangle = \langle CD EF \rangle = \langle EFGH \rangle = \dots = 0 \, .
\end{equation}
In fact, if any of these cut conditions is not satisfied, then the line configuration is equivalent to a product of lower-loop line configurations. All LS configurations for $L\leq 3$ ladder negative geometries are as follows. At $L=1$ there is only one possible LS configuration, shown in Fig.~\ref{LS_L1}(a). At $L=2$ there are three configurations, shown in Fig.~\ref{LS_L1}(b)-(d). 

\begin{table}[h]
\centering
\includegraphics{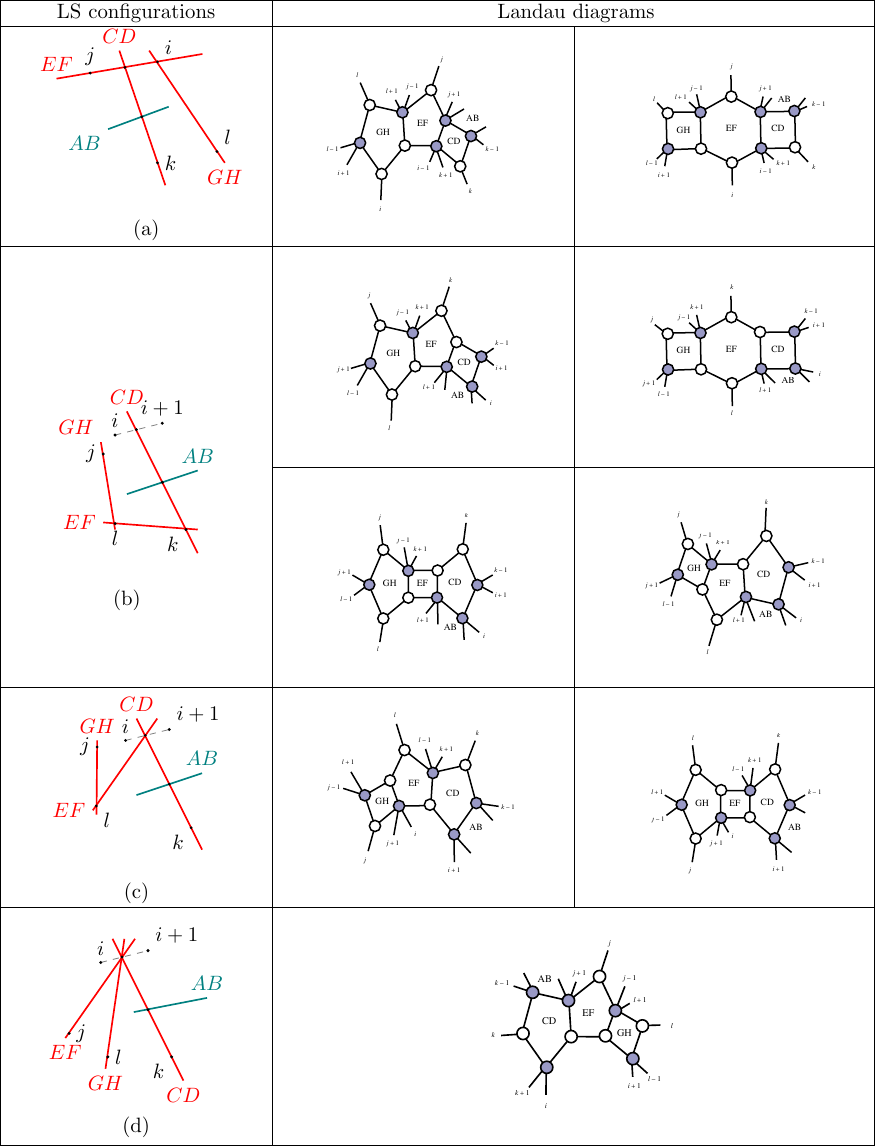}
\caption{All leading singularity line configurations and Landau diagrams for the three-loop ladder. For each leading singularity configuration, its Landau diagrams are related by merging three-point vertices and re-expanding them in different channels.}
\label{tab:11}
\end{table}
\begin{table}[h]
\centering
\includegraphics{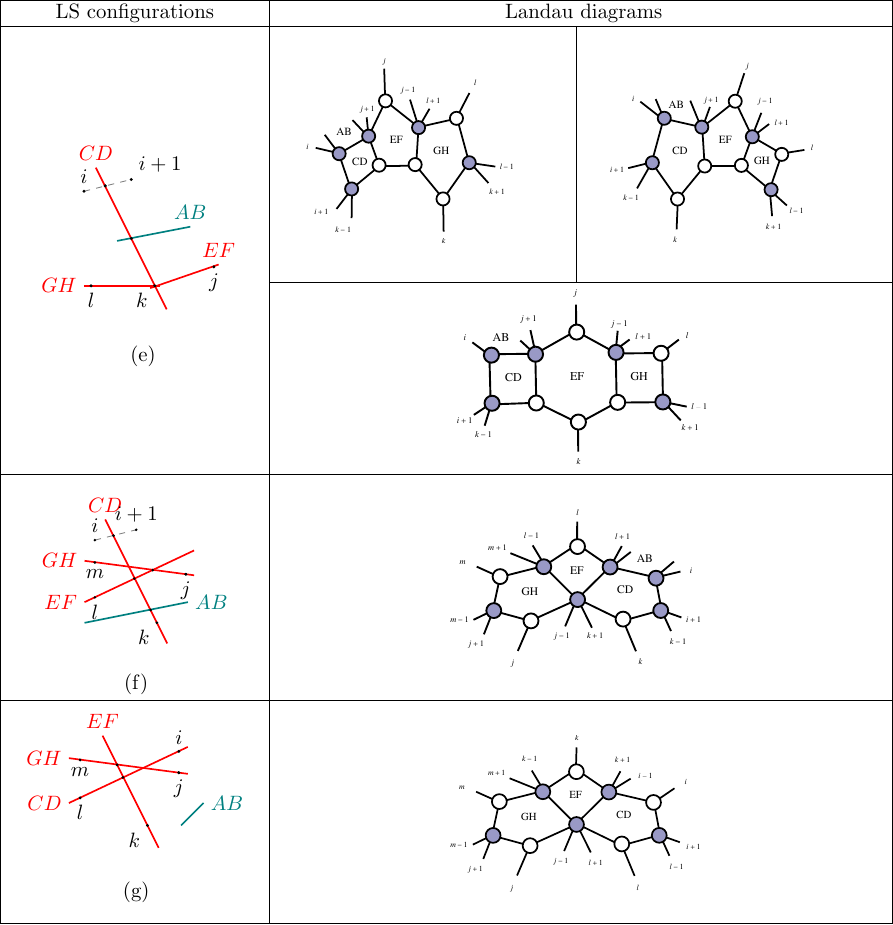}
\caption{All leading singularity line configurations and Landau diagrams for the three-loop ladder.}
\label{tab:12}
\end{table}

All LS configurations at $L=3$ are summarized in the first column of Table~\ref{tab:11} and \ref{tab:12} . Note that the indices in Table~\ref{tab:12} (f) satisfy $i< j<k<l<m \leq i \ , \ |j-m| >1$, and in Table~\ref{tab:12} (g) they satisfy $i< j<k<l<m < i $.

These results, together with the classification of LS configurations for ladders at any number of loops, are proved in Appendix~\ref{app:ladder proof}. The method used in the proof is analogous to those in~\cite{Brown:2025plq}; here we briefly outline the main idea. Each loop line in $\mathbb{P}^3$ has four degrees of freedom. In a LS configuration for the ladder, each loop line other than $AB$ is fully localized by imposing the conditions of the form ${\langle AB_\ell \, ii+1 \rangle} = 0$ and ${\langle AB_\ell \, AB_{\ell + 1} \rangle = 0}$, where $ AB = AB_0$. By counting the degrees of freedom, each loop line is already localized to some lower-dimensional boundary of the respective one-loop Amplituhedron before imposing any multiloop cut. These boundaries have been classified in \cite{Ranestad:2024svp}. We use a positive parameterization for each loop in such a boundary, and impose the multiloop cut conditions. External positivity together with the one-loop MHV conditions restricts the allowed solutions. In particular, this allows for an inductive argument for the ladders of arbitrary number of loops.

\subsection{All-loop leading singularity values of ladders }
\label{section_LS_L1}
We then proceed to \textit{values} of the LS for ladders at $L$ loops, as in eq. \eqref{F_integral_in_LS}, and the linear space generated by them, which are computed from maximal residues associated to all LS configuration we discussed in the last section. This knowledge follows from the explicit construction of the  ladder integrands, see Appendix \ref{app:integrand}. 

According to \eqref{F_integral_in_LS}, and thanks to the explicit knowledge of the integrand for ladders, this integral can be decomposed as
\begin{equation}\label{ladder_function}
    F_{n,ladder}^{(L)} = \sum_{(ij)} \Omega_{n,ij} \, f_{n,ij}^{(L)} \ ,
\end{equation}
where $f^{(L)}_{n,ij}$ are pure functions of transcendental weight $2L$ and $\Omega_{n,ij}$ are the LS. In this case, the indices $(ij)$ run over the arcs of an $n$-gon, i.e. $1 \leq i < j \leq n $ and $i$, $j$ are not adjacent. 
Hence, there are $n(n-3)/2$ LS for ladders, and one checks that these are all linearly independent. 
The rational function $\Omega_{n,ij}$ is the canonical function of the $n$-point one-loop MHV Amplituhedron in $AB := (AB)_0$, with a single extra condition $\langle AB ij \rangle <0$. This defines a positive geometry, whose canonical form can be obtained by triangulating into Kermit spaces \cite{Brown:2025plq}:
\begin{align}\label{Omega_ij}
    \Omega_{n,ij} &= \sum_{\substack{\Delta_1, \,  \Delta_2  \, \subset \, T(ij)  }} [\Delta_1;\Delta_2] \,,
\end{align}
where the sum is over pairs of non-overlapping triangles $\Delta_1 = \{a_1,b_1,c_1\}$ and $\Delta_2 = \{a_2,b_2,c_2\}$, with $i \leq a_1<b_1<c_1 \leq j$ and $j \leq a_2<b_2<c_2 \leq i$,
and with arcs in a triangulation $T(ij)$ of the $n$-gon containing the arc $(ij)$. The function $[\Delta_1;\Delta_2]$ denotes the \textit{Kermit forms}, which are given by
\begin{equation}\label{six_invariant}
    [a_1 b_1 c_1 ; a_2 b_2 c_2] = \frac{\langle AB  (a_1 b_1 c_1) \cap (a_2 b_2 c_2) \rangle ^{2}}{\langle AB a_1 b_1 \rangle \langle AB b_1 c_1 \rangle \langle AB a_1 c_1 \rangle \langle AB a_2 b_2 \rangle \langle AB b_2 c_2 \rangle \langle AB a_2 c_2 \rangle } \ ,
\end{equation}
with the special case 
\begin{equation}\label{four_invariant}
    [abcd] := [a b c ; c d a] =  -\frac{\langle abcd \rangle ^{2}}{\langle AB ab \rangle \langle AB bc \rangle \langle AB cd \rangle \langle AB da \rangle } \ ,
\end{equation}
respectively. Eq. \eqref{Omega_ij} also computes one-loop integrand $\Omega_n^{(1)}$ of the MHV amplitude, by the same formula with $\Delta_1,\Delta_2$ running over all non-overlapping triangles with arcs in any fixed triangulation $T$ of the $n$-gon.

Consider for example $n=5$. At all loops, we only have five LS for ladder negative geometries and its four (twisted) cyclic shifts, see \eqref{twist_symm}. All five LS are labeled by five diagonals of a pentagon. In terms of equations, it reads
\begin{equation}\label{eq:5ptLSladders}
    \raisebox{-3em}{\begin{tikzpicture}[scale = 1]

    \draw[thick] (90:1) 
        \foreach \x in {162,234,306,18} 
            { -- (\x:1) } -- cycle;

    \coordinate (D) at (18:1);
    \coordinate (C) at (90:1);
    \coordinate (B) at (162:1);
    \coordinate (A) at (234:1);
    \coordinate (E) at (306:1);

    \draw[red, thick] (A) -- (C);

    \node[below] at (A) {1};
    \node[left] at (B) {2};
    \node[above] at (C) {3};
    \node[right] at (D) {4};
    \node[below] at (E) {5};\end{tikzpicture}}\qquad \ \  \Omega_{5,13} = [1234] + [123;145] \, ,
\end{equation}
which is equal to the combination $\Omega_{5,13} = r_2-r_0$ in \cite{Chicherin:2024hes}, after identifying $AB\to I_\infty $.  
At $n=6$, ladder negative geometries have nine LS, which are organized into two dihedral orbits labeled by two kinds of diagonals of a hexagon, as follows,
\begin{equation}\label{six-point LS}
\begin{aligned}
    \raisebox{-3.5em}{\begin{tikzpicture}
    
    \draw[thick] (0:1) 
        \foreach \x in {60,120,180,240,300} 
            { -- (\x:1) } -- cycle;

    \coordinate (E) at (0:1);
    \coordinate (D) at (60:1);
    \coordinate (C) at (120:1);
    \coordinate (B) at (180:1);
    \coordinate (A) at (240:1);
    \coordinate (F) at (300:1);

    \draw[red, thick] (A) -- (C);
    
    \node[below] at (A) {1};
    \node[above ] at (C) {3};
    \end{tikzpicture}} &\qquad \ \ \Omega_{6,13}= [1234] + [123;145]+[123;156] \, , \\
    \raisebox{-3.5em}{\begin{tikzpicture}
    
    \draw[thick] (0:1) 
        \foreach \x in {60,120,180,240,300} 
            { -- (\x:1) } -- cycle;

    \coordinate (E) at (0:1);
    \coordinate (D) at (60:1);
    \coordinate (C) at (120:1);
    \coordinate (B) at (180:1);
    \coordinate (A) at (240:1);
    \coordinate (F) at (300:1);

    \draw[red, thick] (A) -- (D);
    
    \node[below] at (A) {1};
    \node[above] at (D) {4};
    \end{tikzpicture}} &\qquad \ \ \Omega_{6,14}= [123;145] +[123;156]  + [1345] + [134;156] \, .
\end{aligned}
\end{equation}

From the perspective of LS configurations, the value of the LS is completely determined by the location of $CD$.  If $CD=ij$, the LS is equal to $\Omega_{n,ij}$, while if $CD=k,AB \cap (ijk)$, as all LS in Fig.~\ref{LS_L1} and as Table \ref{tab:11} and \ref{tab:12}  (a)... (f), the LS can be expressed as a sum of Kermit forms, according to~\cite{Brown:2025plq}. In particular, according to this analysis, the values of the LS of ladders saturate at the latest $L=2$. In fact, thanks to the explicit construction of the ladder integrands, see Appendix~\ref{app:integrand}, we know that the LS saturate at $L=1$, and can be represented by $\Omega_{n,ij}$ for $1 \leq i<j \leq n$.

In this section, we provided a complete classification of leading singularity configurations of negative geometries with the topology of a ladder with a marked node at one end, see Fig.~\ref{fig:laddersgeneral}, together with their associated values. In the next section, this is the starting point for studying the singular loci of eq.~\eqref{F_integral_in_LS} via the geometric Landau analysis procedure.

\section{Landau analysis for negative geometries}
\label{sec:landa and schubert}

In the previous section, we classified and evaluated the leading singularities (LS) of integrals as in eq. \eqref{F_integral_in_LS} for negative geometries with the topology of a ladder. Our goal is now to compute all singular loci of the transcendental functions $f_{n,ij}^{(L)}$ in \eqref{ladder_function}.

The strategy we adopt is that of a \textit{geometric Landau analysis}, which combines the usual Landau analysis for determining the singular loci of a Feynman integral with the underlying positive geometry description for the integrand. This method has been proposed in the context of amplitudes, making use of the geometry of Amplituhedron in \cite{Dennen:2015bet,Dennen:2016mdk,Prlina:2017azl,Prlina:2017tvx}. We follow the same logical steps as in those references, but adapt them to the setting of our interest, namely to integrated negative geometries. The algorithm can be summarized as follows \cite[Section 4]{Dennen:2016mdk}:

\begin{itemize}
    \item {\it \underline{Step 1:}} Classify all leading singularity (LS) configurations of the integral. These correspond to the maximal codimension boundaries of the geometry.
    \item {\it \underline{Step 2:}} To each LS configuration associate \textit{leading} Landau diagrams, whose edges represent (an independent subset of) the cut conditions satisfied on that configuration. 
    \item {\it \underline{Step 3:}} Consider all lower codimension
    boundaries that can be obtained by omitting various subsets of these cut conditions, and eliminate those which do not have full-dimensional intersection with the geometry. We call the latter \textit{residual}, and the other ones \textit{physical}. The selection of physical boundaries can be implemented either via the knowledge of the integrand's numerator, or via on-shell diagrams, as explained in Appendix~\ref{sec: sel rules and bipartite}.
    \item {\it \underline{Step 4:}} For every physical boundary, solve the corresponding Landau equations. The solutions to these equations correspond to the locations of branch points of the integral. Moreover, one would recover all branch points by considering all physical boundaries.
\end{itemize}

Steps 3 and 4 have been conjectured in~\cite{Prlina:2017azl,Prlina:2017tvx}, where the authors provided a heuristic explanation related to the Cutkosky rules. In our setting, we use this algorithm and successfully compute ladder negative geometries in Section~\ref{sec: bootstrapping integrated negative geometries}. The validity of this algorithm, as well as the details on how to implement it, in particular in the setting of negative geometries, are explained in Section~\ref{sec: sel rules and bipartite}.

\subsection{Illustrative example of the method}

Let us work through this procedure with an explicit example for a ladder negative geometry at two loops. To best illustrate the power of the algorithm above, we follow the first step, and then solve the Landau equations for various relaxations, both physical and residual ones. We then explain how and why we select only the physical Landau singularities, and check the procedure on the explicit results computed in this paper. 

\begin{figure}[t]
    \centering 
    \subfigure[]{\label{fig:L2.II_LD.a}\includegraphics[width=0.4\linewidth]{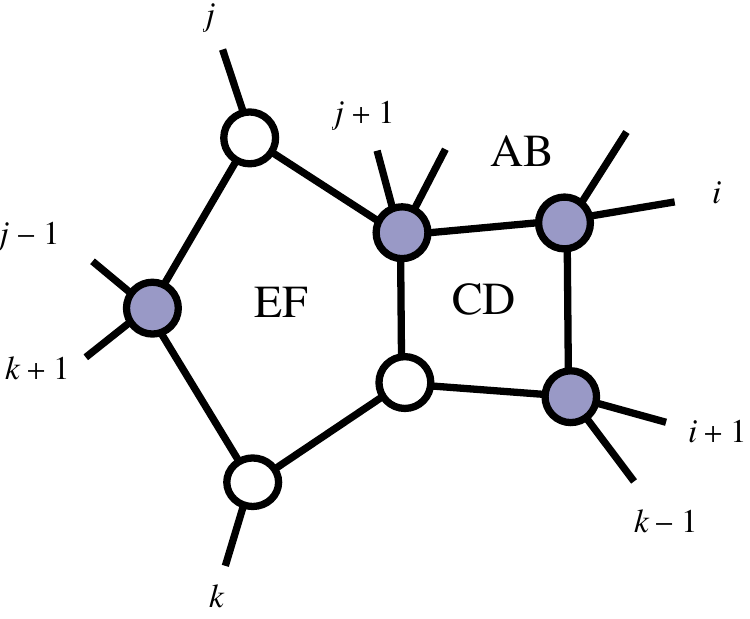}}\quad \quad \subfigure[]{\label{fig:L2.II_LD.b}\includegraphics[width=0.4\linewidth]{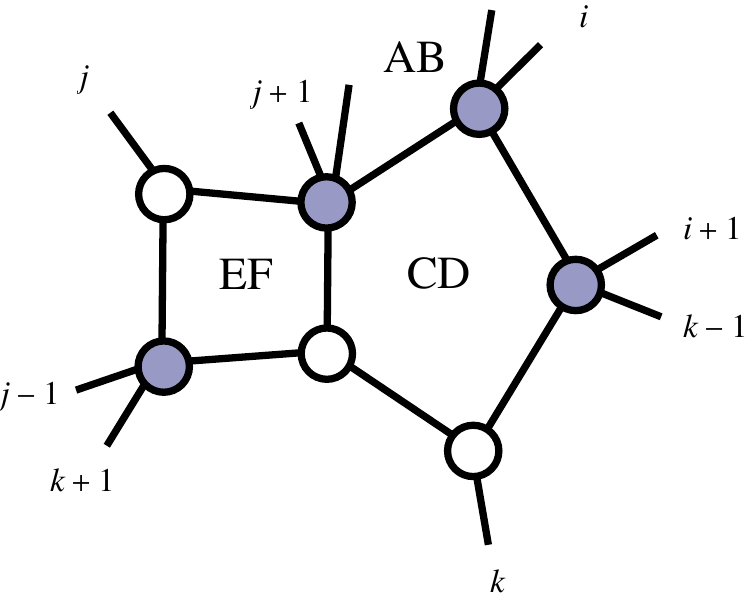}  }
    \caption{Leading Landau diagrams associated to the leading singularity configuration Fig.~\ref{LS_L1}(c). 
    Diagrams (a) and (b) are obtained from eq.~\eqref{exp_boundary} pinching the propagators $\langle CD \, kk{+}1 \rangle$ and $\langle EF \, k{-}1 k \rangle$, respectively. They are related by merging two white three-point vertices and re-expanding them in different channels. The Landau diagrams are drawn with the help of Mathematica package {\tt positroids}~\cite{Bourjaily:2012gy}.}
    \label{fig:L2.II_LD}
\end{figure}

The first step is to start with a leading singularity configuration, and to associate Landau diagrams to it. 
As an example, we choose the LS configuration Fig.~\ref{LS_L1}(c).  This configuration involves nine propagator cuts,
\begin{equation}\label{exp_boundary}
\begin{aligned}
    \langle EF k{-}1k\rangle &=\langle EFkk{+}1\rangle=\langle EF j{-}1j\rangle=\langle EFjj{+}1\rangle=0 \, , \\
    \langle CD k{-}1k\rangle &=\langle CD kk{+}1\rangle=\langle CDii{+}1\rangle=\langle CDAB\rangle=\langle CDEF\rangle=0 \, .
\end{aligned} 
\end{equation}
Note that every loop line has four degrees of freedom, so $CD$ satisfies more conditions than its number of degrees of freedom. In fact, one can easily show that on the support of $\langle CDEF\rangle=0$, only three out of the four vanishing conditions associated to
\begin{equation}\label{CD_EF_cut}
    \langle EF k{-}1k\rangle \, , \  \langle EFkk{+}1\rangle \, , \  \langle CD k{-}1k\rangle \, , \ \langle CDkk{+}1\rangle\ , 
\end{equation}
are independent. This follows because each loop line, $CD$ and $EF$, lives on an MHV Amplituhedron, and therefore an $\overline{{\rm MHV}}$-like solution to 
\begin{equation}
    \langle EF k{-}1k\rangle =\langle EFkk{+}1\rangle = 0 \, ,
\end{equation}
given by $EF$ lying in the plane $(k-1,k,k+1)$, it is not physical as long as $EF$ satisfies at least another cut in $\langle EF   j{-}1j\rangle = 0  $. Therefore, any subset of three conditions among those in eq. \eqref{CD_EF_cut} together with $\langle CDEF\rangle=0$ forces both $CD$ and $EF$ to pass through $k$.
Up to relabeling, the LS configuration Fig.~\ref{LS_L1}(c) is then the maximal cut solution of one of the two leading Landau diagrams in Fig.~\ref{fig:L2.II_LD}.
The other two two-loop LS configurations in Fig.~\ref{LS_L1}, and the three-loop LS configurations for ladders, yield leading Landau diagrams which are displayed in Fig.~\ref{fig:L2ladderLD} and Table~\ref{tab:11} and \ref{tab:12}, respectively.

Equipped with the leading Landau diagrams, we systematically apply the Landau analysis to obtain a prediction for the set of singularities of the corresponding integrated negative geometry. We can write the Landau equations~\cite{Landau:1959fi} for the diagram in Fig.~\ref{fig:L2.II_LD.a}:
\begin{equation*}
\begin{split}
    &\alpha_{1}(y_{CD}-x_k)^2=\alpha_{2}(y_{CD}-x_{i{+}1})^2=\alpha_{3}(y_{CD}-x_{AB})^2=\beta(y_{CD}-y_{EF})^2 =\\
    &\alpha_{4}(y_{EF}-x_k)^2=\alpha_{5}(y_{EF}-x_{k{+}1})^2=\alpha_{6}(y_{EF}-x_j)^2=\alpha_7(y_{EF}-x_{j{+}1})^2 = 0 \, ,\\[0.3cm]
    &\alpha_{1}(y_{CD}-x_k)^\mu+\alpha_{2}(y_{CD}-x_{i{+}1})^\mu+\alpha_{3}(y_{CD}-x_{AB})^\mu+\beta(y_{CD}-y_{EF})^\mu {=} 0 \, , \\[0.3cm]
    &\alpha_{4}(y_{EF}{-}x_{k})^\mu{+}\alpha_{5}(y_{EF}{-}x_{k{+}1})^\mu
    {+}\alpha_{6}(y_{EF}{-}x_{j{-}1})^\mu{+}\alpha_{7}(y_{EF}{-}x_{j{+}1})^\mu{+}\beta(y_{CD}{-}y_{EF})^\mu = 0 \, ,
\end{split}
\end{equation*}
where the first two lines are the cut conditions, and the last two lines are the pinch conditions for $y_{CD}$ and $y_{EF}$, respectively.
The leading Landau singularity lies in the branch where all $\alpha_i,\beta\neq0$. Let us first look at the Landau equations for $y_{EF}$. Note that we are dealing with equations in spacetime dimension four, therefore any five vectors are always linear dependent.
Consequently, $\{\alpha_4,\cdots,\alpha_7,\beta\}$ always admit a non-trivial solution to the pinch condition for $y_{EF}$, for every $y_{CD}$ and $y_{EF}$. The pinch condition for $y_{EF}$ therefore does not lead to constraints on the external kinematics. 
From the four conditions in the second line of the cut conditions, $y_{EF}$ is fully localized, and two solutions are obtained for $y_{EF}$. In momentum twistors, these two 
solutions are $(EF)_1=jk$, $(EF)_2=\bar j\cap\bar k$ \cite{Arkani-Hamed:2010pyv}, where $\bar{i} = (i-1,i,i+1)$.

According to our geometric picture, we should select the MHV solution $EF=jk$, which is the solution in Fig.~\ref{LS_L1}(c). In fact, the numerator of the integrand vanishes on the support of the other solution, since the underlying one-loop geometries are MHV individually in each loop.
Plugging this solution in the conditions involving $y_{CD}$, the remaining cut conditions in momentum twistors are
\begin{equation}\label{eq:pre}
  \beta\langle CDjk\rangle =\alpha_1\langle CD k{-}1k\rangle=\alpha_2\langle CDii{+}1\rangle=\alpha_3\langle CDAB\rangle =0 \, , 
\end{equation}
and the $y_{EF}$ in pinch condition for $y_{CD}$ is also replaced by $y_{EF}^*$ which is dual to $EF=jk$. Again, to guarantee that non-trivial solutions exist for $\{\beta,\alpha_1,\alpha_2,\alpha_3\}$, $CD$ should be localized at the maximal cut solutions of \eqref{eq:pre}. However, the pinch condition of $y_{CD}$ requires further constraints on the external kinematics. To express this condition, we take the dot product of the LHS of the pinch equation for  $y_{CD}$ with the four vectors $(y_{CD}-x_k)^\mu$, $\dots$, $(y_{CD}-y_{EF})^\mu$ respectively. Each of these vectors is light-like, and some elementary manipulations we obtain the linear equation
\begin{equation}\label{eq:pinchmat}
    \left(\begin{matrix}
        0 & (y_{EF}^*{-}x_k)^2 &(y_{EF}^*{-}x_{i{+}1})^2 & (y_{EF}^*{-}x_{AB})^2\\
        (y_{EF}^*{-}x_k)^2 & 0& (x_{k}{-}x_{i{+}1})^2&(x_k{-}x_{AB})^2\\
        (y_{EF}^*{-}x_{i{+}1})^2 & (x_{k}{-}x_{i{+}1})^2&0& (x_{i{+}1}{-}x_{AB})^2\\
        (y_{EF}^*{-}x_{AB})^2 & (x_k{-}x_{AB})^2 & (x_{i{+}1}{-}x_{AB})^2&0
    \end{matrix}\right) \, \cdot\left(\begin{matrix}
        \beta\\\alpha_1\\\alpha_2\\\alpha_3
    \end{matrix}\right)=0 \, ,
\end{equation} 
Hence there exists a non-trivial solution for $(\beta,\alpha_1,\alpha_2,\alpha_3)$ if and only if this matrix has vanishing determinant. We can rewrite the matrix in eq.~\eqref{eq:pinchmat} in momentum twistors using the relation $(x_i-x_j)^2\langle i{-}1i \, I_\infty\rangle\langle j{-}1j \, I_\infty\rangle=\langle i{-}1ij{-}1j\rangle$, as 
\begin{equation}
    \left(\begin{matrix}
        0 & 0 &\langle jk ii{+}1\rangle & \langle jk AB\rangle\\
        0 & 0& \langle k{-}1kii{+}1\rangle&\langle k{-}1k AB\rangle\\
        \langle jk ii{+}1\rangle & \langle k{-}1kii{+}1\rangle&0& \langle ii{+}1 AB\rangle\\
        \langle jk AB\rangle & \langle k{-}1k AB\rangle & \langle ii{+}1 AB\rangle &0
    \end{matrix}\right) \, .
\end{equation}
and the vanishing of its determinant yields the condition 
\begin{equation}\label{sing_a}
    \langle k(k{-}1j)(ii{+}1)(AB)\rangle  \, .
\end{equation}
Therefore, the singular locus of eq.~\eqref{sing_a} is a candidate singularity of the integrated two-loop ladder.

\begin{figure}
    \centering
    \subfigure[]{\label{fig:MHV-2L}\includegraphics[width=0.4\linewidth]{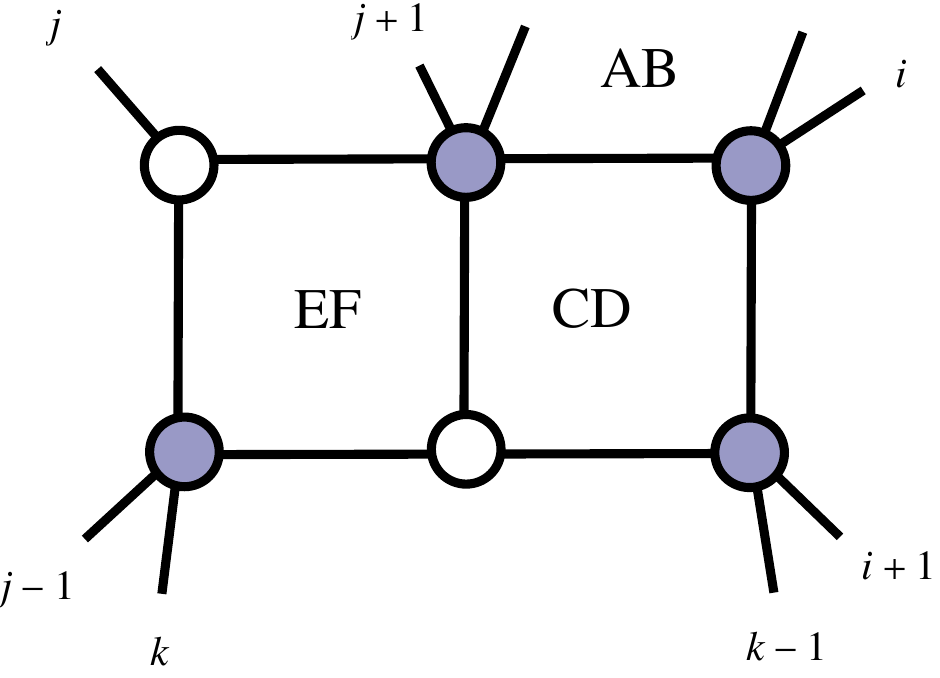}}\quad\quad
     \subfigure[]{\label{fig:wrongshrink}\includegraphics[width=0.4\linewidth, trim= 0 -0.2cm 0 0]{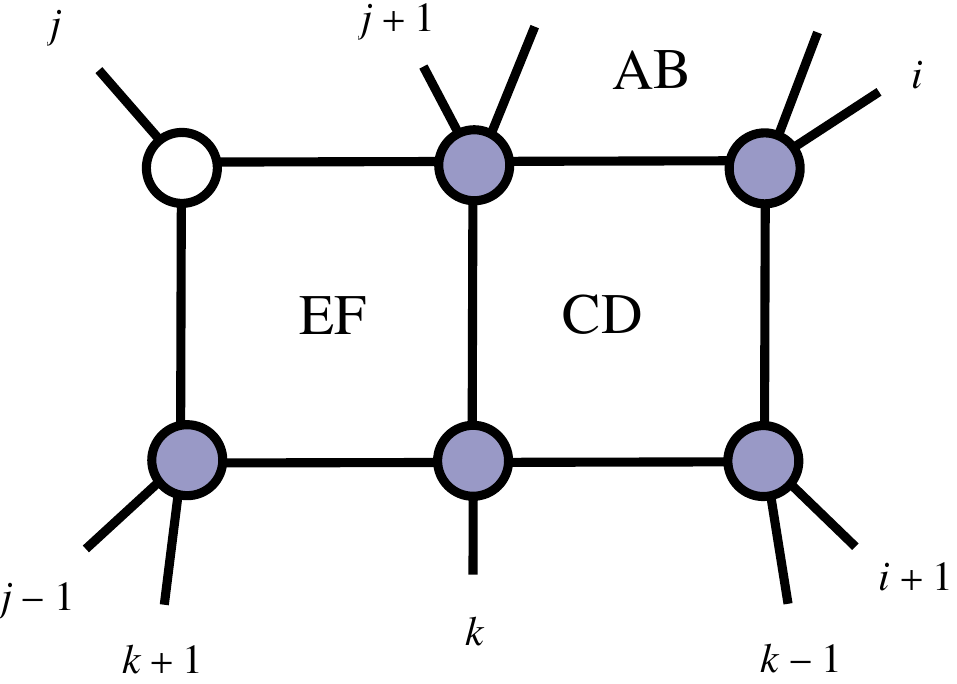}}\caption{Sub-topologies obtained from shrinking one propagator in Fig.~\ref{fig:L2.II_LD}, together with the corresponding bi-coloring of minimal helicity weight. In diagrams (a) and (b), we shrink propagators $\langle EFk{+}1k\rangle$ and $\langle EFk{-}1k\rangle$ in Fig.~\ref{fig:L2.II_LD}, respectively.}
\end{figure}
In addition to the leading Landau singularities, according to \cite{Dennen:2016mdk} we have to consider all possible relaxations with one or more $\alpha_i=0$. This corresponds to omitting some cut conditions, i.e. to pinch some propagators in the leading Landau diagrams. In our running example, we consider the sub-leading Landau diagram obtained from Fig.~\ref{fig:L2.II_LD} by shrinking the propagator $\langle EFkk{+}1\rangle$, and obtain the diagram in Fig.~\ref{fig:MHV-2L}. There are only seven cut conditions for this double-box diagram. Therefore, the loop momenta $CD$ and $EF$ are localized to a one-dimensional space parametrized by $\alpha$ as
\begin{equation}\label{eq:partphy}
    (EF)=j \star \, , \quad   CD= (AB\star)  \cap(ii{+}1 \star) \, , \quad \text{with} \ Z_\star=Z_{k{-}1}+\alpha \, Z_k \, .
\end{equation}
To obtain the associated Landau singularities, we also have to solve the pinch conditions for $EF$ and $CD$. Following a similar procedure as we did for~\eqref{eq:pinchmat} and~\eqref{sing_a}, the pinch conditions are equivalent to 
\begin{align}\label{CD bracket}
\langle CDj (k{-}1k)\cap\bar{j}\rangle=\langle EFii{+}1\rangle\langle k{-}1kAB\rangle - \langle EFAB\rangle\langle ii{+}1k{-}1k\rangle=0 \, .
\end{align}
Imposing this equation fixes $\alpha$ such that $Z_\star=(k{-}1k)\cap\bar{j}$, and the corresponding Landau singularity is given by the vanishing of
\begin{equation}\label{sing_b}
    \langle \star (k{-}1j)(ii{+}1)(AB)\rangle  \, .
\end{equation}

Another interesting sub-leading case is obtained by shrinking the propagator $\langle EFk{-}1k\rangle$ in Fig.~\ref{fig:L2.II_LD}, which yields the diagram in Fig.~\ref{fig:wrongshrink}.
Applying the same analysis as above, we find that $EF = j(kk{+}1)\cap\bar{j}$, and $CD$ solves a four-mass box cut: 
\begin{equation}
    \langle CD j(kk{+}1)\cap\bar{j}\rangle=\langle CDk{-}1k\rangle=\langle CDii{+}1\rangle=\langle CDAB\rangle=0 \, .
\end{equation}
It is well-known \cite{Arkani-Hamed:2010pyv} that the leading singularity of such a box is 
\begin{equation}
\frac1{\langle k{-}1k AB\rangle\langle (kk{+}1j)\cap\bar{ j}ii{+}1\rangle\sqrt{(1{-}u{-}v)^2{-}4u v}},
\end{equation}
where
\begin{equation}\label{badshrin}
\begin{aligned}
  u=\frac{\langle k{-}1k ii{+}1\rangle\langle (kk{+}1j)\cap\bar{ j}AB\rangle}{\langle k{-}1k AB\rangle\langle (kk{+}1j)\cap\bar{ j}ii{+}1\rangle} \, , \quad v=\frac{\langle  ii{+}1AB\rangle\langle k{-}1k(kk{+}1j)\cap\bar{j}\rangle}{\langle k{-}1k AB\rangle\langle (kk{+}1j)\cap\bar{j}ii{+}1\rangle} \, .
\end{aligned}
\end{equation}
For generic indices $i,j,k \in \{1,\dots,n\}$, this is a non-trivial square root for momentum twistors.

The Landau analysis is now applied further to all sub$^m$-leading cases with $m \geq 1$, which are obtained from leading Landau diagrams by pinching $m$ propagators. The last two examples illustrated some cases for $m=1$.

\begin{figure}[t]
\centering

\subfigure[]{\label{fig:5a}\begin{tikzpicture}
    \coordinate (A) at (0,0.85);   
    \coordinate (B) at (1.85,2.1); 
    \coordinate (C) at (1.9,0.05); 
    \coordinate (D) at (2,2); 
    \coordinate (E) at (4,1); 
    \coordinate (F) at (4.7,1.7); 

    \fill[gray!30] (A) to[out=45,in=200] (B) 
                   to[out=-65,in=85] (C) 
                    to[out=180,in=-40] (A) ;


    \draw[thick,magenta] (-0.3,0.5) to[out=50,in=200] (2.3,2.3);

    \draw[thick] (1.8,2.3) to[out=-70,in=80] (1.8,-0.4);

     \draw[thick,magenta] (2.3,0) to[out=180,in=-50]  (-0.3,1.2);



    \fill[teal] (A) circle (3pt) node[left] {$x$};
    \fill[teal] (B) circle (3pt);
    \fill[teal] (C) circle (3pt);
    \fill[teal] (A) circle (3pt);



    \node[below] at (-0.2,0.5) {\textbf{$S_1$}};
    \node[left] at (-0.3,1.3) {\textbf{$S_2$}};
    
    \node at (1.2,1) {\Huge $\mathcal{A}$};
\end{tikzpicture}}
\quad
\subfigure[]{\label{fig:5b}\begin{tikzpicture}[xshift=4.2 cm]

    \coordinate (A) at (0,0.85);   
    \coordinate (B) at (1.85,2.1); 
    \coordinate (C) at (1.9,0.05); 
    \coordinate (D) at (3.8,1); 
    \coordinate (E) at (4,1); 
    \coordinate (F) at (4.7,1.7); 

    \fill[gray!30] (A) to[out=45,in=200] (B) 
                   to[out=-65,in=85] (C) 
                    to[out=180,in=-40] (A) ;


    \draw[thick] (-0.3,0.5) to[out=50,in=200] (2.3,2.3);

    \draw[thick] (1.8,2.3) to[out=-70,in=80] (1.8,-0.4);

     \draw[thick,magenta] (4,1.5) to[out=250,in=0] (2.3,0) to[out=180,in=-50]  (-0.3,1.2);

     \draw[thick,magenta] (1.45,2.4) to[out=-40,in=180]  (4.3,1);



    \fill[teal] (A) circle (3pt);
    \fill[teal] (B) circle (3pt);
    \fill[teal] (C) circle (3pt);

    \fill[orange] (D) circle (3pt) node[below right ] {$x$};



    \node[left] at (-0.3,1.3) {\textbf{$S_1$}};
    \node[left] at (1.45,2.4) {\textbf{$\bar{S}_1$}};
    
    \node at (1.2,1) {\Huge $\mathcal{A}$};
\end{tikzpicture}}
\quad
\subfigure[]{\label{fig:5c}\begin{tikzpicture}[xshift=9.5 cm]

    \coordinate (A) at (0,0.85);   
    \coordinate (B) at (1.85,2.1); 
    \coordinate (C) at (1.9,0.05); 
    \coordinate (D) at (3.8,1); 
    \coordinate (E) at (4,1); 
    \coordinate (F) at (4.7,1.7); 

    \fill[gray!30] (A) to[out=45,in=200] (B) 
                   to[out=-65,in=85] (C) 
                    to[out=180,in=-40] (A) ;


    \draw[thick] (-0.3,0.5) to[out=50,in=200] (2.3,2.3);

    \draw[thick] (1.8,2.3) to[out=-70,in=80] (1.8,-0.4);

     \draw[thick] (2.3,0) to[out=180,in=-50]  (-0.3,1.2);

     \draw[thick,magenta] (1.45,2.4) to[out=-40,in=180]  (4.3,1);

     \draw[thick,magenta] (1.3,-0.35) to[out=40,in=250]  (4,1.4);



    \fill[teal] (A) circle (3pt);
    \fill[teal] (B) circle (3pt);
    \fill[teal] (C) circle (3pt);

    \fill[magenta] (D) circle (3pt) node[below right ] {$x$};



    \node[left] at (1.45,2.4)  {\textbf{$\bar{S}_1$}};
    \node[left] at (1.3,-0.35) {\textbf{$S_1$}};
    
    \node at (1.2,1) {\Huge $\mathcal{A}$};
\end{tikzpicture}}
\caption{
Three possible geometric pictures for the location of a Landau singularity with respect to the positive geometry $\mathcal{A}$. The singularities shown in figures (a) and (b) are \textit{physical}, since they lie at least on a boundary component of $\mathcal{A}$. (c), instead, is \textit{spurious}, as it lies on the intersection of only residual strata. }
\label{fig:selections}
\end{figure}
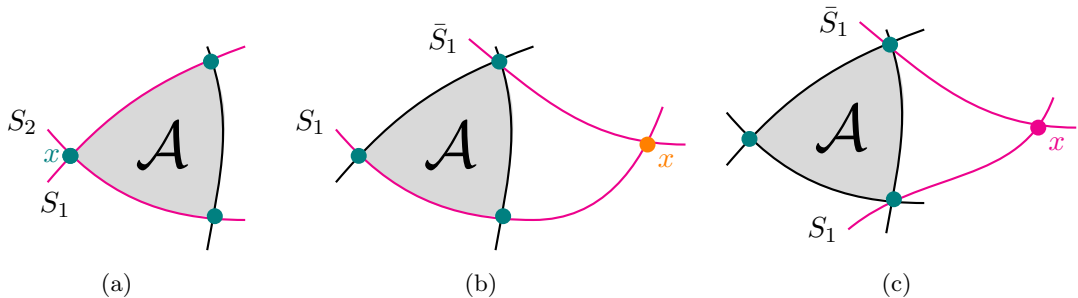

So far, we did not really use the geometric information when performing the Landau analysis. In fact, the geometry entered only for the determination of the LS configurations in section \ref{sec:lead sing}. We now explain how the underlying positive geometric description leads to a selection rule that partitions the Landau singularities into physical and spurious ones. 
By spurious we mean configurations for which a the denominator leads to a singularity, with however is absent thanks to a zero of the numerator (on the cut locus of the Landau diagram)  \cite{Prlina:2017tvx}. This selection rule therefore refines the Landau analysis.
At this stage, we describe the possible scenarios for where a Landau singularity, represented by points $x$ in Fig.~\ref{fig:selections}, is located with respect to the positive geometry~$\mathcal{A}$. We locate each Landau singularity found in the example above to one of the different scenarios.

\begin{enumerate}[label=(\alph*)]
    \item We start from the leading Landau singularity in eq.~\eqref{sing_a}, 
    whose associated Landau diagrams are in Fig.~\ref{fig:L2.II_LD}.
    This corresponds to Fig.~\ref{fig:5a}, which is a vertex of the geometry~$\mathcal{A}$. All such points are in fact leading Landau singularities by construction. On the other hand, sub-leading Landau singularities arise from intersecting the cut conditions, defining a variety $S_1$ in kinematic space, with the pinch conditions, defining another variety $\bar{S}_1$. This can also be understood from the perspective of {\it composite residues} of the integrand. For instance, the double-box cases considered in the last subsection, involve one cut condition, e.g. $\langle CDEF \rangle = 0$ in Fig. \ref{fig:MHV-2L}, that factorizes on the support of the other cut conditions. The Landau singularity is then obtained by equating both factors to zero, therefore fixing all degrees of freedom. This is exactly the notion of a composite residue. Geometrically, the associated variety factorizes as $S_1 \cup \bar{S}_1$, where $S_1$ usually corresponds an MHV-like cut condition, while $\bar{S}_1$ corresponds an $\overline{\text{MHV}}$-like condition. The latter is in fact the same as the pinch condition. The Landau singularity then corresponds to a point in $S_1 \cap \bar{S}_1$.
    \item The prototypical example of a physical sub-leading singularity is given by~\eqref{sing_b}, and it is depicted in Fig.~\ref{fig:5b}. As the figure suggests, the solution to the Landau equations localizes outside the geometry, since \eqref{eq:partphy} with \eqref{CD bracket} yields
    \begin{equation}\label{ineq}
   \langle EF k k +1 \rangle = \langle j(k{-}1k)\cap\bar{j}\ kk{+}1\rangle=
   \langle j \bar{k}\rangle\langle k\bar j \rangle\leq 0 \, ,
    \end{equation}
    indicating that $EF$ localizes outside the geometry if the inequality in \eqref{ineq} is strict, i.e. if $j \notin \{k-1,k,k+1\}$ and $k \notin \{j-1,j,j+1\}$.
    Nevertheless, we claim that this singularity is physical, because the variety~$S_1$ defined by the cut conditions of the diagram in Fig.~\ref{fig:MHV-2L} is a boundary of~$\mathcal{A}$, i.e. it intersects~$\mathcal{A}$ in maximal real dimension. 
    
     This can be checked via on-shell diagrams as explained in Appendix~\ref{app:ladder proof}. Then, by the definition of a positive geometry, the fact that $S_1$ is a boundary is equivalent to the numerator of the integrand not vanishing on $S_1$. In turn, this indicates that the singularity is physical~\cite{Prlina:2017tvx}. In fact, it appears in the symbol of the six-point two-loop ladder, whose computation is presented in~\ref{sec: bootstrapping integrated negative geometries}.
     \item The case of a spurious sub-leading singularity is given by eq. \eqref{badshrin} and depicted as Fig. \ref{fig:5c}. In this case, both $S_1$ and $\bar{S}_1$ are not boundaries of $\mathcal{A}$, which means that the numerator vanishes on the cut locus $S_1$. Other than  the combinatorial implementation via on-shell diagrams, one can check if a solution ot the Landau equations is spurious by checking if the numerator vanishes on the support of the associated cut equations. We checked this explicitly while performing the Landau analysis for the ladder at six points and two loops, and at five points and three loops.
\end{enumerate}

This concludes our walkthrough of the algorithm presented at the beginning of this section via the specific example for the ladder negative geometry at two loops given by Fig.~\ref{LS_L1}(c). We now present the result of this geometric Landau procedure applied to ladder negative geometries at $L=2$.

\begin{figure}[t]
    \centering
    \subfigure{\label{fig:L2ladderLD1}\includegraphics[width=0.375\linewidth]{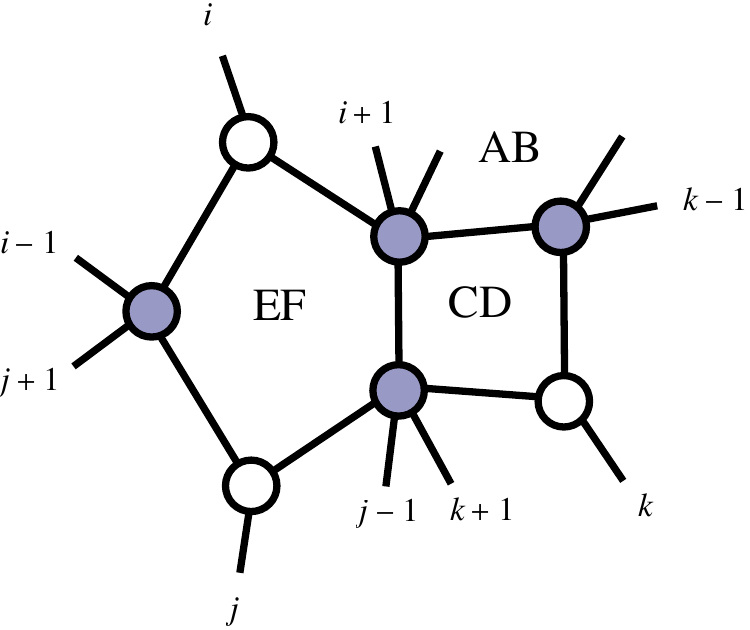}}     \subfigure{\label{fig:L2ladderLD3}\includegraphics[width=0.375\linewidth]{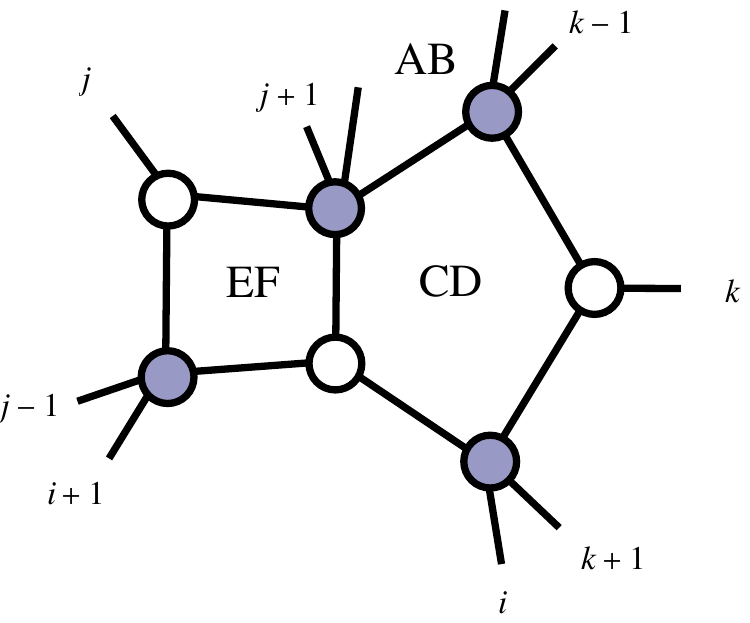}}
    \caption{Leading Landau diagrams for leading singularity configurations Fig.~\ref{LS_L1}(b) and Fig.~\ref{LS_L1}(d) of the two-loop ladder respectively.}
    \label{fig:L2ladderLD}
\end{figure}

\subsection{Singularities of ladders from Landau analysis }
\label{sec:all_letter_ladder}

At two loops, we perform the same analysis also for the other leading singularity configurations in Fig.~\ref{LS_L1}. The corresponding leading Landau diagrams are given in Fig.~\ref{fig:L2.II_LD}, and we provide now a conjecturally complete list of physical singularities for ladders at two loops and any multiplicity $n$. 
These results will be the starting point for our bootstrap in the following computation.

{\it \underline{1. Singularities of two-loop ladders at all points.}} 
We apply the geometric Landau analysis to every two-loop leading singularity configuration of ladders in Fig.~\ref{LS_L1}. This is amount to perform the Landau analysis for diagrams in Fig.~\ref{fig:L2.II_LD}, as well as the two diagrams from Fig.~\ref{LS_L1}(b) and Fig.~\ref{LS_L1}(d) as Fig.~\ref{fig:L2ladderLD}.
This analysis provides a conjecturally complete list of singularities for integrated ladders $F_n^{(\laddtwopic)}$ at two loops and any multiplicities $n \geq 4$.

In the following, all $i,j,k\in\{1,\cdots,n\}$ are taken modulo $n$ but without any assumption on the ordering. We identify $(n{+}1\, n{+}2):=(AB)$ for simplicity. The list of singularities is the following.
\begin{itemize}
    \item [1.] All twistor brackets $\langle i{-}1ij{-}1j\rangle$ and $\langle AB i{-}1i  \rangle$. 
    \item [2.] Square roots of all four-point determinants $\det\{\langle a{-}1ab{-}1b\rangle\}_{a,b\in I}$, with $I$ being the set $\{i,j,k,k{+}1\}$ or $\{i,j,k,n{+}2\}$. Note that in the second case we encounter genuine square roots and associated odd letters for $n\geq6$. The first two types of singularities can be computed from sub$^m$-topologies of all Landau diagrams such as double boxes or double triangles.
    \item [3.] All five-point determinants $\det\{\langle a{-}1ab{-}1b\rangle\}_{a,b\in I}$, with $I$ being the set $\{i,i{+}1,j,j{+}1,k\}$ or $\{i,i{+}1,j,k,n{+}2\}$, from sub$^2$-leading Landau singularities.
    \item [4.] All possible
    \begin{equation}
    \begin{aligned}
    &\langle k(k{-}1k{+}1)(ij)(AB)\rangle \ ,  \ \langle k(k\pm1j)(ii{+}1)(AB)\rangle \ , \\
    &\langle j(j{-}1j{+}1)(kk\pm1) ((ii{+}1k)\cap(ABk))\rangle \ , \ \langle j(j{-}1j{+}1)(ii{+}1)((ii{+}1k)\cap(ABk))\rangle \ ,
    \end{aligned}
    \end{equation}
     from two-loop leading Landau singularities.
    \item [5.] All possible
    \begin{equation}
    \begin{aligned}
    &\langle k(k{-}1k{+}1)((ijj{\pm}1)\cap\bar{i})(AB) \rangle \ , \  \langle k(k{\pm}1 (jj{\pm}1)\cap\bar k)(ii{+}1)(AB)\rangle \ , \\
    & \langle (kk{\pm}1)\cap\bar{j}\ (k{\pm}1 j)(ii{+}1)(AB)\rangle \ ,
    \end{aligned}
    \end{equation}
    from two-loop sub-leading Landau singularities.
    \item [6.] 
    All possible five-point determinants $\det\{\langle X_i X_j\rangle\}$ with the bi-twistors $X_i$ from 
    \begin{equation}
        \{(j{-}1j),\ (jj{+}1),\ (ii{+}1),\ (ii{+}1k)\cap\bar{k},\ (AB)\} \ .
    \end{equation}
    We offer more comments on this last type of singularities in Appendix \ref{app:super}. 
\end{itemize}
Based on the general singularities, we can easily get the alphabet for five- and six-point two-loop ladders by degeneration. Note that to finally get the actual alphabet for symbol bootstrap, one should implement the transition from singular locus to symbol letters, which is performed by \texttt{Effortless.m}.

{\it \underline{2. Alphabet of two-loop ladder at five points.}} 
After implementing $1\leq i,j,k\leq5$ in the general list and passing to Mandelstam variables by
\begin{equation}
   s_{12} = \frac{\langle5123\rangle}{\langle51 AB\rangle\langle23 AB\rangle} \ ,\ \&\  \text{cycl.},
\end{equation}
the Landau singularities for five-point two-loop ladder $F_5^{(\laddtwopic)}$ correspond to 15 two-loop planar pentagon letters from pentagon functions \cite{Gehrmann:2015bfy,Gehrmann:2018yef}, which following the notation in \cite{Gehrmann:2015bfy} are
\begin{equation}\label{20_5_pt_planar_alphabet}
    \{W_1,\cdots, W_5,W_{11},\cdots, W_{20},W_{26},\cdots, W_{30}\} \ .
\end{equation}
Computing the symbol letters from physical singularities at this case is straightforward, since the only square root of Mandelstam variables in two-loop planar pentagon alphabet is the parity-odd scalar
\begin{equation}\label{eq:epsilon}
\varepsilon(i,j,k,l):=4i \, \varepsilon_{\mu\nu\rho\sigma}p_i^\mu p_j^\nu p_k^\rho p_l^\sigma,
\end{equation} 
with $\{i,j,k,l\}=\{1,2,3,4\}$, and will be rationalized when employing momentum twistor variables. Therefore the construction of symbol letters from singularities is just a monomial transformation. This result agrees exactly with the alphabet of $F_5^{(\laddtwopic)}$ computed in \cite{Chicherin:2024hes}. It is worth noting that the planar two-loop letter $W_6=s_{12}{+}s_{23}$ and its cyclic images, i.e. $\{W_{6{+}i}\}_{i=0,\cdots,4}$ defined in \cite{Gehrmann:2018yef}, are absent from \eqref{20_5_pt_planar_alphabet} since the Landau diagrams in Fig.~\ref{fig:L2ladderLD} contain no double-triangle cuts.

{\it \underline{3. Alphabet of two-loop ladder at six points.}}
The alphabet of $F_6^{(\laddtwopic)}$ can be obtained by implementing $1\leq i,j,k\leq6$.  In the following, we switch to the Mandelstam variables
\begin{equation}\label{eq:6ptMand}
s_{12} = \frac{\langle6123\rangle}{\langle61AB\rangle\langle23AB\rangle} \, , \quad  s_{123}= \frac{\langle6134\rangle}{\langle61AB\rangle\langle34AB\rangle} \, ,
\end{equation}
together with their cyclic images.
In total, the alphabet involves $157$ independent letters. We observe that our alphabet is a subset of 245 planar two-loop hexagon letters from \cite{Henn:2025xrc}. The latter have been obtained from the canonical differential equations for all six-point two-loop planar families of Feynman integrals. We should emphasize again that although we have non-planar topologies from Landau diagrams, they do not contribute any new non-planar symbol letters.  
We also caution the reader that in the following entries $W_i$ denotes the six-point letters, following from the notation in \cite{Henn:2025xrc}, and it should not be confused with its earlier usage for the five-point alphabet letters 
 eq.~\eqref{20_5_pt_planar_alphabet}.

In our alphabet, 95 letters are parity even and 62 letters are parity odd with respect to the sign flip of the square-roots. We organize them as follows.  $41$ of the parity-even letters appeared in the one-loop six-point differential equations, (cf. \cite{Henn:2022ydo}). Following the labeling in \cite{Henn:2025xrc}, they are
\begin{align}
   \{W_1,\cdots, W_9,W_{16},\cdots,W_{33},W_{46},\cdots,W_{51},W_{88},\cdots,W_{93},W_{118},W_{119}\} \,.
\end{align}
The remaining $54$ parity-even letters are from the genuine two-loop integrals (cf. \cite{Henn:2025xrc}). We organize them into five dihedral orbits as follows,
\begin{align}
\{W_{i{+}52},W_{i{+}70}\}_{i=0,\dots,5}\ &\cup \{W_{i{+}58}\}_{i=0,\dots,11}\ \cup \{W_{i+76}\}_{i=0,\dots,11}\ \nonumber\\
&\cup \{W_{i+100}\}_{i=0,\dots,5}\ \cup \{W_{i+106}\}_{i=0,\dots,11} \,.\label{eq:oddfrom}
\end{align}
Moreover, only the letters in first four orbits in the previous list belong to the alphabet of the weight-four function space \cite{Henn:2025xrc}, while letters from the last orbit do not. However, for the sake of completeness, we include the letters from the last orbit when constructing the symbol ansatz. 

Among the parity-odd alphabet letters, 48 letters involve only $\varepsilon(i,j,k,l)$
for $i,j,k,l \in \{1,\dots,6\}$, which have the form of square roots if written in the Mandelstam variables. These letters are
\begin{equation}
\{W_{182},\cdots,W_{190},W_{194},\cdots,W_{211},W_{218},\cdots,W_{220},W_{230},\cdots,W_{247}\}
\end{equation}
Taking products of their denominators and numerators as in eq.~\eqref{fact_sr}, we find that $30$ of them correspond to the two-loop singularities, which are the latter three groups in \eqref{eq:oddfrom}. These 48 odd letters, like the case at five points, are rationalized in momentum twistor variables.

\begin{figure}[t]
    \centering
    \subfigure[]{\includegraphics[width=0.30\linewidth]{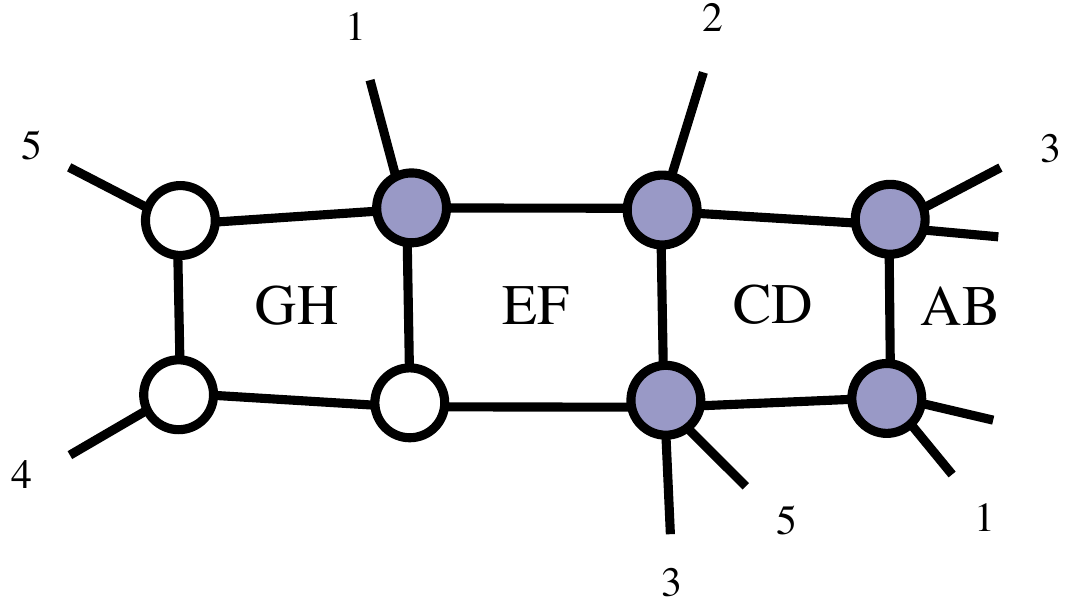}}\quad    \subfigure[]{\includegraphics[width=0.30\linewidth, trim= -1.5cm 0.5cm 0 0]{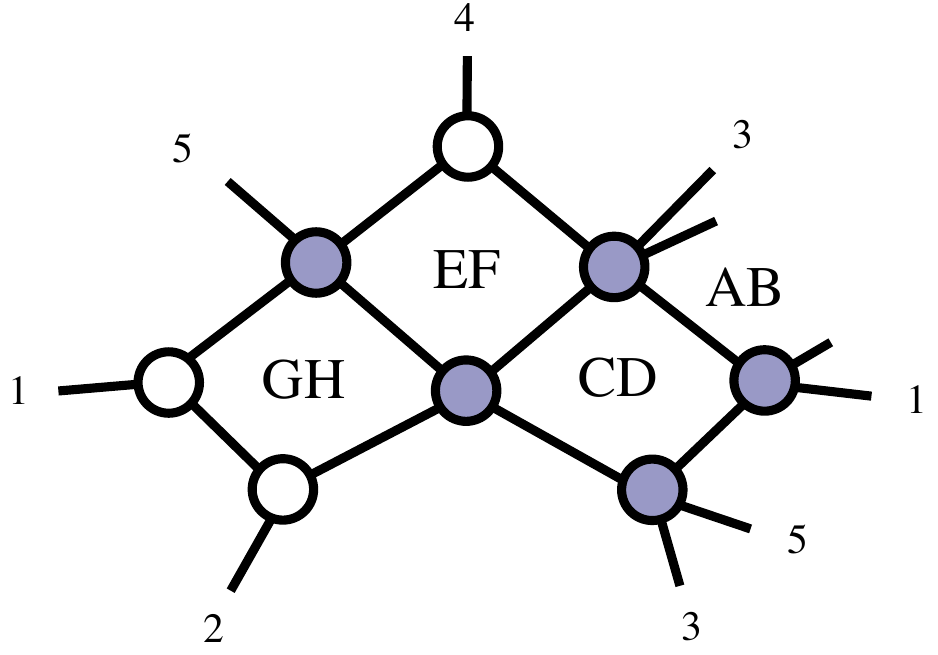}}
    \subfigure[]{\includegraphics[width=0.25\linewidth]{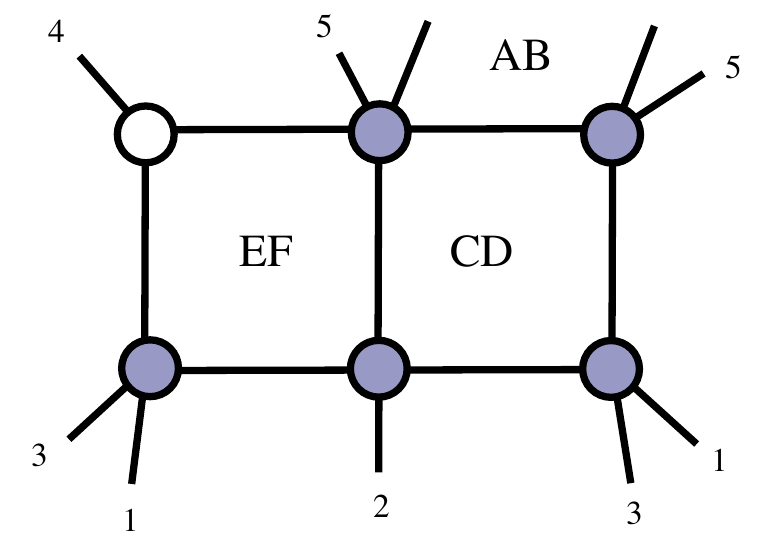}}
    \caption{Sub$^2$-leading Landau diagrams (a) and (b), and the equivalent two-loop non-planar topology (c) for the five-point three-loop ladder-type negative geometry yielding \eqref{eq:d11}.} 
    \label{fig:d11diagrams}
\end{figure}

Finally, note that $W_{118}$ and $W_{119}$ in the even letter list are three-mass-triangle square roots $\sqrt{\Delta_3(s_{12},s_{34},s_{56})}$ and $\sqrt{\Delta_3(s_{23},s_{45},s_{16})}$, where $\Delta_3(a,b,c):=a^2+b^2+c^2-2ab-2bc-2ac$ is the Källén function. They are genuine square roots for momentum twistor variables, therefore will lead to extra non-trivial algebraic letters on top of the physical singularities. By constructing odd letters from the singularities, there are $7+7$ further parity-odd letters involving 
    \begin{align} \label{eq:nonrational}
        \{W_{157},\cdots,W_{166},W_{275},\cdots,W_{278}\} \,.
            \end{align}
Note that these letters also appeared in the one-loop differential equations (cf. \cite{Henn:2022ydo}), and the last four of them contain $\sqrt{\Delta_3}$ and $\epsilon(i,j,k,l)$ simultaneously.

\begin{figure}[t]
    \centering
    \includegraphics[width=0.4\linewidth]{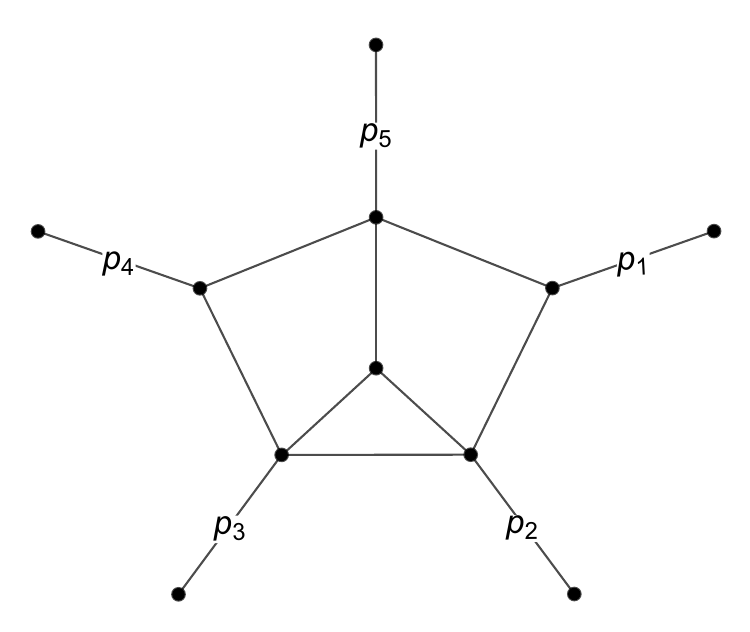}
    \caption{A three-loop planar Feynman integral yielding $1/d_{1}$ as its leading singularity.}
    \label{fig:d11planar}
\end{figure}

{\it \underline{4. Alphabet of three-loop ladder at five points.}} 
Let us discuss the alphabet for $F_5^{(\laddthreepic)}$. For three-loop ladders, there are seven different types of Landau diagrams, see Table~\ref{tab:11} and \ref{tab:12} . A detailed summary of all symbol letters for these topologies at arbitrary multiplicity would be too lengthy to present here. However, we find that, at five points, most of the physical singularities are the planar two-loop letters. The only exception is the square root 
\begin{equation}\label{eq:d11}
   d_1=\sqrt{(s_{12}s_{15}{-}s_{12}s_{23}{+}s_{23}s_{34}{+}s_{15}s_{45})^2{-}4s_{12}s_{15}s_{45}(s_{15}{-}s_{23})} \ ,
\end{equation}
together with its four cyclic images. This appears in several sub-topologies of the Landau diagrams, such as the three-loop ones in Fig.~\ref{fig:d11diagrams}.
The Landau equations of these diagrams are actually equivalent to that of a two-loop non-planar integral as Fig.~\ref{fig:d11diagrams}(c), which is exactly the sub-topology of the two-loop Landau diagram in Fig.~\ref{fig:wrongshrink}. 
This singularity is spurious at two loops because the helicity degree is incorrect. However, at three loops, the square root~\eqref{eq:d11} and the associated parity-odd letters are physical. It is worth mentioning that these letters can also be obtained from planar topologies at five points and three loops, and they were discovered in the canonical differential equation calculations in~\cite{Chicherin:2024hes}. In fact, the square root singularity \eqref{eq:d11} is present in the leading Landau equation of the wheel topology in Fig.~\ref{fig:d11planar}, associated to its maximal cut and corresponding residue value. Therefore, these new letters are still five-point three-loop planar letters.  
However, there is no non-trivial cut like that shown in Fig.~\ref{fig:d11planar} among the Landau diagrams for the negative geometries with ladder topology. Instead, $d_1$ appears in Landau diagrams with a non-planar distribution of indices, and therefore, in some sense, captures genuine non-planar information. Note in fact that expanding the Wilson loop with Lagrangian insertion in negative geometries introduces non-planar singularities in individual terms. These must nevertheless cancel out in the full result as shown in \cite{Chicherin:2024hes}.

The odd letters associated to \eqref{eq:d11} can be constructed by either \texttt{Effortless.m} or by performing the Schubert analysis as in \cite{He:2024fij}. Starting from the $15$ rational letters \eqref{20_5_pt_planar_alphabet} together with the square roots \eqref{eq:d11}, we obtain extra four odd letters for each square root~$d_i$:
\begin{equation}
\begin{aligned}\label{eq:new5ptletters}
   & \frac{s_{12}s_{15}{-}s_{12}s_{23}{-}s_{23}s_{34}{-}s_{15}s_{45}{+}d_1}{s_{12}s_{15}{-}s_{12}s_{23}{-}s_{23}s_{34}{-}s_{15}s_{45}{-}d_1} \ ,\quad  \frac{s_{12}s_{15}{+}s_{12}s_{23}{-}s_{23}s_{34}{-}s_{15}s_{45}{+}d_1}{s_{12}s_{15}{+}s_{12}s_{23}{-}s_{23}s_{34}{-}s_{15}s_{45}{-}d_1} \ , \\
   & \frac{s_{12}s_{15}{-}s_{12}s_{23}{-}s_{23}s_{34}{-}s_{15}s_{45}{+}2s_{23}s_{45}{+}d_1}{s_{12}s_{15}{-}s_{12}s_{23}{-}s_{23}s_{34}{-}s_{15}s_{45}{+}2s_{23}s_{45}{-}d_1} \ , \quad \frac{a+\epsilon(1,2,3,4) d_1}{a-\epsilon(1,2,3,4) d_1} \ , 
\end{aligned}
\end{equation}
with $\epsilon(1,2,3,4)$ from \eqref{eq:epsilon}, and
\begin{align}
    a=&(s_{12}s_{15}{-}s_{12}s_{23}{+}s_{23}s_{34}{+}s_{15}s_{45})^2{+}s_{34} s_{45} (s_{12} s_{15} {+} s_{12} s_{23} {-} s_{23} s_{34} {-} s_{15} s_{45})\nonumber\\
    &+4 s_{12} s_{15} s_{45} (s_{23}-s_{15}) \,.
\end{align}
Considering all their cyclic images and the $d_i$'s, these are 25 new letters. 
Together with the $20$ letters from the planar two-loop alphabet \eqref{20_5_pt_planar_alphabet}, these constitute our conjectural alphabet for bootstrapping the five point three-loop ladder.

\section{Bootstrapping integrated negative geometries}
\label{sec: bootstrapping integrated negative geometries}
In this section, we turn to the computation of the integrated ladder-type negative geometries. Relying on the presentation from the previous sections, we classify the leading singularities and compute the symbol alphabet from a targeted Landau analysis. This information specifies the function space of the observable. In what follows we work at symbol level and construct a symbol ansatz for the integrated ladders. We further restrict the symbol ansatz by conditions such as integrability of the symbol, dihedral symmetry, collinear limits, the d'Alembertian equation, etc. Ideally, a unique solution is obtained once sufficiently many conditions are imposed.

In the following we are concerned with the negative geometries given by the six-point two-loop ladder $F_6^{(\laddtwopic)}$ and five-point three-loop ladder $F_5^{(\laddthreepic)}$. Concretely, we determine a complete set of letters $\{W_i\}$ for the transcendental functions $f_{n,ij}^{(L)}$ in \eqref{ladder_function}, and we write an ansatz for their symbol as
\begin{align}
{\cal S}\left( f_{n,ij}^{(L)} \right) = \sum_{i_1,\ldots,i_{2L}} c_{i_1,\ldots,i_{2L}} [W_{i_1},\ldots,W_{i_{2L}}] \label{eq:symb2L} \ ,
\end{align} 
where the summation indices run over the size of the alphabet, and the coefficients $c$'s are rational numbers to be determined via the constraints mentioned above. For example, a basic requirement is that the symbol is integrable, such that it can be lifted to a well-defined function. The integrability conditions require that for each pair of adjacent entries the following equation holds:
\begin{align}\label{eq:integrable}
\sum_{i_1,\ldots,i_{2L}} c_{i_1,\ldots,i_{2L}} [W_{i_1},\ldots,W_{i_{k-1}},W_{i_{k+2}},\ldots,W_{i_{2L}}] \, d\log W_{i_k} \wedge d\log W_{i_{k+1}} = 0 \ , 
\end{align}
for every $k \in \{1,\ldots,2L-1 \}$. 

Following the bootstrap procedure, we successfully find symbols of the integrated six-point two-loop and five-point three-loop ladders. Surprisingly, although the Landau diagrams contain non-planar sub-topologies, we find that the alphabets for both ladder negative geometries contain only planar letters. For the six-point two-loop ladder, the result localizes in the six-point two-loop planar function space from \cite{Henn:2024ngj,Henn:2025xrc}, as was the case at five points \cite{Chicherin:2024hes}. 
For the five-point three-loop ladder, we need more symbol letters on top of two-loop planar pentagon letters, 
which belong to the planar five-point three-loop alphabet, and are present in the state-of-the-art calculations via differential equation \cite{Chicherin:2024hes}.  
However, for negative geometries of more complicated topology, it is plausible that non-planar cuts may actually contribute to the integrated result. In this paper, we focus on the symbol-level bootstrap only and leave the uplifting of the symbols to functions for future work.

\subsection{The six-point two-loop ladder}
\label{subsec: the six point two loop ladder}

We first turn to the six-point two-loop ladder. According to \eqref{ladder_function}, we write it as 
\begin{equation}\label{ladder_6pts_decomp}
\begin{aligned}
F_6^{(\laddtwopic)} =\,&\Omega_{6,13} \, f_{6,13}^{(2)}+\Omega_{6,24} \, f_{6,24}^{(2)}+\Omega_{6,35} \, f_{6,35}^{(2)}+\Omega_{6,46} \, f_{6,46}^{(2)}+\Omega_{6,15} \, f_{6,15}^{(2)}+\Omega_{6,26} \, f_{6,26}^{(2)}\\
&+\Omega_{6,14} \, f_{6,14}^{(2)} + \Omega_{6,25} \, f_{6,25}^{(2)} +\Omega_{6,36} \, f_{6,36}^{(2)} \,.
\end{aligned}
\end{equation}
The explicit expressions for the leading singularities $\Omega_{6,13}$ and $\Omega_{6,14}$ are given in \eqref{six-point LS}. The pure functions $f_{6,13}^{(2)}$ and $f_{6,14}^{(2)}$ are the main objects of study in this section. 
The six-point ladder $F_6^{(\laddtwopic)}$ in \eqref{ladder_6pts_decomp} is completely determined by these two pure functions because of the dihedral symmetry. Beginning with $157$ letters obtained from the geometric Landau analysis in Section~\ref{sec:all_letter_ladder}, we break up the bootstrap procedure in several steps as follows.

{\it \underline{1. Integrable symbols.}} 
The space of integrable weight-four symbols built from 157 letters is of high dimension.
Therefore, we benefit significantly from imposing some simple conditions early on when constructing the integrable symbols. We impose first/last-entry conditions and employ dimensionless combinations of the alphabet letters.

Firstly, the ladder integrand is dual conformal invariant in the $Z_i$'s and $AB_\ell$'s, for $1 \leq i \leq n$ and $0 \leq \ell \leq L$. After integrating out $L$ loop momenta, the resulting functions $f_{n,ij}^{(L)}$ are dual conformal invariant in the $Z_i$'s and in the unintegrated loop $(AB)_0 = AB$.  
After the identification in \eqref{eq:6ptMand}, these functions depend only on dimensionless ratios of the Mandelstam variables. Consequently, their symbol alphabet involves only dimensionless letters. Imposing this requirement, we eliminate one letter, which is a common scale. Thus, we work with 156 alphabet letters in the following. 

Secondly, in constructing the integrable symbols recursively by weight, we begin with only the eight weight-one symbols 
\begin{equation}
    \left\{\left[\frac{s_{i,i+1}}{s_{12}} \right], \left[\frac{s_{i,i+1,i+2}}{s_{12}} \right] \right\} \, ,
\end{equation}
since the first entries of the symbol specify the physical discontinuities. 

Finally, as discussed in \cite{Chicherin:2024hes}, the last entries of the integrated ladder negative geometries are also constrained by the d'Alembertian differential equation, see Appendix \ref{app:da}. This constraint is implemented recursively in the loop order according to eq.~\eqref{DE}. Differentiating the weight-$2L$ ansatz by the second order differential operator $\mathcal{D}_{ij}$, we naively obtain
\begin{equation}
    \mathcal{D}_{ij} \, f^{(L)}_{n,ij}=S_{2L-1}+S_{2L-2} \, ,
\end{equation}
where $S_{2L-1}$ and $S_{2L-2}$ are symbols   
of weights $2L{-}1$ and $2L{-}2$, respectively. Because of \eqref{DE}, $S_{2L-1}$ has to vanish. This imposes constraints on any last entry $S$, being itself a multiplicative  combination of the 157 symbol letters, which must therefore satisfy
\begin{equation}
    \mathcal{D}_{ij}\log(S)=0 \, . \label{DijlogS}
\end{equation}
This condition significantly restricts the space of last entries. For $f^{(2)}_{6,13}$ and $f^{(2)}_{6,14}$, we find 63 and 58 allowed combinations, respectively. 
The number of integrable symbols for $f^{(2)}_{6,13}$ and $f^{(2)}_{6,14}$ after taking into account these constraints are recorded in Table \ref{table:6}. Then we impose further physical constraints on our symbol ansatz.

\begin{table}[t]
    \centering
    \begin{tabular}{|l|c|l|l|l|}
\hline
  weight & \multicolumn{1}{l|}{1} & 2                   & 3                    & 4   \\ \hline
\# symbols for $f^{(2)}_{6,13}$ & \multirow{2}{*}{8}     & \multirow{2}{*}{53} & \multirow{2}{*}{343} & 675 \\ \cline{1-1} \cline{5-5} 
\# symbols for $f^{(2)}_{6,14}$ &                        &                     &                      & 540 \\ \hline
\end{tabular}
\caption{Number of integrable symbols for $f^{(2)}_{6,13}$ and $f^{(2)}_{6,14}$.}\label{table:6}
\end{table}

{\it \underline{2. Physical conditions.}} 
The first condition we consider is the spurious pole cancellation. The LS value $\Omega_{6,ij}$ with non-adjacent $i$ and $j$ has a simple pole at $\langle AB ij\rangle = 0$. This singularity is absent from the ladder, so this pole must be canceled by the accompanying pure function:
\begin{equation}
    \mathcal{S}\left(f_{6,ij}^{(2)}\right)\Big|_{s_{ij} \to \, 0}=0 \, .
\end{equation}
This translates into a homogeneous condition on the unknown coefficients in the symbols of $f^{(2)}_{6,13}$ and $f^{(2)}_{6,14}$, reducing the degrees of freedom to $100$ and $96$, respectively. 

Next, we require that the collinear limit of the ansatz matches the known five-point two-loop ladder. To apply this condition, it is more convenient to work in momentum twistor space, as e.g. in \cite{Caron-Huot:2011dec}. We consider the following parametrization of the collinear limit
\begin{equation}
    Z_6\to Z_5-\frac{\langle1235\rangle}{\langle1234\rangle}  \epsilon  Z_4+\frac{\langle4523\rangle}{\langle4123\rangle} \epsilon \tau Z_1+\frac{\langle3451\rangle}{\langle3421\rangle}\epsilon^2Z_2 \, ,
\end{equation}
After taking the limit, any divergence caused by it should cancel out, and the resulting expression should be independent of the free parameter $\tau$. Finally, we match the limit of our symbol ansatz with the five-point two-loop ladder. In the limit $\epsilon\to 0$, the six-point leading singularities reduce to five-point LS,
\begin{align}
\Omega_{6,13} &\to \Omega_{5,13} \, , & \Omega_{6,24} &\to \Omega_{5,24} \, , & \Omega_{6,35} &\to \Omega_{5,35} \, , \notag \\
\Omega_{6,46} &\to 0 \, ,           & \Omega_{6,15} &\to 0 \, ,           & \Omega_{6,26} &\to \Omega_{5,25} \, , \notag \\
\Omega_{6,14} &\to \Omega_{5,14} \, , & \Omega_{6,25} &\to \Omega_{5,25} \, , & \Omega_{6,36} &\to \Omega_{5,35} \,.
\end{align}
Thus, the collinear limit of the six-point integrated ladder yields
\begin{align}
     F_6^{(\laddtwopic)}\Big|_{Z_6\to Z_5}&=\Omega_{5,25} \, (f_{6,25}^{(2)} {+} f_{6,26}^{(2)}) \Big|_{Z_6\to Z_5} {+} \Omega_{5,13} \, f_{6,13}^{(2)} \Big|_{Z_6\to Z_5} {+} \Omega_{5,24} \, f_{6,24}^{(2)} \Big|_{Z_6\to Z_5} \nonumber \\
     &\hspace{0.5cm}{+}\Omega_{5,35} \, (f_{6,35}^{(2)} {+} f_{6,36}^{(2)}) \Big|_{Z_6\to Z_5} {+} \Omega_{5,14} \, f_{6,14}^{(2)} \Big|_{Z_6\to Z_5}  \, ,
\end{align}
whose RHS has to agree with the known expression for five-point $F_5^{(\laddtwopic)}$, cf. \cite{Chicherin:2024hes}.
This is the condition we are imposing. This yields five non-trivial inhomogeneous constraints on our symbol ansatz. We find that three conditions among them are sufficient to fix a unique solution. 

{\it \underline{3. Checks.}} 
We performed various other checks of the result, such as correct soft limit \cite{Bianchi:2014gla}, reflection symmetry of $f_{6,13}^{(2)}$ and $f_{6,14}^{(2)}$, as well as cyclic symmetry $\tau^3(f_{6,14}^{(2)})=f_{6,14}^{(2)}$ where $\tau(Z_i)=Z_{i{+}1}$. 
Another strong cross-check of our result is that it satisfies the d'Alembertian differential equation, where the one-loop ladders on the RHS of \eqref{DE} are expressed in terms of chiral pentagons \eqref{eq:6pt_1L}. More explicitly, at two loops we have the recursive relation \eqref{L2_DE}, which at six points yields two non-trivial relations, i.e. nine differential equations cyclically represented by $\mathcal{D}_{13}$ and $\mathcal{D}_{14}$.
We checked that our solution satisfies these differential equations at the symbol level.

\begin{figure}[t]
    \centering
    \subfigure[$W_{102},W_{104}$]{\includegraphics[width=0.35\linewidth]{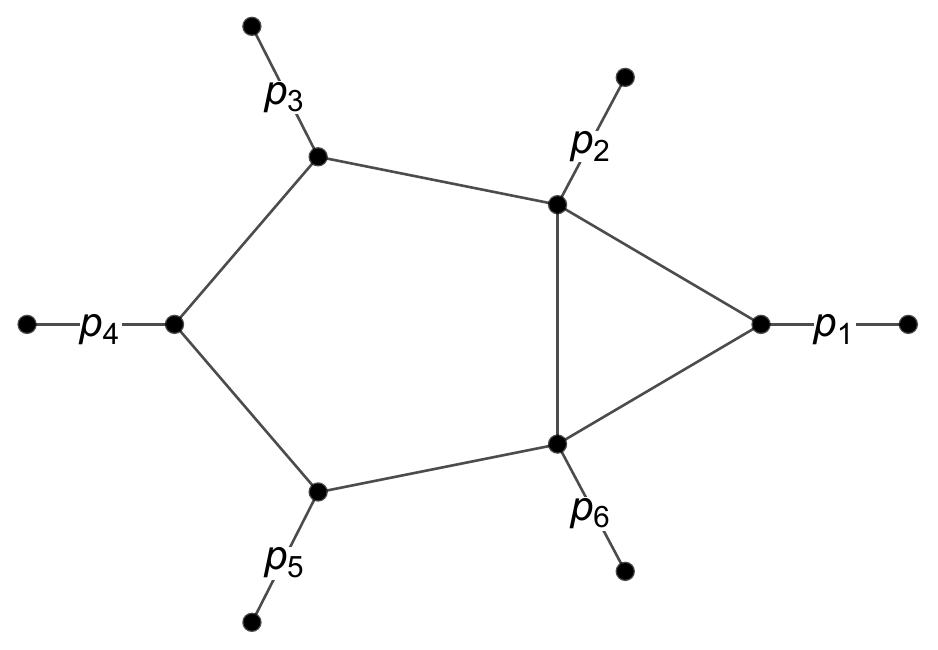}}\quad\quad  
    \subfigure[$W_{81},W_{87}$]{\includegraphics[width=0.35\linewidth]{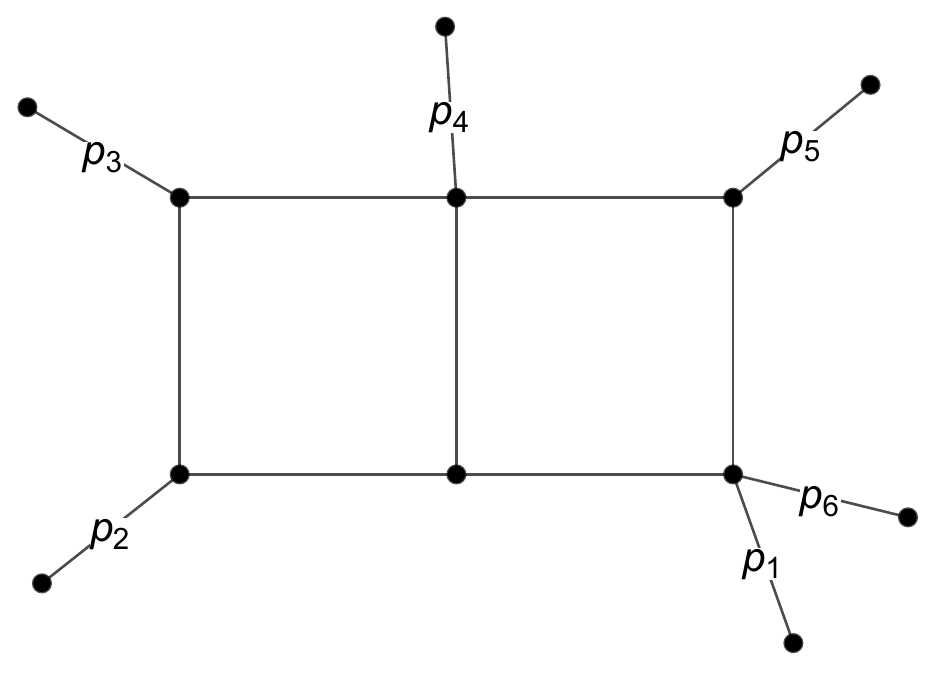}}\\
    \quad\quad  \subfigure[$W_{58},W_{68}$]{\includegraphics[width=0.35\linewidth]{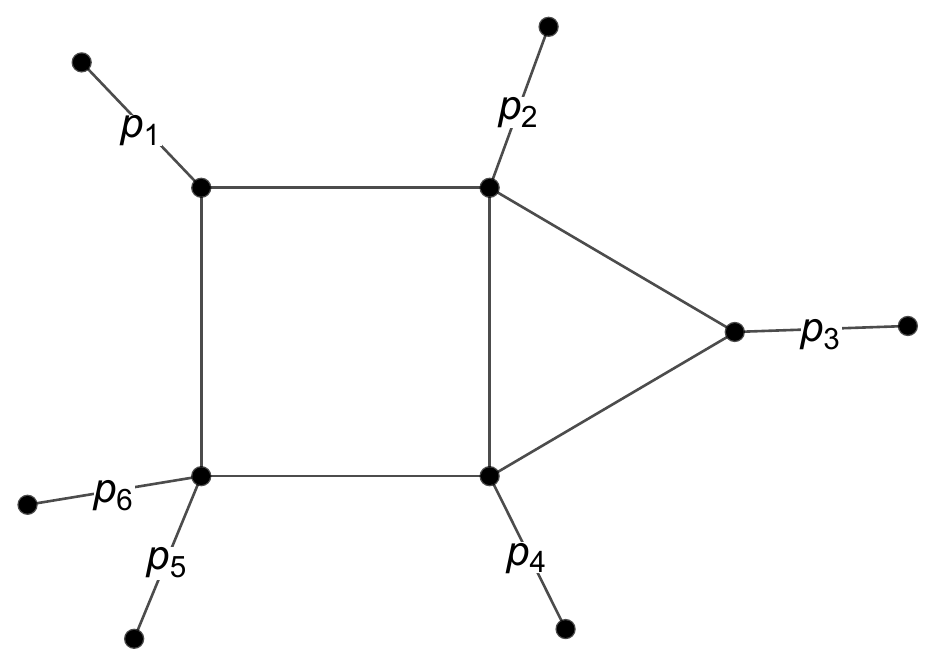}}\quad\quad \subfigure[$W_{60},W_{66}$]{\includegraphics[width=0.35\linewidth]{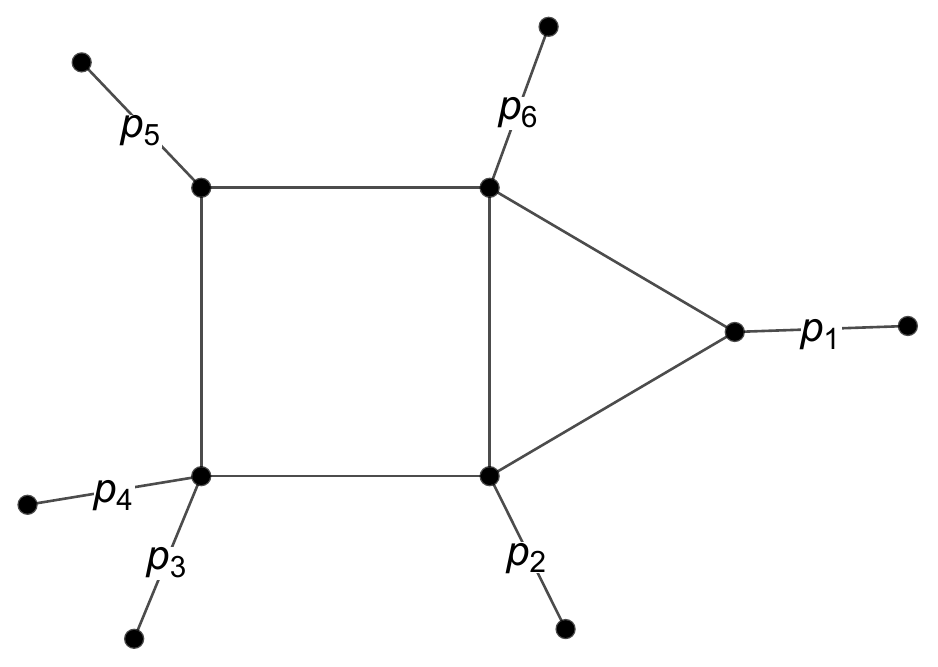}}
    \caption{Four planar Feynman integral sectors and their reflection yielding genuine two-loop singularities in $f_{6,13}^{(2)}$.}
    \label{fig:FD6pt}
\end{figure}

{\it \underline{4. Discussion of properties of the result.}} The alphabet of the function $f_{6,13}^{(2)}$ contains $84$ letters. Among them, eight rational letters come from four leading Landau diagrams and four sub-leading diagrams (Fig.\ref{fig:FD6pt} and their reflection $p_i\to p_{4{-}i}$) at two loops, which following the notation in \cite{Henn:2025xrc} are 
\begin{equation}
    \{W_{58},W_{60},W_{66},W_{68},W_{81},W_{87},W_{102},W_{104}\} \,.
\end{equation}
Comparing with Fig.~1 in \cite{Carrolo:2025pue}, one can see that these singularities form a subset of the alphabet for the full observable $F_6^{(2)}$. Only two of the letters are genuine six-point letters, which originate from penta-triangle integrals. On the other hand, the alphabet of $f_{6,14}^{(2)}$ contains only $58$ letters which are all one-loop singularities. Finally, numbers of independent last-entries for the two functions are 22 and 20, respectively.  We checked that both functions localize in the hexagon two-loop planar Feynman integral function space of \cite{Henn:2025xrc}, analogously to the five-point ladder \cite{Chicherin:2024hes}. In the ancillary files we present the explicit symbol results of two-loop hexagon functions yielding $f_{6,13}^{(2)}$ and $f_{6,14}^{(2)}$.

\subsection{The five-point three-loop ladder}
\label{sec: five point three loop ladder}

Let us proceed to the bootstrap of the five-point three-loop ladder $F_5^{(\laddthreepic)}$. The previous attempt in \cite{Chicherin:2024hes}, based on $26$ planar letters, showed that this integral requires a bigger alphabet. 
We solve this problem by using the 45-letter alphabet found previously via the geometric Landau analysis to make an ansatz for the function space. We bootstrap the symbol of the pure function $f^{(3)}_{5,13}$, which is accompanied by the leading singularity $\Omega_{13}$, and the remaining terms of the three-loop ladder $F_5^{(\laddthreepic)}$ are obtained by cyclic symmetrization
\begin{align} 
F_5^{(\laddthreepic)} = \Omega_{5,13} \,  f^{(3)}_{5,13} + \Omega_{5,24} \,  f^{(3)}_{5,24} +\Omega_{5,35} \,  f^{(3)}_{5,35}+\Omega_{5,14} \,  f^{(3)}_{5,14}+\Omega_{5,25} \,  f^{(3)}_{5,25} \,. \label{F5_3loop}
\end{align}
Since only dimensionless combinations of the alphabet letters are present in the symbol, we eliminate one of the letters and work with the 44-letter alphabet in what follows. In the six-point case, we have pinned down all unknowns in our bootstrap ansatz for the ladder considering its lower-point limits, see Sect.~\ref{subsec: the six point two loop ladder}. The d'Alembertian DE, see Appendix \ref{app:da}, played an auxiliary role restricting the last entries of the symbol. We can attribute this favorable counting to the rich kinematics of the six-particle scattering. In the five-point case, the lower-point limits provide much less information. They are not sufficient for the symbol bootstrap to succeed, so we have to invoke also other properties of the ladder. It turns out that the d'Alembertian DE provides the missing constraints. Moreover, the DE enables us to impose constraints on the last two entries while constructing integrable symbols. These constraints reduce the computational complexity. Also, this bootstrap strategy should be more optimal for higher-loop symbol bootstraps. For this reason, as compared to the six-point two-loop bootstrap, we change the order in which we impose the constraints and rely more extensively on the DE. 

The first symbol entries, which represent the physical branch cuts, are drawn from the set of four letters, $\{s_{i,i+1}/s_{12}\}$. Then we constrain the last two entries of the symbol. Firstly, only 17 multiplicative combinations of the alphabet letters are compatible with the DE, i.e. they are annihilated by ${\cal D}_{13}$, see \eqref{DijlogS}. They are the allowed last entries of the symbol ${\cal S}\left (f^{(3)}_{5,13}\right)$. Secondly,
the second-order differential operator ${\cal D}_{13}$ acts on the second component in the coproduct $(2L-2,2)$ of $f^{(L)}_{5,13}$, which is an integrable weight-two symbol,
\begin{align}
{\cal S} \left(  f^{(L)}_{5,13} \right) = \sum_i S^{(i)}_{2L-2} \otimes S_{2}^{(i)} \,.
\end{align}
This differentiation results in a rational function, provided that the last entry condition is taken into account. The RHS of the DE \eqref{DE} contains five rational coefficients $\overline{\cal C}_{13,kl}$, but only four of them are linearly independent over rational numbers. Let us denote them by $l_a$. Then ${\cal D}_{13} S_2^{(i)}$ should be a linear combination of four $l_a$. We solve the corresponding DE employing an ansatz of weight-two integrable symbols. Their first entries are 44 alphabet letters and their second entries are 17 last entries discussed above. We find 227 weight-two symbols of this sort. Imposing the DE on the weight-two symbol ansatz, we  find four symbol solutions $U_a$ of the inhomogeneous DE, and 135 symbol solutions $V_I$ of the homogeneous DE, 
\begin{align}
{\cal D}_{13} \, U_a = l_a \,,\qquad   
{\cal D}_{13} \, V_I = 0 \,,\qquad a=1,\ldots,4 \,,\quad  I=1,\ldots,135 \,.
\end{align}
Thus, $S_2^{(i)}$ is a linear combination of 4+135 weight-two integrable symbols $U_a$ and $V_I$. Then using the 44-letter alphabet, we recursively construct the integrable symbols up to weight six employing the first entry condition and restricting the second component in the coproduct $(4,2)$ of the weight-six symbols. The counting of integrable symbols constructed this way is given in Table~\ref{tab:symb5pt3L}. 
\begin{table}[t]
\centering
\begin{tabular}{|l|c|c|c|c|c|c|} \hline
weight & 1 & 2 & 3 & 4 & 5 & 6 \\ \hline
\# symbols & 4 & 15 & 61 & 279 & 1380 & 503 \\ \hline
\end{tabular}
\caption{Number of integrable symbols for $f^{(3)}_{5,13}$.}\label{tab:symb5pt3L}
\end{table}
We have not completely exhausted the information contained in the DE. Namely, acting by ${\cal D}_{13}$ on our ansatz of 503 symbols we obtain a linear combination of weight-four integrable symbols accompanied by four rational factors $l_a$,
\begin{align}
{\cal D}_{13} \, {\cal S}\left( f^{(3)}_{5,13} \right) = \sum_a l_a \, S^{(a)}_{4} \,.
\end{align} 
Comparing them with the weight-four symbols of the two-loop ladders $f^{(2)}_{5,kl}$ from the RHS of DE \eqref{DE}, we fix 407 unknowns in our 503-dimensional ansatz. We are able fix  the remaining 96 unknowns demanding the absence of spurious poles in $F_5^{(\laddthreepic)}$, see \eqref{F5_3loop}. That is, the rational prefactor $\Omega_{13}$ has an unphysical pole at $s_{13} \to 0$, which should be compensated by the vanishing symbol,
\begin{align}
{\cal S} \left(  f^{(3)}_{5,13}\right) \bigr|_{s_{13}\to 0} = 0 \,.
\end{align}
Adding the cyclic terms, we have completely fixed the symbol of the five-point three-loop ladder $F_5^{(\laddthreepic)}$.

We have also checked that the soft and collinear limits of the weight-six symbol ${\cal S}\left( F_5^{(\laddthreepic)} \right)$ reproduce the known four-point three-loop ladder expressions ${\cal S}\left( F_4^{(\laddthreepic)} \right)$ \cite{Alday:2013ip,Arkani-Hamed:2021iya}.

{\it \underline{3. Properties of the symbol.}} 
We provide symbol expression for the three-loop ladder $F_5^{(\laddthreepic)}$ in the ancillary file.  The 17 last entries of our symbol ansatz involve 35 alphabet letters including algebraic letters with square roots $d_i$. However, the final result for the ladder turns out to be much simpler. We count 10 last entries for $f^{(3)}_{5,13}$ which involve only planar two-loop letters \eqref{20_5_pt_planar_alphabet},
\begin{align}
& \left[ \frac{W_2}{W_1} \right] , \,
\left[\frac{W_4}{W_1}\right] ,\,
\left[\frac{W_5}{W_3}\right] , \,
\left[\frac{W_{11}}{W_1} \right],\,
\left[\frac{W_{14}}{W_1} \right],\,
\left[\frac{W_{17} W_{26}}{W_{3}} \right],\notag\\
& 
\left[\frac{W_{3}W_{27}}{W_{18}} \right],\,
\left[\frac{W_{3} W_{28}}{W_{19}} \right],\, 
\left[\frac{W_{20} W_{29}}{W_{3}} \right],\,
\left[\frac{W_{1} W_{30}}{W_{16}} \right]\,,
\end{align}
and the last entries of other pure terms $f^{(3)}_{5,ij}$ are obtained by cyclic shifts. Also, the first and second entries of the symbol involve only planar two-loop letters. However, the three-loop letters \eqref{eq:new5ptletters} do appear in the third, fourth and fifth entries. In particular, all 45 alphabet letters are present in the fourth symbol entry.

\section{Conclusions and outlook}
\label{sec:conclusions}

Building upon earlier results \cite{Prlina:2017azl,Prlina:2017tvx,Arkani-Hamed:2021iya,Chicherin:2024hes}, 
our work investigated the connections between various topics in scattering amplitude research, such as positive geometries, Landau analysis and symbology, as well as symbol bootstrap, through a systematic study of integrated negative geometries. In particular, our work provides insights into how the integrand-level positive geometry framework governs the physical singularities of the integrated results. 

We obtained symbol-level results for ladder-type negative geometries, using the symbol bootstrap approach and geometric Landau analysis.
Starting from the primary definition of an individual negative geometry, we enumerated all of its maximal-codimension 
boundaries in the form of intersecting line configurations. 
Based on those line configurations we calculated all leading singularities for the negative geometry.
Furthermore, by treating the line configurations as loop momentum solutions for maximal residues, we explicitly constructed Landau diagrams associated to that negative geometry. By analyzing the Landau diagrams and symbol structures of individual negative geometries via their boundary configurations, we were able to obtain all the singularity information and symbol letters.   Assuming that the function space is given by d$\log$ iterated integrals of these symbol letters, 
we were able to uniquely determine symbol results for ladder negative geometries at six points two loops, and at five points and three loops. We present the explicit symbol results in the following ancillary files:
\begin{itemize}
\item Six points, two loops:
The weight-four symbols for the six-point two-loop ladder, $f^{(2)}_{6,13}$ and $f^{(2)}_{6,14}$ in \eqref{ladder_6pts_decomp}, are provided in {\tt 6pt-2L-ladder.nb}.
    \item Five points, three loops: The weight-six symbol result for the five-point three-loop ladder, $f^{(3)}_{5,13}$ in \eqref{F5_3loop}, are given in {\tt 5pt-3L-ladder.nb}, which loads the data from {\tt 3L-ladder-data.txt}.
    \item The integrand of the ladder negative geometry at any multiplicity and any loop order, cf. \texttt{Ladder-integrand.nb} and Appendix~\ref{app:integrand}.\end{itemize}

We now list some open questions related to our work.

{\it \underline{1. Integrating general positive geometries.}} 
It is an interesting question to apply the algorithm presented in section~\ref{sec:landa and schubert} to more general positive geometries. 
For example, it is an interesting problem to find all singularities for three-loop ladders at any multiplicity. The Landau diagrams we provided in the present case can be a useful starting point for such an investigation.
It is also interesting to identify singular loci for more complicated cases, such as negative geometries beyond ladders, as well as the alphabet for the full observable $F_n^{(L)}$. In particular, such a calculation would test the prediction in \cite{Chicherin:2024hes} for the three-loop planar alphabet.

{\it \underline{2. Leading singularities of negative geometries.}}
In the recent work \cite{Carrolo:2025pue}, the full Wilson loop with a single Lagrangian insertion at six points and two loops has been computed at the symbol level. Combining this result with the symbol result obtained here for the ladder with a marked node at the one end, we can find the symbol of each summand in the two-loop negative geometry decomposition of the Wilson loop with a Lagrangian insertion, 
\begin{equation}\label{WL_L2_decomposition}
F^{(2)}_n = - F_n^{(\laddtwopic)} - \tfrac{1}{2} F_n^{(\laddladdpic)} +\tfrac{1}{2} F_n^{\left(\looppic\right)} \,
\end{equation}
with $n=6$. Note that eq. \eqref{WL_L2_decomposition} follows from eq. \eqref{omega_tilde_in_neg_geom}, which in particular is valid at the integrand level.
The only missing ingredient for determining all individual contributions in eq. \eqref{WL_L2_decomposition} concerns the leading singularities. Although genuine LS of Wilson loops with a Lagrangian insertion were fully classified in \cite{Brown:2025plq}, it is still an open question to classify and evaluate all LS for every individual negative geometry. 
It is therefore interesting to see if for given class of graphs, e.g. by restricting to tree graphs, there are bounds on the complexity of the LS configurations, or at least on the LS values, as well as the count of the linearly independent LS. In Appendix \ref{app: All leading singularity values at two loops}, we discuss all LS of two-loop negative geometries. 

\begin{figure}
    \centering
     \subfigure[]{\label{fig:T3ILandau}\includegraphics[width=0.35\linewidth]{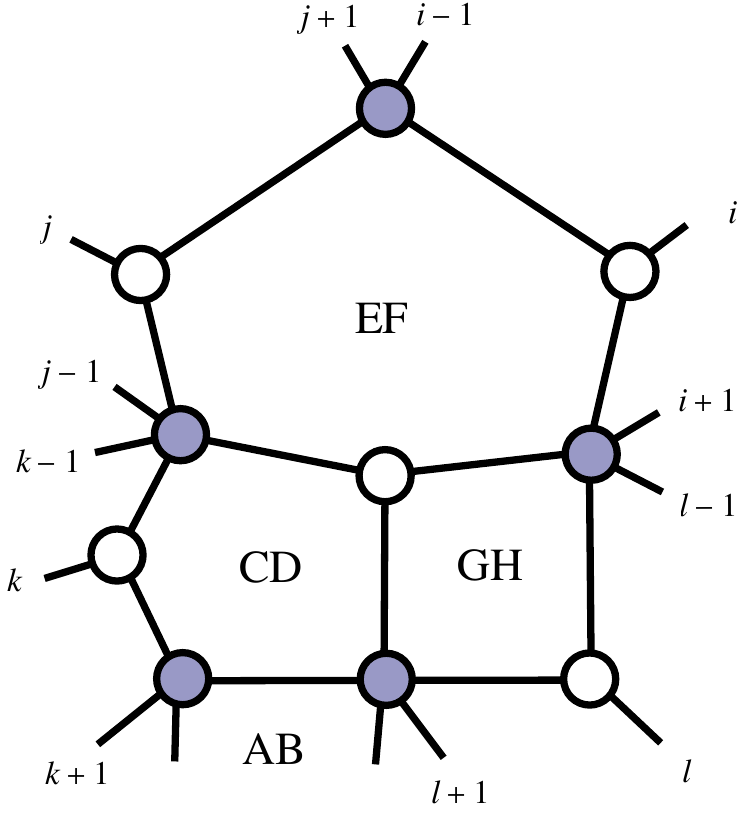}} $\quad\quad$
      \subfigure[]{\label{fig:elliptcishrink}\includegraphics[width=0.35\linewidth]{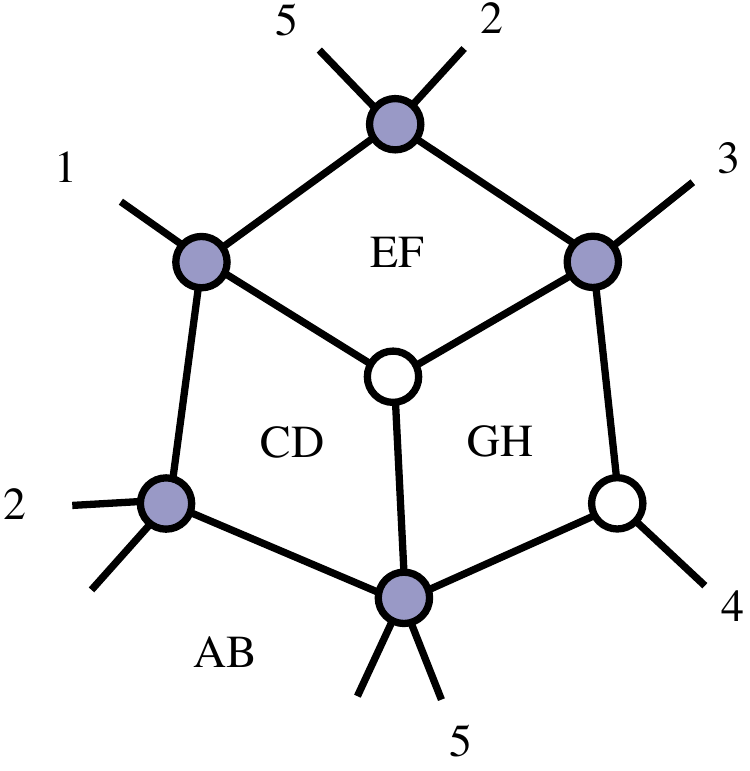}}
    \caption{Diagram \ref{fig:T3ILandau} is leading Landau diagram for the three-loop triangle with one handle, see Fig.~\ref{triangle_LS}, together with its elliptic sub-topology at $n=5$ drawn in  diagram~\ref{fig:elliptcishrink}.}
    \label{fig:elliptic}
\end{figure}

{\it \underline{3. Geometric Landau analysis for individual transcendental pieces.}}
By inspecting the symbols calculated in this work, we observe that only a strict subset of the constructed alphabet is needed to reproduce a single term in the decomposition \eqref{F_integral_in_LS}, e.g. $f_{6,13}^{(2)}$ and $f_{6,14}^{(2)}$ in the six-point two-loop ladder. It is therefore plausible there exist additional selection rules for the alphabet of an individual transcendental piece, such as a geometric one pursued in this paper. Then, a natural question is if to each summand in that equation we can associate a positive geometry, whose canonical form yields exactly the integrand for the corresponding piece. 
This is an open question already for ladders, whose integrand is known explicitly, see eq.~\eqref{ladder_integrand}. This could be useful in principle for applying the same  geometric Landau procedure presented in this paper, but more targeted to an individual summand in eq.~\eqref{F_integral_in_LS}.

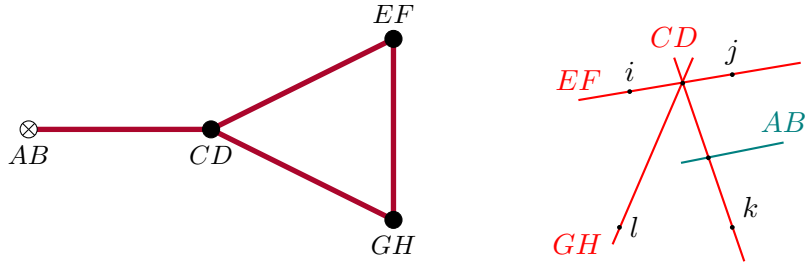
\begin{figure}[t]
\centering
\begin{tikzpicture}[baseline=(current bounding box.south),scale=0.8]

    \draw[mybrown, line width=2pt] (0,0) -- (3,0) -- (6,1.5)--(6,-1.5)--(3,0);

    \draw[fill=white] (0,0) circle (4pt);
    \draw (0,0) -- (45:4pt);
    \draw (0,0) -- (135:4pt);
    \draw (0,0) -- (-45:4pt);
    \draw (0,0) -- (-135:4pt);

   \filldraw (3,0) circle (4pt);
    \filldraw (6,1.5) circle (4pt);
        \filldraw (6,-1.5) circle (4pt);
\node[anchor=north] at (0,-0.1) {\small{$AB$}};
\node[anchor=north] at (3,-0.1) {\small{$CD$}};
\node[anchor=south] at (6,1.6) {\small{$EF$}};
\node[anchor=north] at (6,-1.6) {\small{$GH$}};
\end{tikzpicture}\quad\quad\quad\quad
\begin{tikzpicture}[scale = 0.45]

    \begin{scope}[xshift=0cm]

    \coordinate (i) at (1,-2);
    \coordinate (P) at (0.3,0.05);
    \coordinate (Q) at (-0.45,2.25);
    \coordinate (j) at (-2,2);
    \coordinate (k) at (1,2.5);
    \coordinate (C) at (-0.7,3);
    \coordinate (D) at (1.35,-3);
    \coordinate (E) at (-3.5,1.75);
    \coordinate (F) at (3,2.85);
    \coordinate (A) at (-0.5,-0.1);
    \coordinate (B) at (2.5,0.5);
    \coordinate (G) at (-2.5,-2.5);
    \coordinate (H) at (-0.15,3);
    \coordinate (l) at (-2.3,-2);



    \draw[red, thick] (C) -- (D);
    \node[above, red] at (C) {$CD$};

    \draw[red, thick] (E) -- (F);
    \node[above, red] at (E) {$EF$};

    \draw[red, thick] (G) -- (H);
    \node[left, red] at (G) {$GH$};

    \draw[teal, thick] (A) -- (B);
    \node[above, teal] at (B) {$AB$};

    \fill (i) circle (2pt);
    \fill (j) circle (2pt);
    \fill (k) circle (2pt);
    \fill (P) circle (2pt);
    \fill (Q) circle (2pt);
    \fill (l) circle (2pt);

    \node[above right] at (i) {$k$};
    \node[above] at (j) {$i$};
    \node[above] at (k) {$j$};
    \node[right] at (l) {$l$};

    \end{scope}

\end{tikzpicture}
\caption{The graph associated to the three-loop triangle with one handle and one of its LS line configuration.}
\label{triangle_LS}
\end{figure}

{\it \underline{4. Elliptic cuts in negative geometries.}} As we have seen, Landau diagrams for negative geometries are non-planar in dual space and may lead to non-planar singular loci in general. To explore their singularities, a natural question is: do individual integrated negative geometries involve higher-genus analogues of multiple polylogarithms, such as elliptic functions? 
Currently, we do not have an answer to this question. However, for negative geometries with more complicated topologies, we find elliptic cuts in their associated Landau diagrams. 
The first example of elliptic cut is given by the Landau diagram in Fig.~\ref{fig:T3ILandau} associated to the LS configuration in Fig.~\ref{triangle_LS}, from three-loop triangle with one handle. At $n=5$, we see an elliptic cut from the leading Landau equation of the sub-topology given in Fig.~\ref{fig:elliptcishrink}, with a minimal coloring, by choosing the indices $\{i,j,k,l\}=\{3,5,1,4\}$ in Fig.~\ref{fig:T3ILandau}. It would be an interesting question to explore if such cut and its related singularities will actually show up in the integrated negative geometries, i.e. in eq.~\eqref{F_integral} associated to the three-loop triangle with one handle.

\section*{Acknowledgments}

We thank Chia-Kai Kuo, Song He and Marcus Spradlin for insightful discussions. This work was supported by the European Union (ERC, UNIVERSE PLUS, 101118787). Views and opinions expressed are however those of the authors only and do not necessarily reflect those of the European Union or the European Research Council Executive Agency. Neither the European Union nor the granting authority can be held responsible for them. DC is supported by ANR-24-CE31-7996. JT and DC are grateful to the Max Planck Institute for Physics for hospitality. JT is supported by the DOE grant No.SC0009999 and the funds of the University of California. JH, EM and JT acknowledge the support by the grant NSF PHY-2309135 to the Kavli Institute for Theoretical Physics (KITP).

\appendix

\section{Classification proof for leading singularity configurations of ladders}\label{app:ladder proof}

In this Appendix, we present the proof of the classification of leading singularities for ladder negative geometries. 
A further discussion of this procedure can be found in~\cite{Brown:2025plq}. In contrast to~\cite{Brown:2025plq}, we classify leading singularity configurations of individual negative geometries, rather than those of the Wilson loop with Lagrangian insertion. 

In the following, we use the fact that a loop line $AB$ is an element of ${\rm Gr}(2,4)$, and it can be represented by a $2 \times 4$ matrix modulo left-multiplication by ${\rm GL}(2)$. 
We can therefore parameterize $AB$ using an arbitrary basis of four linearly independent twistors $Z_i,Z_j,Z_k,Z_l$, fixing the ${\rm GL}(2)$ redundancy, as
\begin{equation}
    A = Z_i + \alpha Z_k + \beta Z_l \ , \quad B = Z_j + \gamma Z_k + \delta Z_l \ ,
\end{equation}
for $\alpha, \beta,\gamma,\delta \in \mathbb{C}$.
In the following we use such a parametrisation on the real Grassmannian with $\alpha,\beta,\gamma,\delta \in \mathbb{R}$, and a specific choice of $i,j,k,l$ can simplify or make more transparent some considerations. Note in particular that $AB$ has $4 = \dim({\rm Gr}(2,4))$ degrees of freedom. In the case of $L$-loops, we have $L$ lines $AB_\ell$ in $\mathbb{P}^3$ for $\ell =1, \dots, L$ representing the loop variables.

We now prove the classification results for LS configurations of ladders given in Section \ref{section_LS_L1}. The idea is to use an inductive argument on the number of loops $L \geq 1$. We first perform an explicit analysis for $L=1,2,3$.

{\it{\underline{$L=1$.}}} 
Note that if the LS does not involve the cut $\langle AB CD \rangle = 0$, then the line $CD$ localizes to a zero-dimensional boundary of the one-loop Amplituhedron, i.e. $CD= ij$. The corresponding LS is then given by the canonical function of the one-loop geometry in $AB$ with the single extra constraint $\langle AB ij \rangle < 0$. The value of this LS is equal to $\Omega_{n,ij}$, see eq.~\eqref{Omega_ij}. Assume now that $CD$ intersects $AB$, i.e. that $\langle ABCD \rangle = 0$. Since by definition of LS we do not put any constraint on $AB$, we have that $CD$ must lie in a one-dimensional boundary of the one-loop Amplituhedron. By \cite{Ranestad:2024svp}, we can parametrize this boundary as
\begin{equation}\label{1d_boundary}
    CD = k (i+ \alpha \, (i+1)) \ , \qquad \alpha > 0 \ .
\end{equation}
From the intersection condition with $AB$, it follows that $CD = k, AB \cap (kii+1)$, and in particular that
\begin{equation}\label{alpha}
    \alpha =  \frac{\langle AB ki \rangle}{\langle AB i+1k \rangle}  \ .
\end{equation}
The fact that $CD$ lies in the one-loop Amplituhedron forces $\alpha >0$, which puts constraints on the one-loop geometry in $AB$. According to \cite{Brown:2025plq}, this space can be triangulated into Kermits, and therefore the LS can be expressed as a sum of Kermit forms.

{\it{\underline{$L=2$.}}}  If $CD$ does not intersect $EF$, then $EF=jl$ and $CD$ localizes to one of the $L=1$ configurations seen above. So assume that $\langle CDEF \rangle  = 0$. Since both lines $CD$ and $EF$ must be localized, one of the two must be already fully localized before imposing the intersection condition $\langle CDEF \rangle  = 0$. If $CD$ was already localized, then it must be in one of the $L=1$ configurations. In any case, we can write $CD = k(i+\alpha \, (i+1))$ with $\alpha =0$ or as in \eqref{alpha}. Since $EF$ has at most one degree of freedom before imposing the intersection condition with $CD$, we can write $EF = j(l + \beta \, (l+1))$ for $\beta \geq  0$. Then, 
\begin{equation}\label{1d_1d_cut}
    \langle CDEF \rangle  = 0 \implies \beta = - \frac{\langle ki jl \rangle + \alpha \, \langle ki+1jl \rangle }{\langle ki jl+1 \rangle + \alpha \, \langle ki+1jl+1 \rangle}  \ .
\end{equation}
If $\alpha = 0$, the only way in which $\beta \geq 0$ is if e.g. $l=k$, i.e. $\beta = 0$ and hence $CD=ki$ and $EF = jk$. In this case the LS value is given by $\Omega_{n,ki}$. If $\alpha$ is as in \eqref{alpha}, $\beta \geq 0$ forces either $l=k$, i.e. $\beta = 0$ and $EF = jk$, or $l=i$, i.e. $\beta = \alpha$. These configurations are the second and third ones in Fig. \ref{LS_L1}, respectively. Note that in both cases the LS value is the same as that of Fig. \ref{LS_L1}.

On the other hand, if $EF$ was already localized, than it must be equal to some $ij$. The line $CD$ must now intersect $AB$ and $EF$, and in order to be fully localized it must lie in a two-dimensional boundary of the one-loop Amplituhedron. By \cite{Ranestad:2024svp} there are two types of such, parametrized as
\begin{equation}\label{2d_boundaries}
    C = k \qquad \text{or} \qquad CD = (k + \alpha \, (k+1))(l + \beta \, (l+1)) \quad \text{with $\alpha, \, \beta>0$} \ .
\end{equation}
For the second case, the only solution to 
\begin{equation}\label{2d_0d}
    \langle CDEF \rangle = \langle klij \rangle + \alpha \, \langle k+1lij \rangle + \beta \, \langle kl+1ij \rangle + \alpha \beta \, \langle k+1l+1ij \rangle = 0 
\end{equation}
is when w.l.o.g. $l=i$ and $\alpha = 0$, which is a configuration of the type as the second one in Fig.~\ref{LS_L1}. 
Let us therefore assume that $CD$ has the form of the first equation in (\ref{2d_boundaries}). Then, the intersection condition with $EF = ij$ allows to write $CD = k (i + \alpha \, j)$ and the intersection condition with $AB$ fixes $\alpha$ as 
\begin{equation}\label{alpha2}
    \alpha = \frac{\langle AB ik \rangle }{\langle AB kj \rangle} \ .
\end{equation}
The one-loop Amplituhedron conditions for $CD$ force $\alpha > 0$, which in turn put constraints on the one-loop geometry in $AB$. The situation is analogous to the one at $L=1$, and also in this case the value of the LS can be written as a sum of Kermit forms.

{\it{\underline{$L=3$.}}} 
Assume first that $CD$ and $EF$ are localized. In this case, $CD$ and $EF$ form a configuration among those seen at $L=2$, and we require that the extra loop line $GH$ intersects $EF$. Applying the same argument presented around \eqref{1d_1d_cut} to the loop line $GH$, one checks that the configuration localizes to one among (a)-(e) in~Table~\ref{tab:11}. 

Let us instead assume that $EF$ and $GH$ are localized. Then, by the same argument around \eqref{1d_1d_cut}, we find that $EF=ij$ and $GH=il$. The analysis reduces to that carried out for $L=2$, and yields that $CD=k,AB \cap (ijk)$. This is the configuration appearing in Table~\ref{tab:11}(b).

The new interesting configuration arises by considering the loops $CD$ and $GH$ already localized, i.e. by above $CD=k,AB \cap (kii+1)$ and $GH=jm$, and looking at which configurations for $EF$ are allowed. By counting the degrees of freedom, $EF$ must lie in a two-dimensional boundary of the one-loop Amplituhedron, and the form of these are given in \eqref{2d_boundaries}. The argument around \eqref{2d_0d} shows that $EF$ cannot be of the form as the second equation in \eqref{2d_boundaries}. Therefore, we can take w.l.o.g. $E=l$. Since we assume that $EF$ intersects $GH$, we can take $F$ to lie on $GH$, i.e. we can write $F=j + \gamma \, m$ with some real parameter $\gamma$. Note that in order for $EF$ to lie in the one-loop Amplituhedron, if $l<j<m$ then $\gamma$ must be non-negative. We can now impose the intersection condition between $EF$ and $CD= k(i + \alpha \, (i+1))$, for $\alpha$ as in \eqref{alpha}. This fixes
\begin{equation}\label{2d_crossing}
    \langle CD EF \rangle = 0 \implies  \gamma = -\frac{\langle kilj \rangle + \alpha \, \langle ki+1 lj \rangle}{\langle kilm \rangle + \alpha \, \langle ki+1 lm \rangle} \  .
\end{equation}
One can check that in order for $\gamma $ to be positive, since $\alpha > 0$, there must be \q{crossing} constraints on the indices $i,j,k,l,m$ exactly as for (f) in Table \ref{tab:12}. Degenerate solutions for which some indices coincide and $\gamma = 0$ reduce the configuration to one of those discussed previously. Exactly the same computation shows that the configuration (g) in Table \ref{tab:12} is also allowed. This concludes the three-loop analysis.

{\it{\underline{Higher $L$.}}} As we have seen up to three loops, some of the LS configurations of ladders have a recursive structure. 
In the following we make this observation concrete, and provide a classification for all-loop ladder negative geometries. 

We prove the following. Every LS configuration of a ladder at $L$ loops has the following structure. Either the first two loops $CD$ and $EF$ are as in Fig.~\ref{LS_L1}(b)-(d), or the first three loops $CD$, $EF$ and $GH$ are as (f)  in Table \ref{tab:12}. Then, every loop line $AB_\ell$ for $3 \leq \ell \leq L$ is localized to one of the following configurations:
\begin{enumerate}
        \item $i_\ell j_\ell$, 
        \item $(i_\ell i_{\ell -1}j_{\ell -1}) \cap (i_\ell i_{\ell +1}j_{\ell +1})$, if $i_{\ell -1}<i_{\ell + 1}< j_{\ell -1}<j_{\ell +1}$ and $i_\ell \notin \{i_{\ell -1}, i_{\ell + 1}, j_{\ell -1} , j_{\ell +1}\}$, 
        \item $i_\ell, (ii+1) \cap (kAB)$, if $AB_{\ell -1} = i_{\ell-1} , (ii+1) \cap (kAB)$,  
        \item $i_\ell, AB_{\ell -1} \cap (i_\ell i_{\ell +1} j_{\ell +1})$, if $AB_{\ell -1} = i_{\ell-1} , (ii+1) \cap (kAB)$ with $i<i_{\ell +1} < i_{\ell -1} < i_\ell < j_{\ell +1} \leq i$ and $|j_{\ell +1}-i_{\ell +1}| >1$, 
\end{enumerate}
where $i_\ell, j_\ell \in \{1,\dots,n\}$. Note that 2. means having three consecutive loop lines localized as (g) shown in Table~\ref{tab:12}, and in particular it requires $AB_{\ell-1}$ and $AB_{\ell+1}$ to be localized as in (a). Cases 3 and 4 can happen only if $CD$ and $EF$ belong to configuration Fig.~\ref{LS_L1}(d) shown in Fig.~\ref{LS_L1}~case 3 allows for configurations like (d) shown in Table~\ref{tab:11}, with possibly more loop lines localized as $EF$ and $GH$ in case 4. allows for configurations involving three consecutive loop lines localized as (f) shown in Table~\ref{tab:12}, and in particular it requires that $AB_{\ell-1}$ is localized as in 3 and $AB_{\ell+1}$ as in~1.
    
Moreover, the loops must form a \textit{chain}, meaning that whenever $AB_{\ell+1}$ is localized as in (a), and $AB_\ell$ as in (a) or (c), then w.l.o.g. $i_{\ell+1} = i_{\ell}$.

The proof for this result is by induction over $L$. We already proved the cases $L=1,2,3$ corresponding to points 1,2,3. Let us therefore prove the induction step for $L>3$. Assume first that $AB_{L-1}$ is localized before imposing the multiloop cut condition with $AB_{L}$. This means that the $(L-1)$-ladder consisting in the first $L-1$ loop lines is localized by itself. By induction hypothesis, we have either $AB_{L-1}=i_{L-1}j_{L-1}$ or $AB_{L-1}=i_{L-1},(ii+1) \cap (ABk)$, where the latter can happen only if $CD$ and $EF$ are localized as in configuration (d) of Fig.~\ref{LS_L1}. Then, the allowed locations for $AB_{L}$ are determined by using the same argument as around~\eqref{1d_1d_cut}. The result is the following: in the former case $AB_{L} = i_{L}j_{L}$ with w.l.o.g. $i_L=j_{L-1}$, while in the latter there is the additional solution $AB_{L}=i_{L},(ii+1) \cap (ABk)$. The induction step is therefore concluded for this case.

Assume instead that $AB_L$ is localized before imposing the cut condition with $AB_{L-1}$. Then $AB_L = i_Lj_L$. If $AB_{L-2}$ is also localized before imposing the cut condition with $AB_{L-1}$, then the $(L-2)$-ladder consisting in the first $L-2$ loop lines is localized. As above, by induction hypothesis we have that either $AB_{L-2}=i_{L-2}j_{L-2}$ ot $AB_{L-2}=i_{L-2},(ii+1) \cap (ABk)$, with the latter happening only if the chain started as Fig.~\ref{LS_L1}(d) in Fig.~\ref{LS_L1}. To understand which locations are allowed for $AB_{L-1}$, one follows the same argument as around \eqref{2d_crossing}. The result is the following. In the former case, either $AB_{L-1}=i_{L-1}j_{L-1}$ with w.l.o.g. $i_{L-1}  = j_{L-2}$ and $j_{L-1} = i_{L}$, or $AB_{L-1}$ is localized as in point (b) in the classification result of Section~\ref{section_LS_L1}. In the latter case, we find that solutions (a), (c) and (d) are all possible, if the ordering of the indices allows for them. On the other hand, if $AB_{L-1}$ is also localized before imposing the cut condition with $AB_{L-2}$, by the same argument as around \eqref{1d_1d_cut} we have that $AB_{L-1} = i_{L-1}j_{L-1}$ with w.l.o.g. $j_{L-1}= i_L$, and one can directly use induction with the argument we just gave. The proof is therefore completed.

\section{All leading singularity values for factorized ladders}
\label{app: All leading singularity values at two loops}

\begin{figure}[t]
\centering
\begin{tikzpicture}[baseline=(current bounding box.south),scale=0.8]

    \draw[mybrown, line width=2pt] (0,0) -- (3,0); 
    \draw[mybrown, line width=2pt] (0,0) -- (-3,0); 
    \draw[mybrown, line width=2pt, dashed] (-3,0)--(-6,0);
    \draw[mybrown, line width=2pt, dashed] (6,0)--(3,0);

    \draw[fill=white] (0,0) circle (4pt);
    \draw (0,0) -- (45:4pt);
    \draw (0,0) -- (135:4pt);
    \draw (0,0) -- (-45:4pt);
    \draw (0,0) -- (-135:4pt);

   \filldraw (3,0) circle (4pt);
    \filldraw (6,0) circle (4pt);
     \filldraw (-6,0) circle (4pt);
    \node[anchor=north] at (0,-0.2) {\small{$(AB)_0$}};
     \node[anchor=north] at (3,-0.2) {\small{$(CD)_1$}};
      \node[anchor=north] at (-3,-0.2) {\small{$(AB)_1$}};
    \node[anchor=north] at (6,-0.2) {\small{$(CD)_{L_2}$}};
        \node[anchor=north] at (-6,-0.2) {\small{$(AB)_{L_1}$}};

\end{tikzpicture}
\caption{Graph associated to a ladder at $L$ loops, which factorizes into two ladders of length $L_1$ and $L_2$ with $L_1+L_2=L$.}
\label{factorised_LS}
\end{figure}
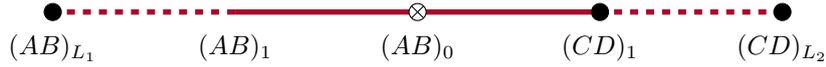

In this section, we provide values for all LS of the $n$-point ladder with a marked node in the middle, see Fig.~\ref{factorised_LS}.

This negative geometry immediately reduces to product of ladders as in Fig.~\ref{fig:laddersgeneral}. Namely, the $(L_1 + L_2)$-loop ladder in Fig.~\ref{factorised_LS} is the product of an $L_1$-loop and an $L_2$-loop ladder, both with a marked node at one end. In the simplest case where $L_1 = L_2 =1$, this classification yields all LS values of the triangle negative geometry, according to the discussion below eq.~\eqref{WL_L2_decomposition}, and hence it completes the classification of LS values of all individual two-loop $n$-point negative geometries.

The factorization of the ladder with a marked node in the middle takes place both at the level of integrands and integrated negative geometries. The integrand of the ladder in Fig. \ref{factorised_LS}, i.e. its canonical form, 
can be computed analogously to that of the ladder~\eqref{ladder_integrand}, 
\begin{align}\label{ladder_L@_marked_integrand_L1L2}
& \widetilde{\Omega}_n^{(L_1+L_2)}\left(\text{fig.~\ref{factorised_LS}}\right) \notag \\ 
&=\sum_{(i_1 j_1), \, (i_2 j_2)} \Omega_n(i_1j_1,i_2j_2) \, \mathcal{C}_{i_1 j_1}(AB,AB_1,\dots,AB_{L_1}) \, \mathcal{C}_{i_2,j_2}(AB,CD_1,\cdots,CD_{L_2}) \, ,
\end{align} 
where the sum runs over all pairs of arcs of an $n$-gon, and $\mathcal{C}_{i j}$ are given in eq.~\eqref{ladder_integrand_C}. Due to the factorized structure of eq. \eqref{ladder_L@_marked_integrand_L1L2}, we can integrate the loops $AB_1,\ldots,AB_{L_1}$ and $CD_1,\ldots,CD_{L_2}$, to obtain
\begin{align}
\label{F_ladder_marked_middle}
F_n\left( \text{fig.~\ref{factorised_LS}} \right) = \sum_{(i_1 j_1), \, (i_2 j_2)} \Omega_n(i_1j_1,i_2j_2) \, f^{(L_1)}_{n, i_1j_1} \,  f^{(L_2)}_{n, i_2j_2}  \,.  
\end{align}
The LS of eq.~\eqref{F_ladder_marked_middle} are the rational functions $\Omega_n(i_1j_1,i_2j_2)$,
since $f^{(L)}_{n,ij}$'s are pure. For example, 
\begin{equation}\label{eq_F_ladder_marked_middle}
    F_n^{(\laddladdpic)} = \sum_{(i_1 j_1), \, (i_2 j_2)} \Omega_n(i_1j_1,i_2j_2) \, I^{cp}_{i_1 j_1} I^{cp}_{i_2 j_2} \, ,
\end{equation}
where the chiral pentagons $I^{cp}_{ij}(AB)$ are given in eq. \eqref{chir_pent}.

For non-crossing arcs, the LS can be expressed as a linear combinations of Kermit forms according to \cite{Brown:2025plq},
\begin{align}\label{LS_simple_formulae}
    \Omega_n(i_1j_1,i_2 j_2) &= \sum_{\substack{\Delta_1, \,  \Delta_2 \subset T \\ \Delta_1 \subset {\rm P}_1 \,, \, \Delta_2 \subset {\rm P}_2 }} [\Delta_1;\Delta_2]  \qquad \text{for non-crossing $(i_1j_1)$ and $(i_2j_2)$}\,,
\end{align}
where the sum is over non-overlapping triangles $\Delta_1 = \{a_1,b_1,c_1\}$ and $\Delta_2 = \{a_2,b_2,c_2\}$ with arcs in a fixed triangulation $T$ of the $n$-gon containing the arcs $(i_1j_1)$ and $(i_2j_2)$.
However, if $(i_1j_1)$ and $(i_2j_2)$ cross, then the LS value has to be determined. These type of LS are called \textit{incompatible} in \cite{Brown:2025plq}, and are absent in the full Wilson loop with Lagrangian insertion, and present only in individual negative geometries, as in~\eqref{F_ladder_marked_middle}.

\begin{figure}[t]
\centering
\begin{tikzpicture}[scale = 0.6]

    \begin{scope}[xshift= 0cm]

    
    \draw[thick] (0,0) circle(3cm);

    \coordinate (i) at (-130:3cm);
    \coordinate (j) at (140:3cm);
    \coordinate (k) at (50:3cm);
    \coordinate (l) at (-40:3cm);

    \node at (i) [below left ] {$i_1$};
    \node at (j) [above left] {$i_2$};
    \node at (k) [above right] {$j_1$};
    \node at (l) [below right ] {$j_2$};


    \fill[black, opacity=0.3] (j) arc (140:230:3cm) -- (i) -- cycle;

    \fill[black, opacity=0.3] (j) arc (140:50:3cm) -- (k) -- cycle;

    \fill[black, opacity=0.3] (k) arc (50:-40:3cm) -- (l) -- cycle;

    \fill[black, opacity=0.3] (l) arc (-40:-130:3cm) -- (i) -- cycle;


    \draw[line width=0.4mm, red, thick] (i) -- (k);
    \draw[line width=0.4mm, red, thick] (j) -- (l);

     \foreach \p in {i,j,k,l} {
        \fill (\p) circle(3pt);
    }

    \node at (180:2.6cm) {$P_1$};
    \node at (0:2.6cm) {$P_2$};
    \node at (90:2.6cm) {$P_3$};
    \node at (-90:2.6cm) {$P_4$};

    \end{scope}

\end{tikzpicture}
\caption{The LS $\Omega_n(i_1j_1,i_2j_2)$ involving the crossing arcs $(i_1j_1)$ and $(i_2j_2)$. This is a LS of the two-loop ladder with marked point in the middle \eqref{F_ladder_marked_middle}. The associated geometric space, can be triangulated into Kermits with possibly a single additional constraint, whose canonical form is given in eq. \eqref{Kermit_intersection2}. The value of the LS is given by eq. \eqref{incompatible_LS1}.}
\label{incompatible_LS}
\end{figure}
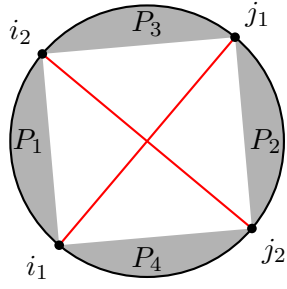

We now give explicit expressions for  $\Omega_n(i_1j_1,i_2j_2)$ in the case of crossing arcs $(i_1j_1)$ and $(i_2j_2)$.
We start by evaluating the canonical form of a Kermit space $[a_1b_1c_1;a_2b_2c_2]$, see eq. \eqref{six_invariant}, with a single additional constraint $\langle AB b_1 d \rangle < 0$ with $c_1 \leq d \leq a_2$. We claim that such a space is a positive geometry, whose canonical function in $AB$ is given by 
\begin{equation}\label{Kermit_intersection2}
    [a_1b_1c_1;a_2b_2c_2|b_1 d]:= \frac{\langle AB (a_1b_1c_1) \cap (a_2b_2c_2) \rangle \, \langle AB (b_1c_1d) \cap (a_2b_2c_2) \rangle}{\langle AB b_1c_1 \rangle \langle AB a_1 c_1 \rangle \langle AB b_1d \rangle \langle AB a_2b_2 \rangle \langle AB b_2c_2 \rangle \langle AB a_2 c_2 \rangle} \ .
\end{equation}
Then, by a triangulation argument along the lines of that presented in Section 4.2 of~\cite{Brown:2025plq}, we compute for crossing arcs $(i_1j_1)$ and $(i_2j_2)$ as in Figure~\ref{incompatible_LS}, 
\begin{equation}\label{incompatible_LS1}
\begin{aligned}
    \Omega_n(i_1j_1,i_2j_2) & = [i_1i_2j_1j_2] + \Omega_n(i_1i_2,j_1j_2) + \Omega_n(i_1j_2,i_2j_1) \\
    & + \sum_{\Delta_1 \subset {\rm P}_1} [\Delta_1;i_1j_1j_2|i_2j_2] + \sum_{\Delta_2 \subset {\rm P}_2} [i_1i_2j_1;\Delta_2|i_2j_2] \\
    & + \sum_{\Delta_3 \subset {\rm P}_3} [\Delta_3;i_1j_1j_2|i_2j_2] + \sum_{\Delta_4 \subset {\rm P}_4} [i_1i_2j_1;\Delta_4|i_2j_2] \,,
\end{aligned}
\end{equation}
where the sums are over triangles $\Delta_r $ inscribed in the polygons ${\rm P}_r$, see Figure~\ref{incompatible_LS}, compatible with an (arbitrary) underlying triangulation of ${\rm P}_r$. For example, for $r=1$ we can triangulate ${\rm P}_1$ by introducing the arcs $(i_1 k)$ for $i_1<k < i_2$. Then, $\Delta_1$ is any triangle with vertices $\{i_1 , k, k+1\}$ for $i_1<k < i_2$. Then, each function $[\Delta_1;i_1j_1j_2|i_2j_2]$ has the form as in eq. \eqref{Kermit_intersection2}, after possibly a dihedral transformation on the indices.

As an example, let us evaluate these crossing LS at $n=5$ and $n=6$. At $n=5$, we have
\begin{equation}
{\tikz[baseline=.1ex]{
  \begin{scope}[xshift = 5 cm]
    \draw[thick] (90:1) 
        \foreach \x in {162,234,306,18} 
            { -- (\x:1) } -- cycle;
    \coordinate (D) at (18:1);
    \coordinate (C) at (90:1);
    \coordinate (B) at (162:1);
    \coordinate (A) at (234:1);
    \coordinate (E) at (306:1);
    \draw[red, thick] (A) -- (C);
    \draw[red, thick] (B) -- (D);
    \node[below] at (A) {1};
    \node[left] at (B) {2};
    \node[above] at (C) {3};
    \node[right] at (D) {4};
    \node[below] at (E) {5};
    \end{scope}
}} \qquad \Omega_5(13,24) = [1234] + [123;145|24]\,,
\end{equation}
together with its four cyclic images. These yield five crossing LS, which are found to be linearly independent from the six-dimensional space generated by the Kermit forms at $n=5$ and $L \geq 2$. These LS are present in both the ladder with a marked node in the middle and in the triangle negative geometry, see eq.~\eqref{WL_L2_decomposition} and~\cite{Chicherin:2024hes}. Hence, the total number of linearly independent LS of individual negative geometries at $n=5$ and $L=2$ is 11. The LS value $\Omega_5(13,24)$ is denoted by $\Omega_0^{13-,24-}(AB)$ in \cite{Chicherin:2024hes}, and also $C_5=[123;145|24]$ in the same reference.

At $n=6$, we have three dihedrally-inequivalent configurations:
\begin{equation}
\begin{aligned}
& {\tikz[baseline=.1ex]{
  \begin{scope}[xshift = 5 cm]
    \draw[thick] (0:1) 
        \foreach \x in {60,120,180,240,300} 
            { -- (\x:1) } -- cycle;

    \coordinate (E) at (0:1);
    \coordinate (D) at (60:1);
    \coordinate (C) at (120:1);
    \coordinate (B) at (180:1);
    \coordinate (A) at (240:1);
    \coordinate (F) at (300:1);
    \draw[red, thick] (A) -- (C);
    \draw[red, thick] (B) -- (D);
    \node[below ] at (A) {1};
    \node[left] at (B) {2};
    \node[above ] at (C) {3};
    \node[above ] at (D) {4};
    \node[right] at (E) {5};
    \node[below ] at (F) {6};
    \end{scope}
}} \qquad \Omega_6(13,24) =  [1234] + [123;145|24] + [123;156|24] \,,\\
& {\tikz[baseline=.1ex]{
  \begin{scope}[xshift = 5 cm]
    \draw[thick] (0:1) 
        \foreach \x in {60,120,180,240,300} 
            { -- (\x:1) } -- cycle;

    \coordinate (E) at (0:1);
    \coordinate (D) at (60:1);
    \coordinate (C) at (120:1);
    \coordinate (B) at (180:1);
    \coordinate (A) at (240:1);
    \coordinate (F) at (300:1);
    \draw[red, thick] (A) -- (C);
    \draw[red, thick] (B) -- (E);
    \node[below ] at (A) {1};
    \node[left] at (B) {2};
    \node[above ] at (C) {3};
    \node[above ] at (D) {4};
    \node[right] at (E) {5};
    \node[below ] at (F) {6};
    \end{scope}
}} \qquad \Omega_6(13,25) = [1235] + [125;345|13] + [156;235|13] \,, \\
& {\tikz[baseline=.1ex]{
  \begin{scope}[xshift = 5 cm]
    \draw[thick] (0:1) 
        \foreach \x in {60,120,180,240,300} 
            { -- (\x:1) } -- cycle;

    \coordinate (E) at (0:1);
    \coordinate (D) at (60:1);
    \coordinate (C) at (120:1);
    \coordinate (B) at (180:1);
    \coordinate (A) at (240:1);
    \coordinate (F) at (300:1);
    \draw[red, thick] (A) -- (D);
    \draw[red, thick] (B) -- (E);
    \node[below ] at (A) {1};
    \node[left] at (B) {2};
    \node[above ] at (C) {3};
    \node[above ] at (D) {4};
    \node[right] at (E) {5};
    \node[below ] at (F) {6};
    \end{scope}
}} \qquad \Omega_6(14,25) = [1245] + [156;234] + [145;234|25] + [124;156|25] \,.
\end{aligned}
\end{equation}

\section{The geometric selection rule from bipartite graphs}
\label{sec: sel rules and bipartite}

In this section we elaborate on the geometric selection rule for the Landau analysis presented previously. This consists in Proposal \ref{proposal_0}, and it is a conjecture stated in \cite{Dennen:2015bet,Dennen:2016mdk,Prlina:2017azl,Prlina:2017tvx}.
We also give a practical combinatorial way of implementing this rule via on-shell diagrams, given in Proposal \ref{proposal}, which extends \cite[Sections 2.4-5]{Prlina:2017azl} to our setting. We point out the subtleties and differences between the amplitude's setting considered in that reference, and that of negative geometries we consider here.
As we have argued in Section~\ref{sec:landa and schubert}, starting from a leading Landau diagram, not all relaxations yield physical singularities. The selection of physical relaxations takes into account the minimal helicity sector in which the singularity from a diagram may appear. In the following, we rely on the following conjecture, stated by the authors in \cite{Prlina:2017azl}.

\begin{tcolorbox}[colback=gray!20, colframe=gray!80, boxrule=0.5mm, arc=4mm, outer arc=4mm]
\begin{proposal}[(Due to \cite{Prlina:2017azl}) for generating all relevant Landau diagrams]\label{proposal_0}
A (sub$^m$-leading) Landau diagram $\mathcal{L}$ should be considered for the singularities of the integral, only if the associated complex variety $S=\mathcal{V}(\mathcal{L})$, defined by the vanishing locus of all propagators involved in $\mathcal{L}$, intersects the positive geometry $\mathcal{A}$ in a set of real dimension equal to the complex dimension of $S$. If this is the case, we say that $\mathcal{L}$ is \textit{physical} and that $S$ is a \textit{boundary}. Otherwise, we call both $\mathcal{L}$ and $S$ \textit{residual}.
\end{proposal}
\end{tcolorbox}
The key property supporting this claim is the following. If a Landau diagram $ \mathcal{L}$ is residual, then the numerator $\mathcal{N}$ of the canonical form of the positive geometry vanishes on $S=\mathcal{V}(\mathcal{L})$. The converse is also true; if a boundary is physical, the numerator does not vanish on it, since by definition the residue of the canonical form on that boundary is the canonical form of the boundary, and it is in particular non-vanishing. Note that $S$ is defined only by the cut conditions from $\mathcal{L}$. In \cite{Prlina:2017azl}, the authors justify the fact that the vanishing of the numerator on $S$ implies that the singularity associated to a solution of $\mathcal{L}$ is spurious, if $\mathcal{L}$ is residual. Note that in principle the Landau analysis is blind to numerators, so including this consideration effectively refines the Landau analysis via the geometric information.

Let us now review how this selection rule is connected to on-shell diagrams, 
following \cite[Section 4]{Prlina:2017tvx}, and connect this discussion with more recent developments in the context of positive geometries.  
A solution to a Landau diagram $\mathcal{L}$ is physical for the N$^k$MHV amplitude, if and only if it corresponds to a (bi)coloring of $\mathcal{L}$, i.e. to an \textit{on-shell diagram}, with helicity weight less or equal to $k$. Hence, in the MHV sector only on-shell diagrams of helicity equal to zero can contribute. The reason for this connection is clear from the geometric point of view. On-shell diagrams are believed to be in bijection with elements in the stratification of the loop Amplituhedron \cite{Arkani-Hamed:2012zlh}. 
An $L$-loop Landau diagram parametrizes an element in the stratification, of codimension equal to $\min\{4L, E \}$, where $E$ is the number of propagators in the diagram, i.e. simple cut conditions. The integrand for the amplitude is conjectured to be the canonical form of the loop Amplituhedron, at fixed specific helicity, number of points and loops. It is conjectured that the \textit{adjoint}, i.e. the numerator of the canonical form, vanishes exactly on the residual arrangement, which is the union of all the strata in the boundary stratification of the geometry, whose intersection with the real geometry is lower dimensional \cite{Ranestad:2024svp,Dian:2024hil}. Therefore, the integrand has vanishing residue at a cut associated to a Landau diagram, if the latter represents a residual stratum. 
This can in turn be detected via on-shell diagrams, as the latter parametrize actual boundaries of the Amplituhedron at a given helicity. Therefore, the task is to exploit the information given by on-shell diagrams to partition the set of Landau diagrams, into physical and residual ones. If a Landau diagram is residual, then the integrand vanishes on \textit{all} the Landau singularities arising from it.

We now apply these ideas to our setting.
We claim that the selection rule provided by the on-shell diagrams works similarly also for negative geometries. There is however one main difference with respect to amplitudes, that have to be taken into account.  
This arises from the fact that among the external kinematical data we have the unintegrated loop $AB$. The presence of cuts involving $AB$ in the Landau diagrams increases the minimal helicity weight, causing a mismatch with the underlying MHV nature of negative geometries. This can be seen from the LS line configuration. For example in the configuration in Fig.~\ref{LS_L1}(a), the line $CD$ intersects the four lines $(k-1k)$, $(kk+1)$, $(ii+1)$ and $AB$. Among these, only two intersect. Therefore, regarding $AB$ as part of the external kinematics, this line configuration would correspond to a one-loop N$^k$MHV configuration with $k \geq 1$.
One is led to think that this issue is fixed by simply requiring that the helicity weight of the diagram obtained after pinching all cuts involving $AB$ must be MHV. However, it turns out that this condition is insufficient to include leading Landau diagrams associated to configurations as the last two rows, i.e. (f) and (g), of Table~\ref{tab:12}. These diagrams carry even higher minimal helicity weight, but correspond nevertheless to boundaries of the ladder negative geometry. In fact, (f) and (g) in Table~\ref{tab:12} are presented with a coloring of minimal helicity weight that is equal to four. 
The procedure we propose for selecting relevant Landau diagrams for a negative geometry at fixed $k$ is therefore the following. 
\begin{tcolorbox}[colback=gray!20, colframe=gray!80, boxrule=0.5mm, arc=4mm, outer arc=4mm]
\begin{proposal}\label{proposal}
Start from any bi-colored leading Landau diagram $\mathcal{L}$ with minimal helicity weight equal to $k$. Discard every sub$^m$-leading diagram obtained from $\mathcal{L}$ by pinching $m$ propagators, if it carries helicity weight equal to $k'$ with $k'>k$. 
\end{proposal}
\end{tcolorbox}
In the context of amplitudes, this rule is equivalent to what is presented in \cite{Prlina:2017tvx}. The modification in the setting of negative geometries takes into account that different LS may have higher minimal helicity weight.

As an illustration of Proposal \ref{proposal}, let us go back to the example discussed in Section \ref{sec:landa and schubert}.
To the leading Landau diagrams in Fig.~\ref{fig:L2.II_LD} we can assign a unique bi-coloring with (minimal) helicity. This coloring corresponds to LS configuration Fig.~\ref{LS_L1}(c) associated to these Landau diagrams. In fact, recall that a white point indicates that three lines intersect in a common point, while a black point is associated to a coplanar condition. According to Proposal ~\ref{proposal}, when we should consider only those sub-leading diagrams whose helicity weight is not higher than that of the leading diagram. As an example of a wrong shrinking, the sub-leading diagram in Fig.~\ref{fig:wrongshrink} has increased helicity weight. Indeed, the latter is obtained by shrinking two vertices of degree three, and this always increases the helicity by one. Therefore, the diagram in Fig.~\ref{fig:wrongshrink} is spurious, and the associated Landau singularity, given in eq.~\eqref{badshrin}, is spurious. Starting from $n=5$, eq.~\eqref{badshrin} can be a non-trivial square root, e.g. for $n = 5$ and $\{i,j,k\}=\{5,4,2\}$. However, as computed in \cite{Chicherin:2024hes}, this singularity is not present in the integrated result for the two-loop five-point ladder negative geometry. We find that the same is true for six-point two-loop and five-point three-loop ladders, In both cases we got extra letters but they all dropped out in the bootstrapped result. These explicit examples support Proposal~\ref{proposal}. 

On the other hand, the Landau singularity \eqref{sing_b} is associated to the Landau diagram in  Fig.~\ref{fig:MHV-2L}. As shown in that figure,
this diagram admits a bi-coloring with helicity weight equal to that of the leading Landau diagram it is obtained from. Geometrically, even though the solution in eq.~\eqref{sing_b} itself lies outside the geometry, as shown by eq.~\eqref{ineq}, its associated cut conditions form a one-dimensional boundary of the geometry. Such boundary is given by the space with $\alpha>0$ in eq.~\eqref{eq:partphy}. This is represented pictorially in Fig.~\ref{fig:selections}(b). As a concrete check, we verified that eq.~\eqref{sing_b} is a letter which is present in the symbol result of six-point two-loop ladder, computed in Section~\ref{subsec: the six point two loop ladder}.

\section{The integrand of the ladder}
\label{app:integrand}

\subsection{The integrand of the ladder at any number of points}

The procedure for computing the integrand of ladder negative geometries at any number of points and loops has been described in \cite{Chicherin:2024hes} and \cite{Glew:2024zoh}. We find it more convenient to work in momentum twistor space, and therefore we follow \cite{Chicherin:2024hes}. The determination of the ladder's integrand at fixed number of points $n \geq 4$ and any loop order $L \geq 1$ boils down into computing certain building blocks $\mathcal{C}_{ij,kl} = \mathcal{C}_{ij,kl}(n)$ for $i,j,k,l \in \{1,\dots,n\}$ with $i<j$ and $k<l$. The pair of indices $(ij),(kl)$ can be thought of as running over the set of pairs of arcs of an $n$-gon. The ladder's integrand is then given by \cite{Chicherin:2024hes}:
\begin{align}\label{ladder_integrand}
    & \widetilde{\Omega}_{n,{\rm ladder}}^{(L)} =   \sum_{(i_1j_1)} \mathcal{C}_{i_1j_1}(AB,AB_1,\ldots,AB_L)\, \Omega_{n,i_1j_1}(AB) \,,
\\ & \mathcal{C}_{i_1j_1}(AB,AB_1,\ldots,AB_L) := \sum_{(i_\ell j_\ell)} P_{i_L j_L}(AB_L,AB_{L-1}) \, \mathcal{C}_{i_{L-1} j_{L-1}, i_{L-2}j_{L-2}}(AB_{L-1},AB_{L-2}) \cdots \notag \\  & \hspace{5cm}\cdots\mathcal{C}_{i_{2} j_{2}, i_{1}j_{1}}(AB_{1},AB) \ ,
\label{ladder_integrand_C}
\end{align}
where the summation runs over arcs of an $n$-gon $(i_\ell j_\ell)$ for $\ell = 1, \dots, L$, and we omitted the dependence of the $\mathcal{C}$'s on $n$. The chiral pentagon integrands are given by
\begin{equation}\label{chir_pent_integrand}
    P_{ij}(AB,CD) := \frac{\langle AB (i{-}1ii{+}1)\cap(j{-}1jj{+}1) \rangle \langle CDij \rangle}{\langle AB i{-}1i \rangle \langle AB ii{+}1 \rangle \langle AB j{-}1j \rangle \langle AB j{-}1j \rangle \langle AB CD \rangle} \ .
\end{equation}
The LS are $\Omega_{n,ij}(AB)$, which are rational functions in twistor coordinates of $AB$, see \eqref{Omega_ij}. Integrations of \eqref{ladder_integrand_C} over $L$ loop momenta result in the pure function $f^{(L)}_{n,i_1j_1}$, see \eqref{ladder_function}. The building blocks $\mathcal{C}_{ij,kl}$ at $n=5$ are displayed in \cite{Chicherin:2024hes}. The latter are computed by first expanding the one-loop Amplituhedron canonical function into Kermit forms, see \eqref{six_invariant}, \eqref{four_invariant} and \eqref{Omega_ij}, and further expanding these into chiral pentagons and boxes. Then, one collects terms with the same LS support $\langle AB kl \rangle$ and subject to $\langle AB ij \rangle <0$. Adding up all such terms yields $\mathcal{C}_{ij,kl}$. This is a hybrid method which uses both the chiral box expansion using local integrands \cite{Arkani-Hamed:2010pyv,Bourjaily:2013mma,Bourjaily:2015jna,Bourjaily:2017wjl,Bourjaily:2020qca} and the direct triangulation of the ladder geometry. We implemented in \texttt{Mathematica} the computation for the $\mathcal{C}_{ij,kl}$'s and the ladder integrand at any $n$ and $L$, and provide the Notebook {\tt Ladder-integrand.nb} in the ancillary files.

\subsection{Explicit expressions for six-point ladders}

Since we present the symbol solution for the integrated six-point two-loop ladder, let us write out the building blocks $\mathcal{C}_{ij,kl}$ at $n=6$. By the cyclic symmetry, we fix $ij=13$ or $ij=14$, all other configurations of indices are obtained by cyclic shifts. For readability, we omit a factor $\tfrac{\langle CD kl \rangle}{\langle AB CD \rangle}$ from each $\mathcal{C}_{ij,kl}=\mathcal{C}_{ij,kl}(AB,CD)$. We provide the code for obtaining these simplifications in the ancillary file {\tt Ladder-integrand.nb}.
\begin{equation}\label{Cs_n6}
\begin{aligned}
    \mathcal{C}_{13,13} &=- \frac{\langle 1234\rangle}{\langle AB13\rangle \langle AB12\rangle  \langle AB 34\rangle}-\frac{\langle 1236\rangle}{\langle AB13\rangle \langle AB16\rangle \langle AB 23\rangle} \, , \quad \\
    \mathcal{C}_{14,14} &= - \frac{\langle 1245\rangle}{\langle AB12\rangle \langle AB14\rangle  \langle AB 45\rangle}-\frac{\langle 1346\rangle}{\langle AB14\rangle \langle AB16\rangle \langle AB 34\rangle} \, , \\
     \mathcal{C}_{13,14} &= -\frac{\langle 3 (12) (45)  (AB) \rangle}{\langle AB13\rangle \langle AB12\rangle  \langle AB 34\rangle \langle AB45 \rangle} \, , \qquad  \mathcal{C}_{14,13} = \frac{\langle 4 (16) (23)  (AB) \rangle}{\langle AB14\rangle \langle AB23\rangle \langle AB 34\rangle \langle AB16 \rangle} \, ,
    \\
   \mathcal{C}_{13,15} &= \frac{\langle AB \cap (123) 456 \rangle}{\langle AB13\rangle \langle AB12\rangle  \langle AB 45\rangle \langle AB56 \rangle} \, , \qquad \ \mathcal{C}_{14,15} = -\frac{\langle 4 (12) (56) (AB)  \rangle}{\langle AB14\rangle \langle AB12\rangle  \langle AB 45\rangle \langle AB56 \rangle} \, , \\
    \mathcal{C}_{13,24} &= \frac{\langle 3 (12) (45) (AB) \rangle}{\langle AB12\rangle \langle AB23\rangle \langle AB 34\rangle \langle AB45 \rangle} \, , \qquad \  \mathcal{C}_{13,24} = \frac{\langle 1 (23)  (45) (AB) \rangle}{\langle AB14 \rangle \langle AB12\rangle  \langle AB 23\rangle \langle AB45 \rangle} \, , \\
    \mathcal{C}_{13,25} &= - \frac{\langle AB \cap (123) 456 \rangle}{\langle AB12\rangle \langle AB23\rangle \langle AB 45\rangle \langle AB56 \rangle} \, , \qquad \mathcal{C}_{14,25} = - \frac{\langle AB \cap (123) 456 \rangle}{\langle AB12\rangle \langle AB23\rangle \langle AB 45\rangle \langle AB56 \rangle} \, ,\\
     \mathcal{C}_{13,26} &= - \frac{\langle 1 (23)(56) (AB)  \rangle}{\langle AB12\rangle \langle AB23\rangle \langle AB 56\rangle \langle AB16 \rangle} \, , \qquad \mathcal{C}_{14,26} = - \frac{\langle 1 (23)(56) (AB)  \rangle}{\langle AB12\rangle \langle AB23\rangle \langle AB 56\rangle \langle AB16 \rangle} \, , \\
     \mathcal{C}_{13,35} &= - \frac{\langle AB \cap (123) 456 \rangle}{\langle AB13\rangle \langle AB23\rangle \langle AB 45\rangle \langle AB56 \rangle} \, , \qquad \mathcal{C}_{14,35} =  \frac{  \langle 4 (23) (56) (AB) \rangle}{\langle AB23\rangle \langle AB34\rangle \langle AB 45\rangle \langle AB56 \rangle} \, ,\\
      \mathcal{C}_{13,36} &= - \frac{\langle 1 (23) (56) ( AB) \rangle}{\langle AB13\rangle \langle AB23\rangle \langle AB 56\rangle \langle AB16 \rangle} \, , \qquad \mathcal{C}_{14,36} = - \frac{ \langle AB \cap (156) 234 \rangle}{\langle AB23\rangle \langle AB34\rangle \langle AB 56\rangle \langle AB16 \rangle} \, , \\
       \mathcal{C}_{13,46} &= 0 \, ,
       \qquad \qquad \qquad \qquad \qquad \qquad \qquad \qquad \mathcal{C}_{14,46} =  \frac{ \langle 1 (34)(56)(AB)    \rangle}{\langle AB14\rangle \langle AB34\rangle \langle AB 56\rangle \langle AB16 \rangle} \, .
\end{aligned}
\end{equation}
Note that the only spurious pole of $\mathcal{C}_{ij,kl}(AB,CD)$ is possibly $\langle AB ij \rangle$, which is canceled by the numerator of the accompanying term in the full expression for the integrand in \eqref{ladder_integrand}.

{\it \underline{L=1 integrated ladder.}}
The six-point one-loop integrated ladder, which is equal to the Wilson loop with single Lagrangian insertion at one-loop order, is given by
\begin{equation}\label{eq:6pt_1L}
    F^{(\laddonepic)}_6 = \Omega_{6,13}\, I^{cp}_{13} + \Omega_{6,14} \, I^{cp}_{14} + \text{cycl.} \ , 
\end{equation}
where the cyclic orbit of the first term contains six terms, while that of the second contains three. The integrated chiral pentagons \eqref{chir_pent_integrand} are pure weight-two  functions,
\begin{equation}\label{chir_pent}
\begin{aligned}
       I_{ij}^{cp} = \int_{CD} P_{ij}(CD,AB) = &\log( u)\log (v)+\text{Li}_2(1{-}u)+\text{Li}_2(1{-}v)+\text{Li}_2(1{-}w) \\      &-\text{Li}_2(1{-}uw)-\text{Li}_2(1{-}vw) \ , 
\end{aligned}
\end{equation}
where the three cross-ratios are defined as
\begin{equation}\label{crpenta}
    u{=}\dfrac{\langle i{-}1iAB\rangle\langle jj{+}1ii{+}1\rangle}{\langle i{-}1ijj{+}1\rangle\langle ABii{+}1\rangle} \ ,\ v{=}\dfrac{\langle jj{+}1AB\rangle\langle i{-}1ij{-}1j\rangle}{\langle jj{+}1i{-}1i\rangle\langle ABj{-}1j\rangle} \ ,\ w{=}\dfrac{\langle i{-}1ijj{+}1\rangle\langle j{-}1jii{+}1\rangle}{\langle i{-}1ij{-}1j\rangle\langle jj{+}1ii{+}1\rangle}.
\end{equation}
The leading singularities in \eqref{eq:6pt_1L} are defined in \eqref{six-point LS}. We write their representations in terms of Mandelstam variables, similarly to the five-point case \cite{Chicherin:2024hes}, which are obtained from \eqref{eq:6pt_1L} after mapping $AB $ to the infinity bitwistor $I_{\infty}$.
More explicitly, we have
\begin{align}
\Omega_{6,13}  \rightarrow\, 
&\frac{1}{2 s_{13}} \Bigl( s_{12} \left(s_{25}
   s_{34}-s_{24} s_{35}+s_{23} s_{45}+{\rm tr}_1\right) +s_{12} \left(s_{16} s_{23}-s_{13}
   s_{26}+s_{12} s_{36}+{\rm tr}_5\right) \notag\\
   &\hspace{1cm} +s_{23}
   \left(s_{16} s_{25}-s_{15} s_{26}+s_{12} s_{56}+{\rm tr}_4\right) + s_{23} \left(s_{14} s_{23}-s_{13} s_{24}+s_{12}
   s_{34}+{\rm tr}_6\right) \notag\\
   &\hspace{1cm} -2
   s_{12} s_{23} s_{234}-2 s_{12} s_{23} s_{345} \Bigr) \,, 
\end{align}
and
\begin{align}
\Omega_{6}^{(0)}{-}\Omega_{6,14}{\rightarrow}\, 
\frac{1}{2\,s_{14}}\,
\Big( s_{23}\left( s_{12} s_{34} {+} s_{23} s_{14} {-} s_{13} s_{24} {+} {\rm tr}_6 \right) {+}s_{56}\left( s_{45}s_{16}{+}s_{56}s_{14}{-}s_{46}s_{15}{+}{\rm tr}_3\right)\Big)\,.
\end{align}
The Mandelstam variables are 
$
    s_{I}=\left(\sum_{i\in I} p_i\right)^2\,, 
$ e.g. the non-adjacent ones are
\begin{align}
    s_{13}&= s_{123}-s_{12}-s_{23}\,,\quad\quad\qquad \ \&\;   \text{cycl.}\,,\nonumber\\
    s_{14}&= s_{23}+s_{56}-s_{123}-s_{234}\,,\;\;\quad \&\;  \text{cycl.}\,,
\end{align}
and the traces ${\rm tr}_i$ are defined as in~\cite{Henn:2022ydo},
\begin{equation}
    {\rm tr}_6=\sqrt{\Delta_3\left(s_{12}s_{34},s_{13}s_{24},s_{23}s_{14}\right)} \,,
\end{equation}
together with its cyclic images $\text{tr}_{i}=\tau^i(\text{tr}_6)$, with $\tau(p_i)=p_{i{+}1}$, where $\Delta_3(x,y,z)$ is the Källén function. The six-point Born level contribution acquires explicitly cyclic symmetric form,
\begin{equation}
\begin{aligned}
    \Omega_{6}^{(0)} \rightarrow \ &\frac{1}{2}\; \Big(
    s_{12} s_{45}+s_{23} s_{56} + s_{16} s_{34}
    -s_{12} s_{23}-s_{23} s_{34}-s_{34}
   s_{45}-s_{45} s_{56}-s_{56} s_{16}-s_{12} s_{16}\\
& -s_{123} s_{234}-s_{345} s_{234}-s_{123} s_{345}
    + 
   {\rm tr}_2 + {\rm tr}_4 + {\rm tr}_6\Big) \, .
\end{aligned}
\end{equation}
\section{The d'Alembertian operator}
\label{app:da}

Here we summarize elements of \cite{Chicherin:2024hes} about the second order differential equations which are relevant to our discussion.
The ladder negative geometries, as well as all negative geometries whose underlying graph contains the unintegrated loop $AB$ as a pending node (i.e. as a vertex of degree one), satisfy a second order differential equation coming from the fundamental solution to the d'Alembert operator
\begin{equation}
    \Box_{x_0} \frac{1}{(x_0-x_{*})^2} = -4i \pi^2 \delta^{4}(x_0-x_{*}) \ ,
\end{equation}
for any dual momentum point $x_{*} \in \mathbb{C}^4$, where $x_0$ is the dual point to $AB$. 
By noting that the integrand of $f^{(L)}_{n,ij}$ contains an overall factor $\frac{\langle AB ij \rangle}{\langle ABCD \rangle}$, and taking $x_{*}$ to be dual to $(ij)$ we obtain $\frac{n(n-3)}{2}$ differential operators
\begin{equation}
    \mathcal{D}_{ij} := (x_0-x_{*})^2 \, \Box_{x_0} \, \frac{1}{(x_0-x_{*})^2} \quad \text{for any arc $(ij)$ of an $n$-gon,} 
\end{equation}
acting recursively on the pure functions which represent loop corrections \cite{Chicherin:2024hes} as
\begin{equation}\label{DE}
\begin{aligned}
    \mathcal{D}_{ij} \, f^{(L)}_{n,ij} &= -4 \sum_{kl} \overline{\mathcal{C}}_{kl,ij} \, f^{(L-1)}_{n,kl} \ , \quad L>1 \ ,\\  
    \mathcal{D}_{ij} \, f^{(1)}_{n,ij} &= 4 \, \overline{\mathcal{C}}_{i+1j+1,ij}  \ ,
\end{aligned}
\end{equation}
where $\overline{\mathcal{C}}_{ij,kl} = \overline{\mathcal{C}}_{ij,kl}(AB)$ are obtained from ${\mathcal{C}}_{ij,kl}(CD,AB) $ by taking the residue at $CD = AB$,
\begin{equation}
    \overline{\mathcal{C}}_{ij,kl}(AB) := \langle AB \rangle^2 \, [\langle ABCD \rangle \, {\mathcal{C}}_{ij,kl}(CD,AB)]\big|_{CD=AB} \ .
\end{equation}
Note that in the frame where $AB \rightarrow I_{\infty}$, we have $\langle AB \rangle^2=1$ and $ \mathcal{C}_{ij,kl}$ at five points and six points can be found in \cite{Chicherin:2024hes} and the expressions \eqref{Cs_n6}.

For instance, the differential equations \eqref{DE} at $L=2$ read
\begin{equation}\label{L2_DE}
\begin{aligned}
    \mathcal{D}_{ij} \, f^{(2)}_{n,ij} &= -4 \sum_{kl} \overline{\mathcal{C}}_{kl,ij} \, I^{cp}_{kl} \  .
\end{aligned}
\end{equation}
where the integrated chiral pentagons $I^{cp}_{kl}$ can be found in \eqref{chir_pent}.

\section{Landau singularities from super-sector integrals}
\label{app:super}
During the calculation of all singularities of two-loop ladders in Section \ref{sec:all_letter_ladder}, there is a distinct type of singularity we want to comment on. For that, we consider the following double-box integral with general legs
\begin{center}
    \begin{tikzpicture}[baseline={([yshift=-0.6ex]current bounding box.center)},scale=0.3]
\draw[black,thick] (0,0)--(10,0)--(10,5)--(0,5)--(0,0);
\draw[black,thick](5,0)--(5,5);
\draw[black,thick](-1,0)--(0,0)--(0,-1);
\draw[black,thick](0,5)--(-0.85,5.85);
\draw[black,thick](4,-1)--(5,0)--(6,-1);
\draw[black,thick](10,-1)--(10,0)--(11,0);
\draw[black,thick](10,6)--(10,5)--(11,5);
\draw[black,thick](4,6)--(5,5)--(6,6);
\filldraw[black] (0,-1) node[anchor=north] {{$i$}};
\filldraw[black] (4,-1) node[anchor=north east] {{$i{+}1$}};
\filldraw[black] (6,-1) node[anchor=north west] {{$k$}};
\filldraw[black] (10,-1) node[anchor=north] {{$k{+}1$}};
\filldraw[black] (-1,0) node[anchor=east] {{$j{+}1$}};
\filldraw[black] (-0.85,5.85) node[anchor=south east] {{$j$}};
\filldraw[black] (4,6) node[anchor=south] {{$j{-}1$}};
\filldraw[black] (11,5) node[anchor=west] {{$l{+}1$}};
\filldraw[black] (11,0) node[anchor=west] {{$l$}};
\filldraw[black] (10,6) node[anchor=south] {{$m$}};
\filldraw[black] (6,6) node[anchor=south west] {{$m{+}1$}};
\filldraw[black] (7.5,2.5) node {{$AB$}};
\filldraw[black] (2.5,2.5) node {{$CD$}};
\end{tikzpicture} \quad$\xrightarrow{\text{super-sector }}$\quad \begin{tikzpicture}[baseline={([yshift=-0.5ex]current bounding box.center)},scale=0.3]
                \draw[black,thick] (0,0)--(0,5)--(4.76,6.55)--(7.69,2.5)--(4.76,-1.55)--cycle;
                \draw[black,thick] (9.43,1.5)--(7.69,2.5)--(9.43,3.5);
                \draw[black,thick] (4.37,8.45)--(4.76,6.55)--(5.97,8.45);
                \draw[black,thick] (4.27,-3.75)--(4.76,-1.55)--(5.67,-3.45);
                \draw[black,thick] (0,0)--(0,5)--(-5,5)--(-5,0)--cycle;
                \draw[black,thick] (-5,5)--(-6.62,6.43);             
                \draw[black,thick] (-6.93,-0.52)--(-5,0)--(-5.52,-1.93);
                \draw[black,thick] (-1.3,6.6)--(0,5)--(1.3,6.6);
                \filldraw[black] (4.37,8.45) node[anchor=south] {{$m$}};
                \filldraw[black] (1.3,6.6) node[anchor=south] {{$m{+}1$}};
                \filldraw[black] (-1.3,6.6) node[anchor=south east] {{$j{-}1$}};
                \filldraw[black] (5.97,8.45) node[anchor=south west] {{$l{+}1$}};
            \filldraw[black](4.27,-3.75) node[anchor=east] {{$i{+}1$}};
            \filldraw[black](-6.93,-0.52) node[anchor=east] {{$j{+}1$}};
            \filldraw[black](-5.52,-1.93) node[anchor=north] {{$i$}};
            \filldraw[black](-6.62,6.43) node[anchor=south east] {{$j$}};
            \filldraw[black] (5.67,-3.45) node[anchor=north west] {{$k$}};
            \filldraw[black] (9.43,1.5) node[anchor=west] {{$k{+}1$}};
            \filldraw[black] (9.43,3.5) node[anchor=west] {{$l$}};
            \filldraw[black] (3.5,2.5) node {{$AB$}};
\filldraw[black] (-2.5,2.5) node {{$CD$}};
            \end{tikzpicture} 
\end{center}
Special cases of this topology played an important role in our discussion in the main text. Note that if the leg $j$ was also massive, then this topology would yield elliptic cut. For this double-box it is not the case. We solve the associated Landau equations with a loop-by-loop approach. On the support of the cut conditions for $CD$, the pinch condition for $CD$ becomes
\begin{equation}\label{eq:super}
    \left(\begin{matrix}
        0&\langle ABii{+}1\rangle &\langle ABj{-}1j\rangle& \langle ABjj{+}1\rangle\\
       \langle ABii{+}1\rangle & 0 &\langle ii{+}1j{-}1j\rangle &\langle ii{+}1jj{+}1\rangle\\
       \langle ABj{-}1j\rangle& \langle ii{+}1j{-}1j\rangle&0&0\\
       \langle ABjj{+}1\rangle&\langle ii{+}1jj{+}1\rangle&0&0
    \end{matrix}\right)\cdot \left(\begin{matrix}
        \alpha_1\\\alpha_2\\\alpha_3\\\alpha_4
    \end{matrix}\right)=0 \ ,
\end{equation}
which means that the determinant of the matrix above vanishes:
\begin{equation}
    \langle AB (ii{+}1j)\cap\bar j\rangle=0 \ .
\end{equation}
Therefore, the cut and pinch conditions for $AB$ form a  ``one-loop box" problem, whose leading singularity yields a square root. One of these special cases gives the square root \eqref{eq:d11}.

However, as discussed in \cite{He:2024fij}, the above singularity only 
ensures the existence of solutions for $\alpha_i$ in \eqref{eq:super}. Requiring all $\alpha_i\neq0$, i.e. a Landau singularity for a leading solution branch, actually yields stronger constraints on the kinematics:
\begin{equation}
    \langle AB ii{+}1\rangle=\langle AB (ii{+}1j)\cap\bar j\rangle=0 \ .
\end{equation}
In other words, we have a singularity from the following conditions
\begin{equation}
    \langle AB kk{+}1\rangle=\langle AB ll{+}1\rangle=\langle AB mm{+}1\rangle=\langle AB ii{+}1\rangle=\langle AB (ii{+}1j)\cap\bar j\rangle=0 \ ,
\end{equation}
which only hold when we have the following five-term Gram-determinant vanishing
\begin{equation}
    \det\{\langle X_i X_j\rangle\} \ , \quad  X_i\in \{ (kk{+}1),\, (ll{+}1),\, (mm{+}1),\, (ii{+}1), (ii{+}1j)\cap\bar j \} \ .
\end{equation}
This is family $6$ of singularities in Section \ref{sec:all_letter_ladder}. Since $\langle ABii{+}1\rangle=0$ is also put on-shell while solving the Landau equations, this singularity can be treated as a leading Landau singularity for a penta-box with an extra propagator added. From the differential equation point of view this is also natural, as integration-by-part identities for the double-box integral family above yields an integral basis which is IBP-related to the super-sector integrals in the penta-box family, which is very important and helpful for constructing UT basis \cite{Dlapa:2021qsl,He:2022ctv}. However, in the data provided from differential equations for one-mass five-point or six-point scattering processes \cite{Abreu:2020jxa,Henn:2025xrc}, all this kind of singularities from super-sectors always appears only beyond weight four. Therefore, they do not appear in the finite observable. In our calculation, we also confirm this from the fact that in the six-point two-loop ladder result this kind of singularities are absent.

\bibliographystyle{JHEP}
\bibliography{refs}

\end{document}